\documentclass[a4paper,11pt]{article}

\usepackage[utf8x]{inputenc}
\usepackage[left=2cm,right=2cm,top=2cm,bottom=3cm]{geometry}
\usepackage[dvipsnames]{xcolor}
\usepackage[linktoc=all,
        hidelinks,
        backref=page,
        colorlinks=true,
        pdfstartview=FitV,
        linkcolor=ColorLink,
        citecolor=ColorCite,
        urlcolor=ColorURL,
        hyperindex=true,
        hyperfigures=false%
    ]{hyperref}
\usepackage{cite}
\usepackage{amsmath}
\usepackage{amsfonts}
\usepackage{amssymb}
\usepackage{amsthm}
\usepackage{appendix}
\usepackage{caption}
\usepackage{tensor}
\usepackage{longtable}
\usepackage{lscape}
\usepackage{tikz}
\usepackage[smalltableaux,centertableaux,centerboxes]{ytableau}
\usepackage{dynkin-diagrams}
\usepackage{graphicx}
\usepackage{afterpage}
\usepackage[shortlabels]{enumitem}
\usepackage{setspace}
\usepackage[nottoc]{tocbibind} 

\definecolor{arXiv}{named}{Maroon}
\definecolor{ColorCite}{named}{BrickRed}
\definecolor{ColorLink}{named}{black}
\definecolor{ColorURL}{named}{RoyalBlue}
\definecolor{ColorMail}{named}{BrickRed}

\newcommand{\myfontbackref}[1]{
    \mbox{\small #1}
}
\renewcommand*{\backref}[1]{}
\renewcommand*{\backrefalt}[4]{%
\ifcase #1 \myfontbackref{no citations}
    \or \myfontbackref{\!\!(Page {\hypersetup{linkcolor=black}#2})}
    \else \myfontbackref{\!\!(Pages {\hypersetup{linkcolor=black}#2})}
\fi
}

\newcommand{\rd}{\mathrm{d}}
\newcommand{\rT}{\mathrm{T}}

\newcommand{\ad}{\mathrm{ad}}
\newcommand{\Ad}{\mathrm{Ad}}

\newcommand{\cA}{\mathcal{A}}
\newcommand{\cD}{\mathcal{D}}
\newcommand{\cE}{\mathcal{E}}
\newcommand{\cF}{\mathcal{F}}
\newcommand{\cG}{\mathcal{G}}
\newcommand{\cL}{\mathcal{L}}
\newcommand{\cH}{\mathcal{H}}
\newcommand{\cM}{\mathcal{M}}

\newcommand{\cO}{\mathcal{O}}

\newcommand{\cR}{\mathcal{R}}

\newcommand{\cV}{\mathcal{V}}

\DeclareRobustCommand{\sscdots}{%
  \vbox{%
    \baselineskip=0.125\normalbaselineskip
    \hbox{.\hspace{-0.25mm}.\hspace{-0.25mm}.}%
    \kern-0.2\baselineskip
  }%
}

\numberwithin{equation}{section}

\usepackage{caption}
\begin{document}

\thispagestyle{empty}
\begin{center}

\vspace*{12mm}
{\Large \bf
Duality-covariant particles and exotic branes
}\\

\vspace{12mm}

\normalsize
{\large
Josh O'Connor${}^a$ and David Osten${}^b$
}

\vspace{8mm}

\href{mailto:jo'connor@irb.hr}{\texttt{jo'connor@irb.hr}}
\;\;
\href{mailto:david.osten@uwr.edu.pl}{\texttt{david.osten@uwr.edu.pl}}

\vspace{8mm}

\!\!\!${}^a${\it Division of Theoretical Physics,
Rudjer Bo{\v{s}}kovi{\'c} Institute\\
Bijeni{\v{c}}ka 54, 10000 Zagreb, Croatia}\\
\vspace{3mm}
\!\!\!${}^b${\it Institute for Theoretical Physics (IFT), University of Wroc\l{}aw\\
pl.~Maxa Borna 9, 50-204 Wroc\l{}aw, Poland}

\vspace{20mm}

{\sc Abstract}

\vspace{4mm}

\begin{tabular}{p{14cm}}

In this paper we construct duality-covariant worldvolume dynamics of particles and branes.
We extend known actions and phase space formulations to include the hidden $E_8$ symmetry of 11D supergravity, analogous to the Ehlers symmetry of 4D gravity.
Making the worldvolume theory manifestly duality-covariant requires the ancillary structure of $E_8$ exceptional field theory to be taken into account.
For zero-branes, we propose an enlarged worldline model with a coadjoint orbit term to encode this.
More generally, we propose worldvolume theories for arbitrary exotic branes in a way that generalises the known gauged sigma model of the Kaluza--Klein monopole.
These are natural in the duality-covariant Hamiltonian formulation employed here.
We discuss the case of zero-branes in eleven dimensions as an illustrative example.

\end{tabular}

\end{center}

\newpage

\hrule

\setcounter{tocdepth}{2}
\tableofcontents

\null\hrule

\section{Introduction}

String theory and M-theory contain extended objects whose charges are organised in terms of duality symmetries.
These symmetries are hidden in supergravity, and they are revealed after compactifying, dualising fields, and arranging the charges into multiplets.
Double field theory (DFT) and exceptional field theory (ExFT) provide a way of making these symmetries manifest before compactification \cite{Hull:2009mi,Hohm:2010pp,Aldazabal:2013sca,Hohm:2013pua,Hohm:2013vpa,Hohm:2013uia,Hohm:2014fxa,Berman:2020tqn}.
The standard spacetime is split into the external dimensions and internal (i.e.~would-be compactified) dimensions.
In DFT and ExFT, the internal coordinates are replaced by generalised coordinates that transform linearly under the relevant T- or U-duality group.
The physical spacetime is recovered by imposing the section constraint to ensure that not all generalised coordinates are physical.
Metric and $p$\hspace{0.4mm}-form degrees of freedom are combined into duality-covariant fields like the generalised metric.

This reformulation is useful not only for supergravity fields.
It also lets us examine the dynamics of probes that carry momentum, winding, or wrapping charges.
The duality-covariant worldline action in \cite{Blair:2017gwn} describes a particle moving around in the extended spacetime of DFT or ExFT.
Upon choosing a physical section, one recovers either the known massless particle actions in ten or eleven dimensions or the massive particles in lower dimensions whose masses are determined by momenta along the dual directions.
In this way, just as DFT and ExFT unify supergravity theories with their compactifications and with other supergravities that can only be obtained via dualities, such as 11D supergravity and type IIB supergravity, the duality-covariant worldline model unifies all particles inside these theories.
Similarly, the duality-covariant worldsheet model \cite{Arvanitakis:2017hwb,Arvanitakis:2018hfn} is a string that is coupled to a DFT/ExFT background, reducing on different physical sections to all the known strings and one-branes in ten and eleven dimensions.

A class of objects for which the worldvolume interpretation remains much less developed is that of exotic branes, i.e.~objects of low codimension whose supergravity descriptions are often only locally geometric \cite{deBoer:2012ma}.
After travelling in a loop around an exotic brane, the fields may return to themselves only up to a duality transformation rather than an ordinary diffeomorphism or gauge transformation.
This makes them natural objects for DFT and ExFT, where duality transformations act as part of the (generalised) geometry of generalised spacetime \cite{West:2003fc,Hull:2007zu,Hohm:2010jy,Coimbra:2011ky,Coimbra:2011nw,Coimbra:2012af}.
Exotic brane backgrounds have already been investigated in the context of DFT and ExFT \cite{Berkeley:2014nza,Berman:2014jsa,Berman:2014hna,Bakhmatov:2016kfn,Bakhmatov:2017les,Berman:2018okd,Berman:2019zcq}.
This suggests that exotic branes should admit a more intrinsic duality-covariant description as dynamical probes.

The aim of this paper is to take a step in that direction.
We propose duality-covariant worldvolume descriptions for exotic branes.
The characteristic smearing of an exotic brane should be visible directly in the probe action.
Here we will focus mainly on zero-branes.
For a standard Kaluza--Klein particle, momentum along an internal circle becomes a charge after compactification.
For an exotic object, the relevant isometry directions are gauged on the worldvolume along the lines of \cite{Bergshoeff:1997gy,Kimura:2014upa,Kimura:2016anf}.
The brane is delocalised along these directions, yet their radii still enter the effective tension.
This leads to a simple form of our exotic brane Hamiltonian:~physical momenta are constrained along the gauged isometries, and the mass term contains appropriate powers of the corresponding radii.

A second goal is to express exotic brane dynamics in terms of a duality-covariant phase space.
For strings, the relationship between worldsheet currents and $\mathrm{O}(d,d)$-covariance plays an important role in doubled formulations \cite{Tseytlin:1990nb,Siegel:1993bj,Alekseev:2004np}.
Here we consider the same for $p$\hspace{0.4mm}-branes and exceptional U-duality groups.
The phase space carries currents valued in the tensor hierarchy, i.e.~a tower of representations of the duality group, and their Poisson brackets reproduce the algebraic structures appearing in DFT and ExFT.
Two main advantages of this formulation are:~(1) the natural duality-covariant Hamiltonian that is quadratic in these currents gives rise to the usual $p$\hspace{0.4mm}-brane actions \cite{Hatsuda:2012uk,Hatsuda:2012vm,Hatsuda:2013dya,Osten:2021fil,Osten:2024mjt}, and (2) a hierarchy of constraints arises that is equivalent to the $\frac{1}{2}$-BPS brane scan \cite{Blair:2017gwn,Arvanitakis:2017hwb,Osten:2021fil,Osten:2023iwc,Osten:2024mjt}.

The case of $E_8$ is slightly trickier.
In addition to the usual gauge parameter $\Lambda^M$ in $E_8$ ExFT, the generalised diffeomorphism algebra requires a constrained ancillary parameter $\Sigma_M$.
Correspondingly, the background contains not only the one-form gauge field $A_\mu{}^M$ but also a constrained one-form $B_{\mu M}$.
This new parameter is part of a St\"uckelberg mechanism that shifts away any components of the dual graviton.
A worldline action that couples only through a fixed momentum $P_M$ cannot see the full $E_8$ gauge structure.
We therefore introduce an internal worldline degree of freedom $Q^M$ that is realised by a coadjoint orbit term
\begin{equation}
    S_\mathrm{orb} = \int\!\rd\tau \, Q^M \big( B_{\mu M}\dot X^\mu + W_M \big) \,,
\end{equation}
where $W_M$ is a new constrained worldline field.
This term is analogous to a particle with colour or spin coupling to a gauge connection, but now adapted to $E_8$ ExFT.
We test this construction against known backgrounds, such as the M0-brane and the exotic zero-brane $0^{(1,7)}$ in eleven dimensions.
The former is governed by the usual momentum charge and reduces to the standard point particle, while the latter is controlled by a dual gravity scalar and has a non-trivial monodromy.

The paper is organised as follows.
In Section~\ref{sec:review_exceptional_fields_particles}, we review DFT and ExFT, the tensor hierarchy, and the known duality-covariant particle actions.
Section~\ref{sec:branes_overview_conjecture} outlines the standard and exotic branes in M-theory, and proposes a general worldvolume description for their dynamics.
We then develop the duality-covariant phase space formalism in Section~\ref{sec:Hamiltonian}, and we derive the current algebra underlying the Hamiltonians.
In Section~\ref{sec:E8_worldline_action} we extend the duality-covariant action of \cite{Blair:2017gwn} to the $E_8$ case, where we explain how the constrained fields and parameters are incorporated.
This is applied to the M0-brane and $0^{(1,7)}$-brane backgrounds in Section~\ref{sec:from_E8_to_11D}.
A discussion of our results and important open questions is found in Section~\ref{sec:conclusion}.
Lastly, in Appendix~\ref{app:deformation}, we work out some details on a quadratic deformation of the coadjoint orbit term, whose phase space is enlarged to a cotangent-bundle over the original orbit.

\section{Review of exceptional fields and particles}
\label{sec:review_exceptional_fields_particles}

In order to make manifest the hidden duality symmetries in theories of gravity and supergravity, both double field theory and exceptional field theory were developed \cite{Hull:2009mi,Hohm:2010pp,Hohm:2013pua,Hohm:2013vpa,Hohm:2013uia,Hohm:2014fxa}.
These become visible global symmetries after compactifying on a torus, say, and by dualising certain fields.
For example, type II supergravity exhibits T-duality, while maximal supergravity in eleven dimensions leads to exceptional U-duality groups.
One can understand these hidden symmetries as originating from transformations of momenta and string winding modes or brane wrapping charges, and in order for them to become manifest in the associated low-energy (super)gravity theories, DFT and ExFT are formulated in terms of an enlarged duality-covariant generalised spacetime on which the symmetry groups act linearly.

In DFT the internal coordinates $y^m$ are supplemented by dual coordinates $\tilde{y}_m$ so that the doubled coordinate $y^M=(y^m,\tilde{y}_m)$ transforms in the fundamental representation of $\mathrm{O}(d,d)$.
In ExFT, the $y^M$ transforms in a representation of an exceptional Lie group, for instance the $\mathbf{56}$ of $E_7$ or the adjoint $\mathbf{248}$ of $E_8$.
One imposes the section constraint to restrict the dependence of the fields and parameters on the generalised coordinates so that not all of them are physical.
Solving the section constraint in different ways, DFT reduces either to type II supergravity or to its compactifications.
Similarly, ExFT reduces to 11D supergravity, type IIB supergravity, or their compactifications.
In this way, DFT and ExFT unify supergravity theories in a way that makes T-duality and U-duality manifest.

These generalised theories replace the usual internal metric and gauge potentials by a generalised metric that parametrises a coset $G/K$ with $G$ the relevant duality group and $K$ its maximal compact subgroup.
As such, in DFT, $E_7$ ExFT, and $E_8$ ExFT, one has
\begin{align}
  \cH_{MN} &\in \frac{\mathrm{O}(d,d)}{\mathrm{O}(d)\times\mathrm{O}(d)} \,,&
  \cM_{MN} &\in \frac{E_7}{\mathrm{SU}(8)} \,,&
  \cM_{MN} &\in \frac{E_8}{\mathrm{SO}(16)} \,.
\end{align}
These objects package the internal metric $g_{mn}$ together with form-field degrees of freedom, such as the Kalb--Ramond two-form $B_{mn}$ in DFT.
In the $E_7$ theory, the generalised metric encodes the internal components of the metric, the three-form field, and the six-form dual potential.
In the $E_8$ theory, the scalar coset is even larger and includes dual graviton components.

\subsection{Review of $E_8$ exceptional field theory}
\label{subsec:E8_ExFT}

To begin, we review the dynamics and gauge structure of $E_8$ exceptional field theory \cite{Hohm:2014fxa,Baguet:2016jph} that we will need in order to write down an $E_8$-covariant particle dynamics in Sections~\ref{sec:Hamiltonian} and ~\ref{sec:E8_worldline_action}.

Let $M,N,\ldots=1,\ldots,248$ denote adjoint indices of $\mathfrak{e}_8$ with structure constants $f^{MN}{}_P=-f^{NM}{}_P$ and Killing metric $\kappa^{MN}$.
A projector that will be used later is $\mathbb{P}^{M}{}_{N}{}^{P}{}_{Q}:=\frac{1}{60}\,f^{M}{}_{NL}f^{LP}{}_{Q}$\,.
For the $E_8$ generalised Lie derivative we will also need the projector onto the $\mathbf{3875}\subset(\mathbf{248}\otimes\mathbf{248})_\mathrm{sym}$ denoted by $(\mathbb{P}_\mathbf{3875})^{MN}{}_{PQ}$ whose explicit expression can be found in \cite{Hohm:2014fxa}.

The fields of $E_8$ ExFT depend on an extended spacetime with two types of coordinates: external coordinates $x^\mu$ ($\mu,\nu,\ldots=0,1,2$) and internal coordinates $y^M\in\mathbf{248}$.
Derivatives with respect to $y^M$ are denoted by $\partial_M$.
One imposes the section constraint for the generalised diffeomorphism algebra to close.
This tells us that the projection of two internal derivatives onto the $\mathbf{1}\oplus\mathbf{248}\oplus\mathbf{3875}$ vanishes:
\begin{equation}
    \big(\mathbb{P}_{\mathbf{1}\,\oplus\,\mathbf{248}\,\oplus\,\mathbf{3875}}\big)^{MN}{}_{PQ}\,\partial_M\otimes\partial_N = 0 \,.
\label{eq:SC_proj}
\end{equation}
More explicitly, for any two fields $\Phi$ and $\Psi$ one has
\begin{align}
    \kappa^{MN} \partial_M \Phi\,\partial_N \Psi &= 0 \,,&
    f^{MN}{}_P\,\partial_M \Phi\,\partial_N \Psi &= 0 \,,&
    (\mathbb{P}_\mathbf{3875})^{MN}{}_{PQ} \, \partial_M \Phi\,\partial_N \Psi &= 0 \,.
    \label{eq:SC_1_248_3875}
\end{align}
Not only does this hold for pairs of derivatives acting on fields, but $E_8$ ExFT also features fields and parameters that carry a `constrained index' in the sense that they obey the same constraint.
Instead of equation \eqref{eq:SC_proj}, it would be more complete to write
\begin{equation}
    \big(\mathbb{P}_{\mathbf{1}\,\oplus\,\mathbf{248}\,\oplus\,\mathbf{3875}}\big){}_{MN}{}^{PQ}\,C_P\otimes \tilde{C}_Q = 0 \,,
    \label{eq:SC_generic}
\end{equation}
for any pair of objects $C$ and $\tilde{C}$ that carry a constrained index.
This holds for any pair of derivatives acting on one field or for any such pair acting on two different fields.
Sometimes these are referred to as the weak section constraint and the strong section constraint.

A choice of physical spacetime is a solution to the section constraint \eqref{eq:SC_generic}.
Different solutions tell us that the $(3+248)$-dimensional spacetime reduces either to the $3+8=11$ dimensions of maximal supergravity, or to the $3+7=10$ dimensions of type IIB supergravity, or to the spacetime relevant to their compactifications all the way down to $3+0=3$ dimensions.
A larger U-duality group means that even more compactifications can be unified with the original theories, such as $E_9$ ExFT and the supergravity theories in 2D \cite{Bossard:2018utw,Bossard:2021jix}, while the maximal unification is given by $E_{11}$ ExFT \cite{Bossard:2019ksx,Bossard:2021ebg}.

The local transformations of the theory are called generalised diffeomorphisms, and in the $E_8$ case they feature a pair of gauge parameters $(\Lambda^M,\Sigma_M)$ where $\Sigma$ carries a constrained index.
The action of $\Lambda$ transformations on the external and internal coordinates is
\begin{align}
    \delta_\Lambda x^\mu &= 0 \,,&
    \delta_\Lambda y^M &= -\Lambda^M \,,
\end{align}
while the fields transform as follows.
Let $V^M$ be an adjoint vector of generalised density weight $\lambda_{(V)}$.
The $E_8$ generalised Lie derivative acting on $V^M$ is \cite{Hohm:2014fxa}
\begin{align}
    \cL_{(\Lambda,\Sigma)} V^M
    = \Lambda^N \partial_N V^M - 60\,\mathbb{P}^M{}_N{}^P{}_Q (\partial_P \Lambda^Q) V^N + \lambda_{(V)} (\partial_N\Lambda^N) V^M + f^{MN}{}_{P}\,\Sigma_N V^P\,.
    \label{eq:gen_Lie_deriv}
\end{align}
The transformations of all ExFT fields under generalised diffeomorphisms are built from the action of the generalised Lie derivative on the adjoint representation $\mathbf{248}$ and its dual.
For covectors one uses the corresponding contragredient action with the same $\Sigma$ term.
The one-form field $A_\mu{}^M$ and its parameter $\Lambda^M$ carry weight one while the constrained field $B_{\mu M}$ and parameter $\Sigma_M$ carry weight zero.
The `ancillary' parameter $\Sigma_M$ is required for the commutator of two generalised Lie derivatives $\cL_{(\Lambda_1,\Sigma_1)}$ and $\cL_{(\Lambda_2,\Sigma_2)}$ to close into a generalised Lie derivative $\cL_{(\Lambda_{12},\Sigma_{12})}$ with parameters $(\Lambda_{12},\Sigma_{12})$ bilinear in $(\Lambda_i,\Sigma_i)$.

The fields of $E_8$ ExFT include the external dreibein $e_\mu{}^a$, internal `248-bein' $\cV_M{}^A\in E_8/\mathrm{SO}(16)$, the one-form field $A_\mu{}^M$, and the constrained one-form $B_{\mu M}$.
The external metric is $g_{\mu\nu}=e_\mu{}^ae_\nu{}^b\eta_{ab}$ while the scalars are encoded in the coset representative $\cV_M{}^A$ that is used to construct the internal metric $\cM_{MN}=\cV_M{}^A\cV_N{}^B\delta_{AB}$.
Dual graviton components in $E_8$ ExFT that are shifted away with the $\Sigma_M$ St\"uckelberg symmetry were discussed in \cite{Hohm:2014fxa}.
This mechanism had previously been worked out for dual gravity in \cite{Boulanger:2008nd}.
Its relation to $E_8$ ExFT was investigated further in \cite{Hohm:2018qhd} and the non-linear parent Lagrangian of \cite{Boulanger:2008nd} was later recovered from the $E_{11}$ ExFT pseudo-Lagrangian in \cite{Bossard:2021ebg}.

The covariant external derivative of any adjoint tensor $T$ of weight $\lambda_{(T)}$ is defined by
\begin{align}
    D_\mu T := \partial_\mu T - \cL_{(A_\mu,B_\mu)} T \,,\label{eq:cov_deriv}
\end{align}
where $(A_\mu,B_\mu)$ is viewed as a $(\Lambda,\Sigma)$-type pair of parameters in equation \eqref{eq:gen_Lie_deriv}.
This definition ensures that $D_\mu T$ transforms covariantly as a tensor of the same weight.
The transformations of the two gauge fields $A_\mu{}^M$ and $B_{\mu M}$ are given by
\begin{align}
    \delta A_\mu{}^{M} &= D_\mu^{(1)} \Lambda^M \,,&
    \delta B_{\mu M} &= D_\mu^{(0)} \Sigma_M - \Lambda^N \partial_M B_{\mu N} + f^N{}_{PQ}\,\Lambda^P \partial_M \partial_N A_\mu{}^Q \,.\label{eq:delta_A_delta_B}
\end{align}
Here $D_\mu^{(\lambda)}$ indicates that the covariant derivative \eqref{eq:cov_deriv}
acts on an object of weight $\lambda$\,.

The Lagrangian for $E_8$ exceptional field theory reduces to the bosonic part of the 11D supergravity Lagrangian in the $(3+8)$-dimensional split when one chooses to work on the 11D solution to the section constraint \cite{Hohm:2014fxa}.
The supersymmetric completion of $E_8$ exceptional field theory was worked out in \cite{Baguet:2016jph}.
Instead of focusing on the supergravity dynamics written in terms of ExFT, in this paper we focus on worldline dynamics of particles and zero-branes in an exceptional field theory background -- see \cite{Blair:2017gwn}.

\subsubsection*{Remark on Y-tensors}

In DFT and $E_n$ ExFT for $n\leqslant7$ one can write the section constraint as
\begin{equation}
    Y^{MN}{}_{PQ} \, C_M \otimes \tilde{C}_N = 0 \,.
\end{equation}
The so-called Y-tensor $Y^{MN}{}_{PQ}$ is constructed from invariant tensors of the U-duality group \cite{Berman:2012vc}, see also \cite{Cederwall:2015ica,Cederwall:2017fjm}.
Generalised diffeomorphisms take a simple form in terms of the Y-tensor
\begin{equation}
    \cL_\Lambda V^M = \Lambda^N \partial_N V^M - V^N \partial_N \Lambda^M + Y^{MN}{}_{PQ} \, \partial_N \Lambda^P V^Q + (\lambda_{(V)} + \omega) \, \partial_N \Lambda^N V^M \,,
\end{equation}
where $\omega$ is the intrinsic weight (zero in DFT and $\frac{1}{n-9}$ in $E_n$ ExFT), and the relevant weights are
\begin{align}
    \lambda_{(\Lambda)} = \lambda_{(A)} &= -\omega \,,&
    \lambda_{(g)} &= -2\omega \,,&
    \lambda_{(\cM)} &= 0 \,.
\end{align}
For $E_8$ one can also use a Y-tensor\footnote{There is a Y-tensor for any simply-laced Kac--Moody algebra with generalised spacetime coordinates in any highest weight module \cite{Bossard:2017aae}.} to define the section constraint \cite{Berman:2019efr}
\begin{equation}
    Y^{MN}{}_{PQ}
    = -60\,\mathbb{P}^{M}{}_{Q}{}^{N}{}_{P}
    + 2\,\delta^{(M}_{P}\delta^{N)}_{Q}
    = \big( 62\,\mathbb{P}_{\mathbf{1}}
    + 30\,\mathbb{P}_{\mathbf{248}}
    + 14\,\mathbb{P}_{\mathbf{3875}} \big)_{PQ}{}^{MN} \,.
\end{equation}
However, the generalised Lie derivative constructed from $\Lambda^M$ and the $E_8$ Y-tensor alone does not form a closed algebra.
It must be supplemented by the covariantly constrained parameter $\Sigma_M$.

\subsubsection*{The tensor hierarchy}

Another central input is the so-called tensor hierarchy of $E_n$ representations denoted by $\cR_1$, $\cR_2$, etc.
The gauge field $A_\mu{}^M$ belongs to $\cR_1$ which is also the representation of the internal coordinates, the two-form fields belong to $\cR_2$, the three-forms to $\cR_3$, and so on.
Abstractly, the tensor hierarchy is controlled by a nilpotent differential $\hat{\partial}:\cR_{k+1}\to\cR_k$ and by the bullet product $\bullet:\cR_k\times\cR_l\to\cR_{k+l}$.
The field strength $\cF_{\mu\nu}{}^M$ of $A_\mu{}^M$ is not the usual non-abelian curvature.
It must be completed by a $\hat{\partial}$-exact term $\hat\partial B_{\mu\nu}$ with $B_{\mu\nu}\in\cR_2$ that transforms in a way that absorbs the failure of the usual curvature to transform covariantly, as is very familiar from gauged supergravity theories.
Similarly, the three-form $C_{\mu\nu\rho}\in\cR_3$ is introduced to modify the field strength of the two-form, and so on.
In the $E_n$ case, the $(9-n)$-form in $\cR_{9-n}$ in the tensor hierarchy is accompanied by another $(9-n)$-form that carries a constrained $\cR_1$ index.
For example, the $E_7$ tensor hierarchy contains a one-form $A_\mu{}^M\in\mathbf{56}$, a two-form $B_{\mu\nu}{}^\alpha\in\mathbf{133}$, and a constrained two-form $B_{\mu\nu M}\in\overline{\mathbf{56}}$.
The $E_8$ tensor hierarchy includes $A_\mu{}^M\in\mathbf{248}$ as usual and also the aforementioned constrained one-form $B_{\mu M}\in\overline{\mathbf{248}}$.

Sometimes we will express the action of $\hat{\partial}$ and $\bullet$ product in terms of $\eta$- and $D$-symbols as
\begin{align}
    (\hat{\partial} A)^{M_{k-1}} &= D_{N_k}{}^{M_{k-1}P} \partial_P A^{N_k} \,,&
    (A \bullet B)^{M_{k+l}} &= \eta^{M_{k+l}}{}_{N_kP_l} A^{N_k} B^{P_l} \,,
    \label{eq:eta_D_symbols}
\end{align}
for $A\in\cR_k$ and $B\in\cR_l$.
Here, $M_k,N_k,\ldots=1,\dots,\dim\cR_k$ denote the $\cR_k$ representation indices, and as a shorthand we write $M,N,P,\dots$ to denote $\cR_1$ indices.
Note that the Y-tensor can be written in terms of the symbols in \eqref{eq:eta_D_symbols} as $Y^{MN}{}_{PQ}=D_{L_2}{}^{MN}\eta^{L_2}{}_{PQ}$ \cite{Osten:2024mjt}.

For duality groups in DFT and ExFT, the $\eta$- and $D$-symbols have been derived explicitly in a case by case analysis \cite{Hohm:2013vpa,Hohm:2013uia,Hohm:2014fxa,Abzalov:2015ega,Musaev:2015ces,Hohm:2015xna,Wang:2015hca}, or can be understood from other underlying structures \cite{Palmkvist:2013vya,Cederwall:2018aab,Cederwall:2019qnw,Cederwall:2019bai}, as more generally, the tensor hierarchy can be understood as a differential graded Lie algebra -- see \cite{Bonezzi:2019ygf,Lavau:2019oja,Bonezzi:2019bek,Arvanitakis:2018cyo,Osten:2023iwc}.
This means that there are several consistency conditions on $\hat{\partial}$ and $\bullet$ which can be understood as the (graded) Jacobi identities and Leibniz identities of the underlying differential graded Lie algebra.
This structure will be realised on the phase space of branes, as explained in Section~\ref{sec:Hamiltonian}.

\subsection{Duality-covariant particle actions}
\label{subsec:exceptional-particle}

Here we summarise the worldline action for a particle in the extended spacetime of DFT or $E_n$ ExFT for $n\leqslant7$ \cite{Blair:2017gwn}.
The starting point is usual Kaluza--Klein reduction in which a massless particle leads to massive particles upon compactification to lower dimensions, with masses determined by momenta along the internal directions.
The idea was to generalise this from toroidal compactification to some generalised notion of massless particle in the extended spacetime of DFT or ExFT.
Particle states in lower dimensions obtained from wrapped strings and branes are organised into representations of the U-duality group, and their charges are encoded in a generalised momentum $P_M$.
There are additional obstacles to writing down an action for an $E_8$ particle due to the constrained parameter $\Sigma_M$ and the constrained one-form $B_{\mu M}$.
We will propose a generalisation to the $E_8$ case in Section~\ref{sec:E8_worldline_action}.

The particle worldline is a map from $\tau\in[0,1]$ to $(X^\mu(\tau),Y^M(\tau))$, with embedding fields $X^\mu$ and $Y^M$ for the external coordinates $x^\mu$ and the (generalised) internal coordinates $y^M$ in the coordinate representation $\cR_1$.
The $\cR_1$ representation for $E_7$ is the $\mathbf{56}$ while for $E_8$ it is the adjoint $\mathbf{248}$ itself.
The background fields that can be pulled back to the worldline are the external metric $g_{\mu\nu}$, the internal metric $\cM_{MN}$, and the one-form $A_\mu{}^M$.
The $E_8$ tensor hierarchy includes $A_\mu{}^M\in\mathbf{248}$ as usual but also a constrained one-form $B_{\mu M}\in\overline{\mathbf{248}}$ that can also be seen by the worldline.

The exceptional particle action given in \cite{Blair:2017gwn} is, in Nambu--Goto form,
\begin{equation}
    S = \int\!\rd\tau \, \Big( {-}\sqrt{ p_M \cM^{MN} p_N } \, \sqrt{ - g_{\mu\nu} \dot{X}^\mu \dot{X}^\nu } + p_M A_\mu{}^M \dot{X}^\mu \Big) \,,
    \label{eq:S_Blair_compactified}
\end{equation}
describing `massive' particles in the external dimensions, where $\big\{g_{\mu\nu},\cM_{MN},A_\mu{}^M\big\}$ are the background ExFT fields, and $p_M$ is the fixed generalised momentum.
This action reduces to the standard particle actions coming from Kaluza--Klein momentum, string winding, and wrapped branes.

Uplifting to higher dimensions, one incorporates the internal embedding fields $Y^M$ in an action resembling that of a massless particle in extended spacetime.
However, invariance under generalised diffeomorphisms of the target theory (DFT or ExFT) tells us that the naive pullback $\dot{Y}^M+\dot{X}^\mu A_\mu{}^M$ is not enough.
One must introduce\footnote{This additional field was written as $\cA^M$ in \cite{Blair:2017gwn} but we shall write it as $V^M$ to be consistent with \cite{Arvanitakis:2018hfn}.} a new worldline gauge field $V^M$.
In Polyakov form, the action is
\begin{equation}
    S = \frac12 \int\!\rd\tau\, e^{-1} \Big( g_{\mu\nu} \, \dot{X}^\mu \dot{X}^\nu + \cM_{MN} \, \cD_\tau Y^M \, \cD_\tau Y^N \Big) \,,
    \label{eq:S_Blair}
\end{equation}
where the covariantised worldline velocity in the extended internal directions is defined by
\begin{equation}
    \cD_\tau Y^M
    := \dot{Y}^M + \dot{X}^\mu A_\mu{}^M + V^M \,,
    \label{eq:DY_Blair}
\end{equation}
and $V^M$ is the new auxiliary worldline one-form in the $\cR_1$ representation, constrained to obey
\begin{equation}
    V^M \partial_M = 0 \,.
    \label{eq:V_partial_constraint}
\end{equation}
This generalises the gauge connection in the doubled-yet-gauged formulation of DFT, where a physical point is the same as a gauge orbit in doubled internal spacetime $y^M=(y^m,y_{\tilde{m}})$ \cite{Park:2013mpa,Lee:2013hma}.
The constraint \eqref{eq:V_partial_constraint} ensures that $V^M$ gauges the unphysical directions -- see also \cite{Arvanitakis:2018hfn} for the same mechanism within a duality-covariant worldsheet sigma model.

The equation of motion for the einbein is the generalised mass shell constraint
\begin{equation}
    g_{\mu\nu} \dot{X}^\mu \dot{X}^\nu + \cM_{MN} \, \cD_\tau Y^M \, \cD_\tau Y^N = 0 \,.
\end{equation}
This supports the interpretation as a massless particle.
A tempting-yet-naive generalised line element
\begin{equation}
    \rd s^2 = g_{\mu\nu} \, \rd X^\mu \rd X^\nu + \cM_{MN} \, \big( \rd Y^M + \rd X^\mu A_\mu{}^M \big) \big( \rd Y^N + \rd X^\nu A_\nu{}^N \big)
\end{equation}
does not transform as a scalar under generalised diffeomorphisms.
The obstruction is tied to the existence of certain trivial gauge parameters.
Equivalently, the section constraint implies that $Y^M$ are identified under $Y^M\mapsto Y^M + (\hat{\partial}\lambda)^M$, where $\lambda$ is $\cR_2$-valued and we recall that $\hat{\partial}:\cR_k\to\cR_{k-1}$ is the differential on the tensor hierarchy.

The role of $V^M$ is to gauge shifts along unphysical directions, giving a worldline gauge symmetry
\begin{align}
    \delta_\lambda Y^M &= (\hat{\partial} \lambda)^M \,,&
    \delta_\lambda X^\mu &= 0 \,,&
    \delta_\lambda V^M &= -\frac{\rd}{\rd\tau}(\hat{\partial} \lambda)^M \,,
    \label{eq:delta_lambda_Y,X,V}
\end{align}
under which $\cD_\tau Y^M$ is invariant.
Importantly, the constraint \eqref{eq:V_partial_constraint} is automatic if $V^M=(\hat\partial U)^M$ for $U\in\cR_2$ with $\delta_\lambda U=-\dot{\lambda}$\,.

It was discussed in \cite{Blair:2017gwn} that \eqref{eq:V_partial_constraint} is empty in the trivial solution $\partial_M=0$ to the section constraint, and the momentum $P_M$ in the compactified theory may in principle be arbitrary.
However, in brane reductions one expects that $P_M$ satisfies something like the $\frac12$-BPS or section constraint, and these are now understood to follow from linear constraints on $V^M$.

\subsubsection*{Canonical momenta and the momentum constraint}

Define the momentum conjugate to $Y^M$ as
\begin{equation}
    P_M := \frac{\partial\cL}{\partial \dot{Y}^M}
    = e^{-1} \cM_{MN} \, \cD_\tau Y^N \,,
    \label{eq:Blair_internal_momentum}
\end{equation}
or equivalently
\begin{equation}
    \cD_\tau Y^M = e\,\cM^{MN} P_N \,.
    \label{eq:DY=MP}
\end{equation}
Since $V^M\in\cR_1$ is not arbitrary but rather constrained to lie in the image of $\hat{\partial}$, i.e.~to generate only trivial coordinate shifts, its equation of motion does not set $P_M=0$\,.
Instead, varying \eqref{eq:S_Blair} with respect to $V^M$ gives
\begin{equation}
    \delta S = \int\!\rd\tau \, P_M \, \delta V^M \,, \label{eq:Blair_eomV}
\end{equation}
and since $\delta V^M$ ranges only over unphysical directions, one obtains that $P_M$ is orthogonal to all these directions, and the Euler--Lagrange equation for $V^M$ is equivalent to the unphysical (dual) components of $P_M$ being set to zero.
In practice, this is the same as the standard section constraint in the sense that $\cR_2$-valued objects are in the kernel of $Y^{MN}{}_{PQ} \, P_M \partial_N$\,:
\begin{equation}
    Y^{MN}{}_{PQ} \, P_M \partial_N \lambda^{(PQ)} = 0 \,.
\end{equation}
For DFT, the constrained momentum $P_M$ satisfies the BPS-like constraint $P_M P^M = 2\,p_m\hspace{0.4mm}\tilde{p}^{\hspace{0.3mm}m}=0$ for $P_M=(p_m,\tilde{p}^{\hspace{0.3mm}m})$.
Essentially, the worldline symmetry removes non-BPS momentum components that would have violated the section constraint.

If the $Y^M$ enter \eqref{eq:S_Blair} only via $\dot{Y}^M$ then one can perform a partial Legendre transform (Routhian reduction) in the cyclic $Y^M$ coordinates.
Using \eqref{eq:Blair_internal_momentum} and \eqref{eq:DY=MP}, the action in first-order form is
\begin{align}
    S^{(1)}
    = \int\!\rd\tau \bigg( \frac{1}{2e} g_{\mu\nu} \dot{X}^\mu \dot{X}^\nu + P_M \cD_\tau Y^M - \frac{e}{2} \, \cM^{MN} P_M P_N \bigg) \,.
    \label{eq:S_kinetic_P}
\end{align}
The equations of motion for $Y^M$ tell us that $P_M$ is constant, and the theory reduces to a relativistic particle in the external spacetime with an effective mass squared proportional to $p_M\cM^{MN}p_N$ where $p_M$ are the conserved internal momenta.

If instead the coordinates $Y^M$ are not cyclic, then the equation of motion for $Y^M$ does not give $\dot{P}_M=0$ but it does give a force law
\begin{equation}
    \dot{P}_M = \frac{1}{2e} \partial_M g_{\mu\nu} \dot{X}^\mu \dot{X}^\nu - \frac{e}{2} \partial_M \cM^{NP} P_N P_P + \dot{X}^\mu \partial_M A_\mu{}^N P_N \,.
    \label{eq:P_dot_force}
\end{equation}
In this case, $P_M$ is not constant and no fixed-charge action follows.
The $V^M$ equation of motion still imposes the constraint $P_M\,\delta V^M=0$ such that, after choosing a section, components of $P_M$ along the unphysical directions are removed.
Of course, the complete Legendre transform is given by
\begin{equation}
    S^{(1)}
    = \int\!\rd\tau \bigg( \pi_\mu \dot{X}^\mu + P_M \cD_\tau Y^M - \frac{e}{2} \Big( g^{\mu\nu} \pi_\mu \pi_\nu + \cM^{MN} P_M P_N \Big) \bigg) \,,
    \label{eq:Blair_Legendre}
\end{equation}
or equivalently
\begin{equation}
    S^{(1)}
    = \int\!\rd\tau \bigg( P_\mu \dot{X}^\mu + P_M \big( \dot{Y}^M + V^M \big) - \frac{e}{2} \Big( g^{\mu\nu} \big( P_\mu - A_\mu{}^M P_M \big) \big( P_\nu - A_\nu{}^N P_N \big) + \cM^{MN} P_M P_N \Big) \bigg) \,,
\end{equation}
where the conjugate momentum for $X^\mu$ is
\begin{equation}
    P_\mu := \frac{\partial L}{\partial \dot{X}^\mu}
    = \frac{1}{e} g_{\mu\nu} \dot{X}^\nu + P_M A_\mu{}^M \,,
    \label{eq:Blair_external_momentum}
\end{equation}
and the covariant external momentum is given by
\begin{equation}
    \pi_\mu := P_\mu - P_M A_\mu{}^M
    = \frac{1}{e} g_{\mu\nu} \dot{X}^\nu \,.
\end{equation}
Note that $P_\mu=\pi_\mu+P_MA_\mu{}^M$ was also the conjugate momentum of $\dot{X}^\mu$ in \eqref{eq:S_kinetic_P}.
Varying with respect to $P_M$ gives \eqref{eq:Blair_internal_momentum}.
Substituting back reproduces the second-order action \eqref{eq:S_Blair}.

\subsubsection*{Summary of the symmetries of the action}

First of all, Blair's action \eqref{eq:S_Blair} respects worldline reparametrisation invariance.
The most important new symmetries are generalised diffeomorphisms pulled back from the ExFT target space.
The gauge parameter is a vector $\Lambda^M$ of weight $-\omega$ and the background fields transform by the generalised Lie derivative.
Moreover, the worldline coordinates transform as in \eqref{eq:gen_diffeo_Y_X} below.
The quantity in brackets in \eqref{eq:S_Blair} transforms as a density of weight $-2\omega$ and the variation of the worldline multiplier $e^{-1}$ must transform as a scalar of weight $+2\omega$ in order to compensate it.

The generalised diffeomorphisms of the background fields are given by
\begin{align}
    \delta_\Lambda g_{\mu\nu} &= \cL_\Lambda g_{\mu\nu} \,,&
    \delta_\Lambda \cM_{MN} &= \cL_\Lambda \cM_{MN} \,,&
    \delta_\Lambda A_\mu{}^M &= D_\mu \Lambda^M = \partial_\mu\Lambda^M - \cL_{A_\mu}\Lambda^M \,,
\end{align}
where the only parameter is $\Lambda^M\in\cR_1$.
The worldline coordinate fields transform as
\begin{align}
    \delta_\Lambda Y^M &= -\Lambda^M \,,&
    \delta_\Lambda X^\mu &= 0 \,.
    \label{eq:gen_diffeo_Y_X}
\end{align}
The transformation of $V^M$ is defined by the requirement that $\cD_\tau Y^M$ transforms covariantly:
\begin{equation}
    \delta_\Lambda (\cD_\tau Y^M) = \cL_\Lambda (\cD_\tau Y)^M \,.
\end{equation}
This uniquely fixes $\delta_\Lambda V^M$ up to trivial parameters \eqref{eq:delta_lambda_Y,X,V}.
In the absence of Scherk--Schwarz twists,
\begin{equation}
    \delta_\Lambda V^M
    = \Lambda^N \partial_N V^M - V^N \partial_N \Lambda^M + Y^{MN}{}_{PQ} \, \partial_N \Lambda^P \, \big( \dot{Y}^Q + V^Q \big) \,.
    \label{eq:gen_diffeo_V^M}
\end{equation}

In addition, we have the usual external diffeomorphisms with parameter $\xi^\mu$ that act on background fields by the ordinary Lie derivative in the external spacetime, with $\partial_\mu$ replaced by $D_\mu$.
Now, for the generalised particle this is essentially manifest:~$X^\mu$ are embedding fields for the external spacetime coordinates, $g_{\mu\nu}$ is the external metric, and $\dot{X}^\mu A_\mu{}^M$ is the pullback of the external one-form.
Hence the action is also invariant under external spacetime diffeomorphisms.

Note that $\lambda$ in \eqref{eq:delta_lambda_Y,X,V} is a worldline parameter.
In target space, there is another gauge parameter $\Xi_\mu\in\cR_2$ for which
\begin{equation}
    \delta_\Xi A_\mu{}^M = (\hat{\partial}\Xi_\mu)^M = Y^{MN}{}_{PQ} \partial_N \Xi_\mu{}^{(PQ)} \,.
\end{equation}
In DFT with T-duality group $\mathrm{O}(d,d)$, the tensor hierarchy is very simple:~$\cR_1=\mathbf{2d}$ and $\cR_2=\mathbf{1}$.
This means that $\Xi_\mu$ is an $\mathrm{O}(d,d)$ singlet, and one has $\delta_\Xi A_\mu{}^M=\partial^M\Xi_\mu$\,.
Pulling back to the worldline, this fixes $\delta_\Xi V^M=-\dot{X}^\mu\partial^M\Xi_\mu$ so that $\cD_\tau Y^M$ is invariant.
Similarly, in ExFT one has
\begin{equation}
    \delta_\Xi V^M
    = - \dot{X}^\mu \delta_\Xi A_\mu{}^M
    = - \dot{X}^\mu \, Y^{MN}{}_{PQ} \partial_N \Xi_\mu{}^{(PQ)} \,.
\end{equation}
In $E_7$ ExFT, there is a two-form $B_{\mu\nu}{}^\alpha$ in the $\cR_2$ and a second two-form $B_{\mu\nu M}$ carrying a constrained $\cR_1=\mathbf{56}$ index.
Both two-forms appear in the field strength of $A_\mu{}^M$, leading to $\Xi_\mu{}^\alpha$ in the $\cR_2$ and a second parameter $\widetilde{\Xi}_{\mu M}$ that carries a constrained index.
The worldline field $V^M$ transforms as
\begin{equation}
    \delta_\Xi V^M
    = - \dot{X}^\mu \big( \delta_\Xi A_\mu{}^M + \delta_{\widetilde{\Xi}} A_\mu{}^M \big) \,,
\end{equation}
where explicit expressions can be found in \cite{Hohm:2013uia}.

\subsubsection*{Reduction to known particle actions}

Choose a solution to the section constraint in which the fields depend on a physical subset $y^m$ of the internal coordinates $y^M$ and not on the unphysical dual coordinates $y^{\tilde{m}}$.
The constraint $V^M\partial_M=0$ then implies that only the dual components $V^{\tilde{m}}$ survive.
Integrating out the cyclic $Y^{\tilde{m}}$ coordinates, the equations of motion for $V^{\tilde{m}}$ set the conjugate momenta to zero.
The generalised massless particle action reduces to the standard action for a massless particle in ten or eleven dimensions
\begin{equation}
    S = \frac12 \int\!\rd\tau \, \hat{e}^{-1} \, \hat{g}_{\hat{\mu}\hat{\nu}} \dot{X}^{\hat{\mu}} \dot{X}^{\hat{\nu}} \,,
\end{equation}
where hatted objects are ten/eleven-dimensional.
It is useful to think of the generalised particle \eqref{eq:S_Blair} as a U-duality covariantisation of the massless particle solutions in IIA, IIB, and M-theory.

If the background is instead independent of all extended coordinates $Y^M$ then all of them can be integrated out.
One then recovers the compactified action in lower dimensions with constant conjugate momenta $p^M$ that appears in the definition of the mass.
So, the extended particle action \eqref{eq:S_Blair} unifies the massless particles in M-theory and IIB string theory with (U-duality covariantised) massless and massive Kaluza--Klein particles that are produced upon compactification to lower dimensions.

After a Scherk--Schwarz twist and reducing the twisted exceptional particle action, the action of a massive D0-brane in Romans IIA theory was recovered in \cite{Blair:2017gwn}.
This included a term with mass $m$ and an additional worldline vector originating from $V^M$ in the exceptional particle action.
In other words, despite being introduced as a technical fix, $V^M$ has a direct role in massive IIA theory.

\subsubsection*{Some comments on exceptional generalised geometry}

The constraint \eqref{eq:V_partial_constraint} has an interpretation in terms of generalised geometry \cite{Hull:2007zu,PiresPacheco:2008qik,Coimbra:2011ky,Coimbra:2011nw,Coimbra:2012af}.
In a Lie algebroid, and more generally a Leibniz or an exceptional algebroid, one has a vector bundle $E\to M$ over the physical internal spacetime manifold $M$ together with an anchor map $\rho:E\to \rT M$ projecting generalised vectors.
For example, in the $E_{7(7)}\times\mathbb{R}^+$ exceptional tangent bundle one has
\begin{equation}
    E = \rT M \oplus \Lambda^2 \rT^* M \oplus \Lambda^5 \rT^* M \oplus (\Lambda^7 \rT^*M \otimes \rT^*M) \,,
\end{equation}
and the anchor projects onto $\rT M$.
As a result, $\ker\rho$ is the subbundle of unphysical directions \cite{Bugden:2021wxg,Hulik:2023aks}:
\begin{equation}
    \ker\rho = E \ominus \rT M
    = \Lambda^2 \rT^*M \oplus \Lambda^5 \rT^*M \oplus (\Lambda^7 \rT^*M \otimes \rT^*M) \,.
\end{equation}
From this perspective, $V^M$ belongs to $\ker\rho$ pulled back to the worldline.
In the $E_8$ case, a description of the geometry in terms of Leibniz algebroids or Y-algebroids is obstructed by the presence of dual gravity \cite{Hulik:2023aks}, but the analogous idea that $V^M$ belongs to the kernel of some anchor map $\rho$ would make sense in an $E_8$ generalised geometry.

This is not the same as a generalised Killing vector $\xi^M$ that is fixed and would satisfy \cite{Grana:2008yw,Coimbra:2016ydd}
\begin{equation}
    \cL_\xi \cM_{MN} = 0 \,.
    \label{eq:generalised_Killing}
\end{equation}
The case of $E_8$ would be different since the transformations would be generated by $(\xi^M,\sigma_M)$ with $\sigma_M$ some fixed constrained parameter.
In contrast, $V^M$ is not a fixed vector that generates symmetries in the same way.
It is simply a worldline gauge field whose components along the physical directions are removed by the constraint \eqref{eq:V_partial_constraint}.

\section{Exotic branes and worldvolume actions} 
\label{sec:branes_overview_conjecture}

\subsection{An overview of zero-branes in M-theory}
\label{subsec:overview_of_branes}

First, let us introduce some standard notation for the branes in M-theory \cite{deBoer:2012ma,Kimura:2016xzd,Berman:2018okd}.
A brane $b^{(\dots,d,c)}_a$ is an extended object with $b$ spatial dimensions that is smeared along $a+c+d+\dots$ circles with radii
\begin{equation}
    R_{i_1}\,,\,\dots\,,\,R_{i_a}\,,\, R_{j_1}\,,\,\dots\,,\,R_{j_c}\,,\,
    R_{k_1}\,,\dots\,R_{k_d}\,,\,
    \dots\,,
\end{equation}
where $a,b,c,\dots$ describe the scaling of tension (or mass) with the respective radii (and the 11D Planck length $l_P$):
\begin{equation}
    T \sim \frac{\dots (R_{k_1} \dots R_{k_d})^3 (R_{j_1} \dots R_{j_c})^2 (R_{i_1} \dots R_{i_b})}{(R_{l_1} \dots R_{l_a}) \, l_P^{1-a+b+2c+3d+\dots}} \,.
    \label{eq:branetensionscaling}
\end{equation}
The $a$ directions with inverse scaling in the radii correspond to `momentum' modes.
This is the setting in which brane tensions scale under standard Kaluza--Klein reductions along $a$ circles.
For the present paper, we set $a=0$ and omit the subscript, as we are concerned with the exotic behaviour of the other types of smearing.

These branes are the natural objects that couple electrically to the mixed-symmetry gauge fields in M-theory, e.g.~$b^c$ couples to $A_{1+b+c,c}$, $b^{(d,c)}$ couples to $A_{1+b+c+d,c+d,d}$, and so on.
For example, the $5^3$-brane couples to the higher dual three-form $A_{9,3}$ and the $0^{(1,7)}$-brane couples to the higher dual graviton $h_{9,8,1}$.
Subscripts indicate the heights of the columns in the $\mathrm{GL}(11)$ Young tableaux.

\subsubsection*{Supergravity solutions and monodromies}

Let us review some known facts for the exotic branes as supergravity solutions and their geometric properties.
For comparison with the backgrounds in \cite{Berman:2018okd}, we split the coordinates as 
\begin{equation}
    (t,\xi^1, \dots, \xi^b, \zeta^1 , \dots , \zeta^k ,z^1, \dots , z^{c+d+\dots}),
\end{equation}
with longitudinal directions $\vec{\xi}_{(b)}$ and transverse directions $\vec{\zeta}_{(k)}$.
This corresponds to a brane $b^{(\dots,d,c)}$ that is smeared around $c+d+\dots$ other transverse directions $\vec{z}_{(c+d+\dots)}$.
We are using the notation $\rd\vec{x}_{(k)}^{\,2} = (\rd x^1)^2 + \dots + (\rd x^k)^2$, and $H(\zeta)$ denotes a harmonic function with the following properties:
\begin{itemize}[leftmargin=*]
\item
    For asymptotically flat codimension-$k$ branes, one may take $H(r)=1+Q r^{2-k}$, where $r$ is the radial coordinate in the transverse space, and where $Q$ is some coefficient that we will not explain.
\item
    For codimension-two exotic branes it is more natural to define
    \begin{align}
        H(r) &:= h_0 + \sigma \ln\!\Big( \frac{\mu}{r} \Big) \,,&
        K(r,\theta) &:= H(r)^2 + \sigma^2 \theta^2 \,,
        \label{eq:H_K_def}
    \end{align}
    with constants $\sigma$, $h_0$, and $\mu$, and polar coordinates $(\zeta^1,\zeta^2)=(r,\theta)$ on the transverse plane.
\end{itemize}
The first standard object is called the M0-brane or M-wave \cite{Hull:1984vh}.
It is the Kaluza--Klein momentum mode of M-theory.
Let $z$ be the (null) wave direction with transverse directions $\vec{\zeta}_{(9)}$.
A representative is given by
\begin{equation}
    \rd s^2_{\mathrm{M0}}
    = - 2 \, \rd t \, \rd z + H(\zeta) \, \rd z^2 + \rd \vec{\zeta}_{(9)}^{\,2} \,.
    \label{eq:metric_M0}
\end{equation}
Dimensionally reducing along one of the transverse directions, the M0-brane becomes the D0 solution of IIA theory.
Equation \eqref{eq:metric_M0} is a localised solution, meaning that $H$ depends on all the transverse coordinates $\zeta$, i.e.~there are no directions around which the brane is smeared.
In contrast, a smeared brane is uniformly distributed along some of its transverse directions, and $H$ no longer depends on them.
For a comparison with the exotic zero-brane, we smear the M0 along all directions except for $(r,\theta)$, in which case $H$ is the Green's function in \eqref{eq:H_K_def}.
The representative becomes
\begin{align}
\begin{aligned}
    \rd s^2_{\mathrm{M0}}
    &= -2 \, \rd t \, \rd z + H(r) \, \rd z^2 + \rd r^2 + r^2 \rd \theta^2 + \rd \vec{z}_{(7)}^{\,2}  \\
    &= -H^{-1} \, \rd t^2
    + H \big( \rd z - H^{-1} \, \rd t \big)^2
    + \rd r^2 + r^2 \rd \theta^2
    + \rd \vec{z}_{(7)}^{\,2} \,,
    \label{eq:metric_M0_smeared}
\end{aligned}
\end{align}
where for simplicity we have written $z=z_1$ and $\vec{z}_{(7)}=(z_2,\dots,z_8)$.

The landscape of $\frac12$-BPS branes in 11D supergravity completes a U-duality multiplet of M2- and M5-branes \cite{deBoer:2012ma}.
Remarkably, this landscape also contains an exotic particle-like object denoted $0^{(1,7)}$.
The following supergravity solution describing a single back-reacted $0^{(1,7)}$-brane was given in \cite{Berman:2018okd}:
\begin{align}
\begin{aligned}
    \rd s^2_{0^{(1,7)}}
    &= -2 \, \rd t \, \rd z
    + HK^{-1} \rd z^2
    + K \big( \rd r^2 + r^2 \rd \theta^2 \big)
    + \rd \vec{z}_{(7)}^{\,2} \\
    &= -H^{-1}K \, \rd t^2
    + HK^{-1} \big( \rd z - H^{-1}K \, \rd t \big)^2
    + K \big( \rd r^2 + r^2 \rd \theta^2 \big)
    + \rd \vec{z}_{(7)}^{\,2} \,,
    \label{eq:metric_0^(1,7)}
\end{aligned}
\end{align}
with a quadratic special direction $z=z_1$ and seven other special isometry directions
$\vec{z}_{(7)}=(z_2,\dots,z_8)$, so it is essentially a point particle that has been smeared along eight dimensions.

This brane is not charged under the three-form potential or the dual six-form potential.
Instead, some component of the dual graviton is selected, or equivalently a scalar $\varphi_{z_1}$ dual to a Kaluza--Klein vector after compactification to 3D.
One may write the component as $\tilde{h}_{z_2\dots z_8z_1,z_1}=-K^{-1}\sigma\theta$.
The comma separates the block of eight antisymmetric indices from the ninth index.
The dual graviton is a mixed-symmetry tensor, so antisymmetrising all nine indices forces it to vanish identically.

This solution is exotic in the sense that the monodromy for $\theta\rightarrow\theta+2\pi$ is \textit{not} a coordinate transformation but instead a non-trivial duality transformation.
In particular, the non-trivial monodromy is contained in $K=|\tau|^2$ with $\tau=-\sigma\theta+iH$ and $\tau\rightarrow \Omega\cdot\tau:=\frac{a\tau+b}{c\tau+d}$ along the monodromy, where
\begin{equation}
    \Omega = \left(\begin{matrix}
    a&b\\c&d
    \end{matrix}\right) = \left(\begin{matrix}
    1&-2\pi\sigma\\0&1
    \end{matrix}\right) \in \mathrm{SL}(2,\mathbb{R}) \subset E_8 \,.
    \label{eq:Omega_monodromy}
\end{equation}
We will say a bit more about this in the discussion around equation \eqref{eq:unipotent_frame_g_u}.
The extra dimensions $\vec{z}_{(7)}$ factorise so that, effectively, this is a 4D solution $\cM_4\times\mathbb{R}^7$.
When ignoring the extra dimensions, this can also be understood as a solution of 4D gravity, where the duality group is just the Ehlers group $\mathrm{SL}(2,\mathbb{R})$ that appears in the presence of an isometry.

It is useful to contrast this with a smeared M0-brane \eqref{eq:metric_M0_smeared}.
A gravitational wave along an isometry direction $z$ carries momentum $P_z$.
Reducing along $z$, the metric components $g_{\mu z}$ lead to a gauge field $A_\mu$ and the momentum is then interpreted as electric charge under this vector.
On the other hand, the charge for the $0^{(1,7)}$ is encoded in the dual gravity scalar $\varphi_z$.
In a duality-covariant formulation like $E_8$ ExFT, this field appears naturally in the scalar sector.
However, from the 11D point of view, it is non-geometric in that it does not admit a standard local covariant formulation in supergravity.

Lastly, the Kaluza--Klein monopole KK$6^1$ is a geometric solution of 11D supergravity that has six spatial worldvolume dimensions $\vec{\xi}$.
We take the transverse coordinates $\vec{\zeta}=(r,\theta,z_1)$, with $z_1$ the base direction of Taub--NUT and $z_2$ the fibre, around which it is smeared.
The standard codimension-three metric is
\begin{equation}
    \rd s^2_{\mathrm{KK6}^\mathrm{1}}
    = - \rd t^2 + \rd \vec{\xi}_{(6)}^{\,2}
    + H \rd \vec{\zeta}_{(3)}^{\,2}
    + H^{-1} (\rd z_2 + A)^2 \,,
\end{equation}
where $\rd A=\star_3\rd H$.
With orientation $\mathrm{vol}_3=r\,\rd r\wedge\rd\theta\wedge\rd z_1$ on the base, one obtains
\begin{align}
    \rd H &= -\sigma r^{-1} \rd r \,,&
    \star_3 \rd H = -\sigma\,\rd \theta \wedge \rd z_1 \,,
\end{align}
As such, one can choose the local connection $A=-\sigma\theta\,\rd z_1$ and the local metric is
\begin{equation}
    \rd s^2_{\mathrm{KK6}^\mathrm{1}}
    = - \rd t^2 + \rd \vec{\xi}_{(6)}^{\,2}
    + H \big( \rd r^2 + r^2 \rd\theta^2 + \rd z_1^2 \big)
    + H^{-1} \big( \rd z_2 - \sigma\theta\,\rd z_1 \big)^2 \,.
    \label{eq:metric_KK6^1}
\end{equation}
Dependence of $A$ on $\theta$ means that travelling along a loop around the defect gives
\begin{equation}
    \rd z_2 - \sigma\theta\,\rd z_1
    \;\;\longrightarrow\;\;
    \rd z_2 - \sigma(\theta+2\pi)\,\rd z_1
    = \rd z_2 - \sigma\theta\,\rd z_1 - 2\pi\sigma\,\rd z_1 \,,
\end{equation}
which can be undone by a large diffeomorphism $z_2\longrightarrow z_2+2\pi\sigma z_1$ of the fibre coordinate.
In other words, the KK$6^1$ background is patched by $\mathrm{SL}(2,\mathbb{R})\subset E_8$ acting on the $(z_1,z_2)$-torus.
This is still `geometric' even though its codimension-two representative has non-trivial monodromy.
In contrast, the exotic zero-brane $0^{(1,7)}$ depends on $\theta$ in a way that requires patching by transformations that are not the usual diffeomorphisms or gauge transformations of supergravity.

\subsection{A conjecture for exotic brane worldvolume dynamics}
\label{subsec:exotic_brane_dynamics}

Until this point, we have focused on the point of view of supergravity, its dynamics, and its solutions.
In particular, we reviewed its U-duality covariant formulation in ExFT and its exotic brane solutions.
We will now switch to the worldvolume perspective.
This is less established from the point of view of treating exotic branes as dynamical objects propagating in supergravity backgrounds.
For earlier discussions of worldvolume actions and mixed-symmetry couplings for exotic five-branes, see \cite{Chatzistavrakidis:2013jqa,Chatzistavrakidis:2014sua}.
Similar to the string, membrane, or DBI actions, we propose worldvolume actions for all exotic branes.
Here our aim is a general construction in which U-duality covariance is manifest.
This will be a main point of the present paper.

Generalising the construction of the worldvolume action for the Kaluza--Klein monopole in \cite{Bergshoeff:1997gy,Bergshoeff:1998ef}, we conjecture the following action describing the propagation of an exotic brane $b^{(\dots,d,c)}$ in an 11D or 10D supergravity background with coordinates $\hat{x}^{\hat{\mu}}$, admitting a metric $\hat{g}_{\hat{\mu}\hat{\nu}}$:
\begin{align}
    S = T_0 \int\!\rd^{b+1}\sigma \, v_1^2 \dots v_c^2 \ w^3_1 \dots w_d^3 \dots \sqrt{\det\gamma_{\alpha\beta}} \,.
    \label{eq:exoticNGgauged}
\end{align}

Let us summarise our notation as follows.
We use hatted labels and indices for objects in the full physical (11D or 10D) spacetime, and unhatted objects for the split $\hat{x}^{\hat{\mu}}=(x^\mu,y^m)$ into $(11-n)+n$ or $(11-n)+(n-1)$ dimensions.

There are, in principle, also the 11D and 10D $p$\hspace{0.4mm}-form potentials but we only refer to them in the dimensional split, and we neglect them here.
An exotic brane $b^{(\dots,d,c)}$ requires the existence of $c+d+\dots$ special isometry directions, around which the brane is smeared.
We write
\begin{equation}
    V_i^{\hat{\mu}} = v_1^{\hat{\mu}}, \dots, v_c^{\hat{\mu}}\,;
    w_1^{\hat{\mu}}, \dots, w_d^{\hat{\mu}}\,; \dots
\end{equation}
to denote the corresponding Killing vectors for the compact isometry directions with radii given by
\begin{equation}
    2 \pi R_i = \int\!\rd z_i\,V_i \,,
\end{equation}
with $V_i:=\sqrt{V_i{}^{\hat{\mu}} V_{i{\hat{\mu}}}}$\,, in a background described by coordinates $z_i$ adapted to isometry directions $V^{\hat{\mu}}_i$, since the metric in these adapted coordinates takes the form $g_{z_iz_i}=V^2$.
The appearance of the powers $v_i^2$, $w_i^3$, and so on, is the worldvolume analogue of the scaling of the brane tension in \eqref{eq:branetensionscaling}.

The embedding of the (exotic) brane into target space is described by the embedding fields $X^{\hat{\mu}}(\sigma)$, where $\sigma^{\alpha}=(\sigma^0,\dots,\sigma^b)$ are the brane worldvolume coordinates.
The induced worldvolume metric is $\gamma_{\alpha\beta}=\hat{g}_{{\hat{\mu}}{\hat{\nu}}}D_\alpha X^{\hat{\mu}}D_\beta X^{\hat{\nu}}$ with covariant derivatives
\begin{equation}
    D_\alpha X^{\hat{\mu}}
    = \partial_\alpha X^{\hat{\mu}}
    + \sum_{i=1}^{c} C^{(2,i)}_\alpha v_i^{\hat{\mu}} + \sum_{i=1}^{d} C^{(3,i)}_\alpha w_i^{\hat{\mu}} + \dots\,,
    \label{eq:CovariantDerivative}
\end{equation}
where $C^{(I,i)}_\alpha$ are Lagrange multipliers whose equations describe the absence of brane dynamics along the isometry directions, i.e.
\begin{equation}
    V^{\hat{\mu}}_i D_\alpha X^{\hat{\nu}} \hat{g}_{{\hat{\mu}}\hat{\nu}}
    \approx 0 \,.
    \label{eq:LagrangianGaugedConstraints}
\end{equation}
Integrating out these multipliers gives a Nambu--Goto action with respect to the induced metric
\begin{equation}
    \gamma_{\alpha\beta}
    = \partial_\alpha X^{\hat{\mu}} \partial_\beta X^{\hat{\nu}} \Pi_{{\hat{\mu}}{\hat{\nu}}} \,,
\end{equation}
where $\Pi_{{\hat{\mu}}{\hat{\nu}}}$ is an effective (degenerate) metric
\begin{equation}
    \Pi_{{\hat{\mu}}{\hat{\nu}}}
    = g_{{\hat{\mu}}{\hat{\nu}}}
    - \sum_{i=1}^c \frac{v_{i{\hat{\mu}}} v_{i{\hat{\nu}}}}{v^2_i}
    - \sum_{i=1}^d \frac{w_{i{\hat{\mu}}} w_{i{\hat{\nu}}} }{w^2_i} - \dots \,, \label{eq:ReducedMetric}
\end{equation}
orthogonal to the isometry directions.
In the following, we will mostly use a Polyakov-type action
\begin{equation}
    S = \frac{T_0}{2} \int\!\rd^{b+1}\sigma \, \sqrt{\gamma} \: \bigg( \, v_1^{\frac{4}{b+1}} \dots \, v_c^{\frac{4}{b+1}} \, w^{\frac{6}{b+1}}_1 \dots \, w^{\frac{6}{b+1}}_d \, \dots \, \gamma^{\alpha\beta} D_\alpha X^{\hat{\mu}} D_\beta X^{\hat{\nu}} \hat{g}_{{\hat{\mu}}{\hat{\nu}}} - (b-1)  \bigg)
    \label{eq:exoticPolyakovgauged}
\end{equation}
where the covariant derivative was defined in \eqref{eq:CovariantDerivative}.
There is a freedom to rescale the worldvolume metric as $\gamma^{\alpha\beta}$ that gives rise to a number of superficially distinct forms of the action.
We set \eqref{eq:exoticPolyakovgauged} as our convention, choosing the cosmological constant term not to be rescaled by isometry vectors.

As a side comment to complete the landscape of equivalent classical formulations of exotic brane dynamics, the Lagrange multipliers $C^{(I,i)}_\alpha$ can be integrated out to give a Polyakov action
\begin{equation}
    S = \frac{T_0}{2} \int\!\rd^{b+1}\sigma \sqrt{\gamma} \, \Big( \gamma^{\alpha\beta} \partial_\alpha X^{\hat{\mu}} \partial_\beta X^{\hat{\nu}} \Pi_{{\hat{\mu}}{\hat{\nu}}} - (b-1) \, v_1^4 \dots v_c^4 \, w^6_1 \dots w_d^6 \, \dots \Big) \,,
    \label{eq:exoticPolyakovReducedMetric}
\end{equation}
with respect to a degenerate metric $\Pi$ as defined in \eqref{eq:ReducedMetric}.

\subsubsection*{Example: Kaluza--Klein monopole}

The best understood non-trivial example that motivated our conjecture is the worldvolume theory of the Kaluza--Klein monopole in 11D supergravity.
Following \cite{Bergshoeff:1997gy}, the dynamics can be described as a gauged sigma model.
The $\mathrm{U}(1)$ isometry that corresponds to the special isometry direction of the supergravity solution is gauged.
The action of the monopole coupling to a background metric $\hat{g}_{\hat{\mu}\hat{\nu}}$ was proposed as
\begin{equation}
	S = - \frac{T_{\mathrm{KK}6^1}}{2} \int\!\rd^7\sigma \sqrt{-\gamma} \, \Big( v^{4/7} \gamma^{ij} D_i X^{\hat{\mu}} D_j X^{\hat{\nu}} \hat{g}_{\hat{\mu}\hat{\nu}} - 5 \Big) \,,
    \label{eq:KKMaction}
\end{equation}
where $v^2 = - v^{\hat{\mu}} v^{\hat{\nu}} \hat{g}_{\hat{\mu}\hat{\nu}}$ is the norm of the Killing vector $v^{\hat{\mu}}(\hat{x})$, parametrising the isometry.
One observes this very characteristic dependence on $v$ in the action.
This fits into the above conjecture for a brane $b^{(\dots,d,c)}$ for $b=6$, $c=1$, and $d,e,\ldots=0$.
Several cross-checks of this action were undertaken in \cite{Bergshoeff:1997gy}:
\begin{itemize}[leftmargin=*]
\item
    It was shown that \eqref{eq:KKMaction} can source the Kaluza--Klein monopole solution of 11D supergravity.
\item
    Double dimensional reduction from 11D to 10D, i.e.~where the compactification on $S^1$ is both on target space and the worldvolume, has been shown to result in the (non-exotic) DBI action for the D6-brane, or 10D KK monopole action.
    The possibility for these non-trivial cross-checks for exotic branes is quite rare: often the dimensional reduction of an exotic brane will simply result in an exotic brane in lower dimensions.
\end{itemize}
In Section~\ref{sec:Hamiltonian}, we offer an additional check that this works for arbitrary branes $b^{(\dots,d,c)}$.
There we will show that the actions \eqref{eq:exoticPolyakovgauged} are natural in terms of exceptional field theory.

\subsubsection*{Hamiltonian analysis}

In order to connect with Section~\ref{sec:Hamiltonian}, let us present the Hamiltonian formulation of exotic brane dynamics conjectured in \eqref{eq:exoticPolyakovgauged}.
First, we make the spatial-temporal split $\sigma^\alpha=(\sigma^0,\sigma^i)$, where $\sigma^0=\tau$ and with spatial indices $i,j,\ldots=1,\dots,b$.
The canonical momentum $\hat{p}$ for the embedding fields $X$ in \eqref{eq:exoticPolyakovgauged} is
\begin{equation}
    \hat{p}_{\hat{\mu}} = T_0 \sqrt{\gamma} \gamma^{00} v_1^{\frac{4}{b+1}} \dots \, v_c^{\frac{4}{b+1}} \, w^{\frac{6}{b+1}}_1 \dots \, w^{\frac{6}{b+1}}_d \hat{g}_{\hat{\mu}\hat{\nu}} \dots D_0 X^{\hat{\nu}} \,,
\end{equation}
after fixing $b$ of the $b+1$ reparameterisation freedoms by the standard choice $\gamma^{0i}=0$.
Moreover, $\gamma^{00}$ will be fixed accordingly below.
The equation of motion for the unfixed metric components $\gamma^{ij}$ is
\begin{equation}
    h_{ij} = \gamma_{ij}
    = v_1^{\frac{4}{b+1}} \dots \, v_c^{\frac{4}{b+1}} \, w^{\frac{6}{b+1}}_1 \dots \, w^{\frac{6}{b+1}}_d \dots \, D_i X^{\hat{\mu}} D_j X^{\hat{\nu}} \hat{g}_{\hat{\mu}\hat{\nu}} \,.
\end{equation}
In terms of the spatial metric $h_{ij}$, the metric determinant $\gamma$ is
\begin{equation}
    \gamma = \gamma_{00} h
    = v_1^{\frac{4b}{b+1}} \dots \, \frac{1}{b!} \varepsilon^{i_1\dots i_b} \varepsilon^{j_1\dots j_b}
    D_{i_1} X^{\hat{\mu}_1} D_{j_1} X^{\hat{\nu}_1} \dots D_{i_b} X^{\hat{\mu}_b} D_{j_b} X^{\hat{\nu}_b}
    \hat{g}_{\hat{\mu}_1 \hat{\nu}_1} \dots \hat{g}_{\hat{\mu}_b \hat{\nu}_b} \,.
\end{equation}
The Hamiltonian constraint becomes
\begin{equation}
    H = \hat{p}_{\hat{\mu}} D_0 X^{\hat{\mu}} - L
    = - \frac{T_0}{2} \frac{\sqrt{\vert\gamma_{00}\vert}}{\sqrt{h}\,v_1^{\frac{4}{b+1}}\dots} \hat{p}_{\hat{\mu}} \hat{p}_{\hat{\nu}} \hat{g}^{\hat{\mu}\hat{\nu}} - \frac{1}{2} \sqrt{\vert\gamma_{00}\vert}\sqrt{h}
    \approx 0 \,,
\end{equation}
and finally, choosing $\gamma_{00}=hv_1^{\frac{8}{b+1}}\dots$ leads to
\begin{equation}
    H = \frac{1}{2} \hat{p}_{\hat{\mu}} \hat{p}_{\hat{\nu}} \hat{g}^{\hat{\mu}\hat{\nu}}
    + \frac{1}{2} v_1^{4} \dots \varepsilon^{i_1 \dots i_b} \varepsilon^{j_1 \dots j_b}
    \partial_{i_1} X^{\hat{\mu}_1} \partial_{j_1} X^{\hat{\nu}_1} \dots \partial_{i_b} X^{\hat{\mu}_b} \partial_{j_b} X^{\hat{\nu}_b}
    \hat{g}_{\hat{\mu}_1\hat{\nu}_1} \dots \hat{g}_{\hat{\mu}_b\hat{\nu}_b}
    \approx 0 \,,
    \label{eq:exoticBraneHamiltonianConjecture}
\end{equation}
where $v_1^4\dots$ is shorthand for $v_1^4\dots v_c^4\,w_1^6\dots w_d^6\dots$ in \eqref{eq:exoticBraneHamiltonianConjecture}.
In this last step, we worked in the gauge $C^{(I,i)}_\alpha=0$ so that the covariant derivatives have become partial derivatives, i.e.~$D_\alpha X^{\hat{\mu}}\to\partial_\alpha X^{\hat{\mu}}$.
The equations of motion \eqref{eq:LagrangianGaugedConstraints} are retained separately as additional first-class constraints
\begin{equation}
    V^{\hat{\mu}}_i \hat{p}_{\hat{\mu}} \approx 0 \,.
    \label{eq:Vp=0}
\end{equation}
As secondary constraints we have
\begin{equation}
    V^{\hat{\mu}}_i(\hat{x}) \, \partial_{\hat{\mu}} f(\hat{x},\hat{p}) \approx 0
\end{equation}
for all Killing vectors $V=(v,w,\dots)$ and all functions $f$ on the worldvolume phase space.
In practice, we will always choose coordinates adapted to these isometries and assume that they do not depend on the associated coordinates.
Resultingly, there will be similarities with the dimensional reductions, but the forms of the actions do \textit{not} correspond to dimensional reductions of standard brane actions, as we will argue below in the case of zero-branes.

\subsubsection*{Zero-branes} 

Let us mention two central questions regarding the exotic brane actions that we conjectured above.
In the context of this article, we pose and approach them for zero-branes (or particles), but similar questions for higher-dimensional branes can be answered in a very similar manner.

Exotic branes were considered mainly because they complete the U-duality multiplet for standard $\frac{1}{2}$-BPS branes in supergravity \cite{deBoer:2012ma}.
Not all actions of the form \eqref{eq:exoticPolyakovgauged} will fit into U-duality multiplets; only some combinations in $b^{(\dots,d,c)}$ are allowed.
So, one question that we would like to answer is: \textit{how is the form of actions \eqref{eq:exoticPolyakovgauged} directly constrained by U-duality covariance?}
This will be presented in the next section.

We will elaborate briefly on the distinction between Kaluza--Klein reduction of standard particles and exotic zero-branes.
Superficially, both settings share some similarities.
Both rely on the presence of target space isometries, but the meaning of these isometries is different on the worldvolume.

Assume a target space with an isometry, i.e.~Killing vector $\partial_z$ in adapted coordinates $\hat{x}^{\hat{\mu}}=(x^\mu,z)$.
In the standard relativistic point particle action
\begin{equation}
    S \sim \int\!\rd\tau \sqrt{- \hat{g}_{\hat{\mu}\hat{\nu}} \dot{X}^{\hat{\mu}} \dot{X}^{\hat{\nu}}} \,,
\end{equation}
this corresponds to a \textit{global} symmetry.
This is the setting of the Kaluza--Klein reduction of the usual particle action.
The momentum splits as $\hat{\pi}_{\hat{\mu}}=(\pi_\mu,\pi_z)$, where $\pi_z$ is the conserved charge corresponding to the global symmetry and its value is \textit{not} restricted.
Still, there is dynamics along the isometry direction.
In particular, the Hamiltonian is
\begin{equation}
    H = \frac{1}{2} \hat{g}^{\hat{\mu}\hat{\nu}} \hat{\pi}_{\hat{\mu}} \hat{\pi}_{\hat{\nu}}
    = \frac{1}{2} g^{\mu\nu} \pi_\mu \pi_\nu + \frac{1}{2} \hat{g}^{zz} \pi_z \pi_z
    \sim \frac{1}{2} \pi_\mu \pi^\mu + \frac{1}{2} \bigg( \frac{\pi_z}{R_z} \bigg)^2
    \approx 0 \,.
\end{equation}
For the Kaluza--Klein reduced particle, the (originally massless) particle becomes massive, with mass scaling like $m\sim\frac{1}{R_z}$ with respect to the length scale or radius $R_z$ of the isometry direction.
The case $\pi_z=0$ corresponds to the massless particle in one dimension less.
In any case, the particle is still \textit{localised} in the extra dimension.

In contrast, the characteristic feature of exotic objects from the worldvolume perspective is that they are \textit{smeared} (delocalised) along the isometry direction.
The isometry of the target space will be promoted to a local gauge symmetry of the action.
The particle does not have any physical degrees of freedom along this isometry direction.
Mathematically, this is described by the constraint
\begin{equation}
    V^{\hat{\mu}} \pi_{\hat{\mu}} = 0
\end{equation}
for the corresponding Killing vector $V$.
From this perspective, it is similar to the dimensionally-reduced particle -- compare with \eqref{eq:Vp=0}.

The crucial difference from the worldline point of view lies in the Hamiltonian \eqref{eq:exoticBraneHamiltonianConjecture}.
For a given zero-brane $0^{(\dots,d,c)}$ the general form of the Hamiltonian reduces to
\begin{equation}
    H = \frac{1}{2} \big( \pi_\mu \pi^\mu + v^4_1 \ldots v_c^4 \, w_1^6 \ldots w_d^6 \ldots \big) \approx 0 \,,
\end{equation}
where $v,w$ are the length scales of the $c+d+\ldots$ isometry directions.
So, the differences are as follows.
\begin{itemize}[leftmargin=*]
\item
    Despite delocalisation (and hence vanishing momentum) along the isometry direction, the particle generically appears as massive in the lower dimensional theory.
\item
    The scaling of the mass in terms of the radii of the isometry directions is different from what happens in Kaluza--Klein compactification.
 \end{itemize}

\section{Duality-covariant phase space formalism for exotic branes}
\label{sec:Hamiltonian}

Worldvolume dynamics and duality symmetries were first connected in \cite{Duff:1989tf,Tseytlin:1990nb,Tseytlin:1990va,Siegel:1993bj,Siegel:1993th} where string worldsheet dynamics was formulated in an $\mathrm{O}(d,d)$-covariant way.
This is natural in terms of first-order actions or the Hamiltonian formulation.
For the string this led to the development of so-called $\cE$-models that became important for generalised dualities and integrable models \cite{Klimcik:1995ux,Sfetsos:1999cc,Demulder:2018lmj,Hassler:2020xyj,Borsato:2021gma,Borsato:2021vfy}, the connection to non-commutative geometry \cite{Blair:2014kla,Osten:2019ayq}, and more recently the construction of a generalised version of Cartan geometry \cite{Hassler:2024hgq,Hassler:2025rag}.
It was also shown that the classical phase space of the heterotic string is $\mathrm{O}(d,d+n)$-covariant \cite{Hatsuda:2022zpi,Osten:2023cza}.

On the other hand, the literature on worldvolume theories in the context of exceptional field theory for $p$\hspace{0.4mm}-branes in higher dimensions, like D-branes or M-branes in string and M-theory, is sparse.
There are two main reasons for this:~(1) the role of worldvolume theories for individual $p$\hspace{0.4mm}-branes in M-theory is not as well-understood as that of worldsheet models for the strings in string theory, and (2) there is a fundamental obstruction in realising duality symmetries on the higher-dimensional $p$\hspace{0.4mm}-brane dynamics.
The reason for this obstruction is that U-duality relates different worldvolume dynamics to each other.
This obstruction was shown to arise in attempts to realise the exceptional symmetries on their phase space, such as for the M2- \cite{Duff:1990hn,Duff:2015jka,Hatsuda:2012vm,Strickland-Constable:2021afa}, M5- \cite{Hatsuda:2013dya}, and D-branes \cite{Hatsuda:2012uk}.
Related attempts have also been untertaken at the level of QP-manifolds and AKSZ sigma models \cite{Arvanitakis:2018cyo,Arvanitakis:2021wkt,Arvanitakis:2022fvv,Osten:2023iwc} in order to understand the topological parts of such brane actions.
Let us also highlight the proposals for duality-covariant brane dynamics in references \cite{Sakatani:2016sko,Sakatani:2017vbd,Sakatani:2020umt,Hatsuda:2023dwx}, although there the relation to the standard $\frac{1}{2}$-BPS branes of string and M-theory is less clear.
Moreover, the top-down approach in \cite{Osten:2021fil,Osten:2024mjt} which generalises \cite{Arvanitakis:2018hfn} to higher-dimensional $p$\hspace{0.4mm}-branes using the Hamiltonian formulation arranges the phase space variables into representations of the duality group.

As reviewed below, the central statement is the following.
We define a $p$\hspace{0.4mm}-brane phase space and Hamiltonian abstractly from the geometric and algebraic data of ExFT and the tensor hierarchy.
This alone is not restrictive enough to characterise interesting and realistic $p$\hspace{0.4mm}-brane models.
In addition, one needs the so-called \textit{brane charge constraint}.
It arises as a natural (sufficient) condition such that the generalised Lie derivative of ExFT is realised on the brane phase space.
More relevant for us, its solutions are in a one-to-one correspondence with $\frac{1}{2}$-BPS branes, and hence give a natural brane scan from ExFT and a definition of Hamiltonian dynamics.
Our new statements about this approach are:
\begin{itemize}[leftmargin=*]
\item
    We show that for zero-branes this approach is equivalent to Blair's exceptional particle model that was introduced in \cite{Blair:2017gwn} and reviewed here in Section~\ref{subsec:exceptional-particle}.
\item
    This Hamiltonian construction can be generalised to include ancillary transformations of ExFT, i.e.~with constrained gauge parameters.
    With this, one can implement it in the $E_8$ case.
\item
    The Hamiltonians for exotic branes introduced in \eqref{eq:exoticBraneHamiltonianConjecture} emerge from this top-down approach from solutions of the brane charge constraint.
    We demonstrate this for the M-theory point particle and the exotic $0^{(1,7)}$-brane.
\end{itemize}

\subsection{Brane dynamics and charges}

The dynamics of $\frac{1}{2}$-BPS branes can be constructed from ExFT, as described in \cite{Osten:2024mjt} for exceptional U-duality groups up to $E_7$.
In this section, we will review and extend this claim to include $E_8$.

Consider the Hamiltonian description of a $p$\hspace{0.4mm}-brane propagating in an ExFT background that is characterised by the external metric $g_{\mu\nu}$, the internal metric $\cM_{MN}$, and the tensor hierarchy of form fields $A_\mu{}^{M}$, $B_{\mu\nu}{}^{M_2}$, $C_{\mu\nu\rho}{}^{M_3}$, and so on.\footnote{Recall that $M_k,N_k,\dots$ are $\cR_k$ representation indices, and that for $\cR_1$ we write $M,N,\dots$ with no subscript.}
The $p$\hspace{0.4mm}-brane is embedded into the ExFT background via the embedding fields $(X^\mu(\sigma)$,$Y^{M}(\sigma))$, where the dependence on $Y^M$ is subject to the section constraint.
For simplicity, we shall first restrict ourselves to the dynamics in the internal space, as this takes a universal form for all exceptional groups $E_n$.
The essential ingredients can be organised as follows.

The phase space variables can be arranged into a tower of currents $t_{M_k}\in\cR_k$ associated with the tensor hierarchy of the U-duality group, i.e.~typically $E_{n}$ for $p>1$\,.
These spatial $(p-k+1)$-forms on the brane are constructed from the canonical momentum density, embedding coordinates, and the possible worldvolume gauge fields or auxiliary fields.
This tower of currents $\{t_{M_k}\}$ generalises the generalised momentum $P_M$ that appeared in the first order treatment of Blair's particle action.
For a particle or zero-brane we will use the notation $P_M=t_M\in\cR_1$, and for higher-dimensional branes we will use $t_{M_k}$ to emphasise that these phase space currents can be understood as generators of the $\cR_k$ representation.\footnote{Note that these brane charge constraints do not give a `charge' in the traditional sense, like an integral of a conserved current. Instead it is called a charge in the sense that it characterises objects. We mention this because `current' may be a confusing label.}

As an illustration, for the M2-brane one will find \cite{Hatsuda:2012uk,Osten:2021fil}
\begin{align}
    t_{M} &= (P_m , \rd Y^m \wedge \rd Y^n, 0 , \dots) \,,& 
    t_{M_2} &= (\rd Y^m , 0 , \dots) \,,& 
    t_{M_3} &= (1,0,\dots) \,,
    \label{eq:M2braneChargeSolution}
\end{align}
where the split $Y^M=(Y^m,Y^{\widetilde{m}})$ corresponds to a solution $\partial_M = (\partial_m, 0)$ to the section constraint, and $P_m$ is the canonical dual to the embedding field $Y^m(\sigma)$.
Here, $\cR_k$ indices have been decomposed into $\mathrm{GL}(n)$ indices.
    
In order to simplify the expression, we wrote the spatial form part of these currents in index-free form as $\rd Y^m\wedge\rd Y^n\sim\varepsilon^{ij}\partial_iY^m\partial_jY^n$, where $i,j,\dots=1,2$ are spatial indices on the worldvolume of the M2-brane.
In this section, we denote by $\mathrm{d}$ the exterior derivative on a spatial slice of the worldvolume.
The currents $t_{M_k}$ obey a Poisson bracket algebra of the form
\begin{equation}
    \{ t_{M_k}(\sigma) , t_{N_l}(\sigma^\prime) \} 
    = - \eta^{K_{k+l}}{}_{M_kN_l} \, t_{K_{k+l}} (\sigma') \wedge \rd' \delta (\sigma - \sigma') \,.
    \label{eq:CurrentAlgebraSmall}
\end{equation}
Possible total-derivative contributions on the worldvolume will be ignored.
As reviewed in Section~\ref{subsec:E8_ExFT}, the $\eta$-symbols are just the structure constants of the bullet product.
Equivalently, they describe the embedding $\cR_{k+l}\hookrightarrow\cR_k\otimes\cR_l$.
The Jacobi identity of \eqref{eq:CurrentAlgebraSmall} is equivalent to the graded Jacobi identity satisfied by the bullet product of the tensor hierarchy.

The Hamiltonian and spatial diffeomorphism constraints are encoded in the internal metric $\cM_{MN}$ and the $D$-symbols.\footnote{We introduce the $D$-symbols as embeddings $\overline{\cR}_{k+l}\hookrightarrow\overline{\cR}_k\otimes\overline{\cR}_l$. In many cases one expects them to be the canonical duals to the $\eta$-symbols.}
Explicitly, the Hamiltonian is
\begin{equation}
    H = \frac{1}{2} \int \cM^{MN} t_M t_N \approx 0 \,,
    \label{eq:PhaseSpaceFormalismHamiltonian}
\end{equation}
and the diffeomorphism constraint is
\begin{equation}
    D_{M_2}{}^{NK} t_N t_K \approx 0 \,.
    \label{eq:PhaseSpaceFormalismConstraint}
\end{equation}
This is how ExFT data defines brane dynamics.
In the spirit of a top-down definition for worldvolume dynamics in ExFT, one could consider this Hamiltonian description to be very natural.
The tensor hierachy structure defines the Poisson structure and any constraints via the $\eta$- and $D$-symbols, with the Hamiltonian itself given in terms of the internal metric.
On the other hand, one can show that for all usual $p$\hspace{0.4mm}-brane actions the Hamiltonian actually takes this natural form \cite{Osten:2021fil,Osten:2024mjt} subject to the brane charge constraints discussed below.
    
The current $t_M$ takes the role of the `canonical momentum' for the internal embedding field:
\begin{equation}
    \{ t_M(\sigma), f ( Y(\sigma^\prime) ) \}
    = - \partial_M f ( Y(\sigma) ) \, \delta(\sigma-\sigma^\prime) \,.
    \label{eq:BraneChargeConstraintZero}
\end{equation}
As only some components of the coordinate field $Y^M$ are physical after solving the section constraint, only some components of $t_M$ are fixed to be `physical' canonical momenta, the others can be expressed by the embedding fields themselves or potentially additional worldvolume gauge fields.\footnote{This is the case for the M5-brane for example, but will not be relevant in the cases considered here.}
In particular, one can write $t_M=(P_m,t^{\widetilde{m}})$, where $P_m$ is the canonical momentum for the physical component $Y^m(\sigma)$.
It is easy to check that the M2-brane \eqref{eq:M2braneChargeSolution} satisfies this property.

Note that even if one works on a solution to the section constraint these assumptions give rise to a theory that is too `large' in the sense that it possesses too many degrees of freedom.
In particular, it seems that each representation $\cR_k$ in the tensor hierarchy has its own independent physical field.
Unless one uses a particular realisation of the currents, like \eqref{eq:M2braneChargeSolution} for the M2, this larger model does not correspond to a conventional $p$\hspace{0.4mm}-brane worldvolume theory.
This only happens if one imposes the so-called brane charge constraints \cite{Blair:2017gwn,Arvanitakis:2018hfn,Osten:2021fil,Osten:2024mjt}.

\subsubsection*{Brane charge constraints}

From the point of view of this top-down Hamiltonian formulation of exceptional particles, the central assumption is that the local transformations of ExFT, i.e.~the generalised Lie derivative, are realised on the brane phase space.
This is what will give rise to a hierarchy of \textit{brane charge constraints} \cite{Arvanitakis:2018hfn}.

In order for the generalised Lie derivative to act on phase space, we represent a section $\Phi$ of a generalised $\cR_k$-bundle on the worldvolume phase space as
\begin{equation}
    \Phi = \int \Phi^{M_k}(Y(\sigma)) \, t_{M_k}(\sigma) \,,
\end{equation}
where the integration is over the spatial worldvolume.
For generic currents obeying \eqref{eq:CurrentAlgebraSmall}, the Poisson bracket does not directly reproduce the generalised Lie derivative.
In order to obtain the generalised Lie derivative
\begin{equation}
    \{ \Phi,\Lambda \}
    = \cL_{\Lambda} \Phi
\end{equation}
for the parameter
\begin{equation}
    \Lambda = \int \Lambda^M(Y(\sigma)) \, t_M(\sigma) \,,
\end{equation}
one must impose the so-called \textit{brane charge constraints}
\begin{equation}
    {D_{M_{k+1}}}^{N_{k}K} \, t_{N_{k}} \partial_K =  t_{M_{k+1}} \wedge \rd Y^N \partial_N 
    \,,
    \label{eq:BraneChargeSmall}
\end{equation}
for $k\geq 1$\,.
Together with the identification $t_M=(P_m,\dots)$ from \eqref{eq:BraneChargeConstraintZero}, these constraints ensure the desired correspondence.
Remarkably, solutions to all these brane constraints seem to be in one-to-one correspondence to $\frac{1}{2}$-BPS branes in the sense that the Hamiltonian \eqref{eq:PhaseSpaceFormalismHamiltonian} and Poisson structure \eqref{eq:CurrentAlgebraSmall} give rise to standard actions for the worldvolume dynamics of $\frac{1}{2}$-BPS branes, i.e.~actions for the M2- and M5-branes, or the DBI actions for D-branes.
We review solutions to these constraints below.

\textit{Comment 1.}
This conjecture is based on the construction of current algebras \cite{Hatsuda:2012uk,Hatsuda:2012vm,Hatsuda:2013dya} and QP-manifolds \cite{Arvanitakis:2018hfn,Arvanitakis:2018cyo,Arvanitakis:2021wkt} associated with dynamics of $\frac{1}{2}$-BPS branes in supergravity.
The setup introduced above can be considered as the exceptional-covariant version, and solutions to the brane charge constraints will reduce the previous results \cite{Hatsuda:2012uk,Hatsuda:2012vm,Hatsuda:2013dya, Arvanitakis:2018hfn,Arvanitakis:2018cyo,Arvanitakis:2021wkt}.
This has been explicitly demonstrated for $E_n$ ExFT with $n\leq 7$ and for the corresponding branes with up to seven spatial dimensions, including the Kaluza--Klein monopole KK$6^1$ in 11D \cite{Osten:2021fil,Osten:2023iwc,Osten:2024mjt}.
These previous results excluded exotic objects like the $5^3$- and $0^{(1,7)}$-branes, and in general all codimension-two exotic branes.
These require eight dimensions in target space in the 11D case, or seven in the IIA and IIB cases.
In contrast, they should appear only starting from $E_{n}$ with $n>7$.
        
\textit{Comment 2.} The hierarchy of currents for a $p$\hspace{0.4mm}-brane naturally ends at $t^{(p+1)}$ because this is a zero-form.
This leads to the natural constraint ${D_{M_{p+2}}}^{N_{p+1}K} \, t_{N_{p+1}} \partial_K = 0$, or equivalently
\begin{equation}
    t^{(p+1)} \otimes \partial \, \big|_{\cR_{p+2}} = 0 \,.
\end{equation}
This is a generalisation of the statement from \cite{Arvanitakis:2018hfn} for the $p=1$ case, i.e.~for strings or one-branes.

\subsubsection*{Known solutions to the brane charge constraints}

In order to be self-contained in this article, let us collect the known solutions for charge constraints in $E_n$ ExFT for $n<8$ (and hence $p<8$) collected in \cite{Osten:2021fil,Osten:2023iwc,Osten:2024mjt}, building on \cite{Hatsuda:2012uk,Hatsuda:2012vm,Hatsuda:2013dya,Arvanitakis:2018hfn,Arvanitakis:2018cyo,Arvanitakis:2021wkt}.
The (perhaps surprising) conclusion of this programme is that these solutions are in natural correspondence with $\frac{1}{2}$-BPS branes in supergravity.
Solving the brane charge constraints therefore amounts to a brane scan that does not use supersymmetry in any form.
Naturally, all the solutions presented here are also valid for $n \geq 8$, but, of course, there will be additional solutions in these cases.
The solutions for the M-theory section are collected in Table~\ref{tab:solutions_M-theory} on page~\pageref{tab:solutions_M-theory}.
On the other hand, the solutions for the IIB section have not been discussed in detail in the covariant formalism.
These were presented in a non-covariant form in \cite{Hatsuda:2012uk} and later in terms of the associated QP-manifolds in \cite{Arvanitakis:2018cyo,Arvanitakis:2021lwo}.
We present the IIB solutions up to $p=5$ in Table~\ref{tab:solutions_IIB} on page~\pageref{tab:solutions_IIB}.
For all $p$\hspace{0.4mm}-brane charge solutions, the currents $t_{M_k}$ vanish for $k>p+1$.
Let us emphasise that many of the higher-dimensional brane charge solutions require the existence of additional worldvolume gauge field strengths $F=\rd A$.
In this sense the brane charge constraints predict the precise non-trivial form of the $p$-brane actions in supergravity.

\begin{table}[ht]
\centering
\begin{align*}
\renewcommand{\arraystretch}{1.2}
\begin{array}{c|l}
\text{brane} & \text{currents} \\
\hline \hline 
\text{M0}
    & t_{M_1} = ( P_m , 0 , 0 , 0 )
    = (t_m,t^{m[2]},t^{m[5]},t^{m[7];n}) \\
\hline
\text{M2}
    & t_{M_1} = ( P_m, \rd Y^{m_1} \wedge \rd Y^{m_2} , 0 , 0 ) \\
    & t_{M_2} = ( \rd Y^m , 0 , 0 )
    = (t^{m},t^{m[4]},t^{m[6];n}) \\
    & t_{M_3} = ( 1 , 0 , 0 )
    = (t,t^{m[3]},t^{m[5];n}) \\
\hline
\text{M5}
    & t_{M_1} = ( P_m , F \wedge \rd Y^{m_1} \wedge \rd Y^{m_2} , \rd Y^{m_1} \wedge \dots \wedge \rd Y^{m_5} , 0 ) \\
    & t_{M_2} = (F \wedge \rd Y^{m} , \rd Y^{m_1} \wedge \dots \wedge \rd Y^{m_4} , 0 ) \\
    & t_{M_3} = ( F , \rd Y^{m_1} \wedge \rd Y^{m_2} \wedge \rd Y^{m_3} , 0 ) \\
    & t_{M_4} = ( \rd Y^{m_1} \wedge \rd Y^{m_2} , 0 )
    = (t^{m[2]},t^{m[4];n})\\
    & t_{M_5}=(\rd Y^m,0)=(t^m,t^{m[3];n}) \\
    & t_{M_6}=(1,0)=(t,t^{m[2];n}) \\
\hline
\text{KK}6^1
    & t_{M_1} = ( P_m , 0 , 0 , v^{[m_1} \rd Y^{m_2} \wedge \dots \wedge \rd Y^{m_7]} v^n ) \\
    & t_{M_2} = ( 0 , 0 , v^{[m_1} \rd Y^{m_2} \wedge \dots \wedge \rd Y^{m_6]} v^n ) \\
    & \quad \quad \vdots \\
    & t_{M_5} = ( 0 , v^{[m_1} \rd Y^{m_2} \wedge \rd Y^{m_3]} v^n ) \\
    & t_{M_6} = ( 0 , v^{[m_1} \rd Y^{m_2]} v^n ) \\
    & t_{M_7} = ( v^m v^n ) =(t^{m;n})
\end{array}
\end{align*}
    \caption{Solutions to the charge constraints in the M-theory section.
    $F=\rd A=\star F$ is the self-dual three-form field strength on the M5-brane worldvolume, and $v^m$ is a target Killing vector that is needed for the existence of the KK monopole, obeying $v^mP_m\approx0$.
    Note that for the monopole one might expect $t_{M_8}=(t^n)=(v^n)$ from the tensor structure, but from the worldvolume perspective it would be a $(-1)$-form and therefore must vanish.}
    \label{tab:solutions_M-theory}
\end{table}

\begin{table}[ht]
\centering
\begin{align*}
\renewcommand{\arraystretch}{1.2}
\begin{array}{c|l}
\text{brane} & \text{currents} \\
\hline \hline 
\text{F1/D1}
    & t_{M_1} = ( P_m , q_i \, \rd Y^m , 0 , 0 , 0 )
    = (t_m,t_i{}^m,t^{m[3]},t_i{}^{m[5]},t^{m[6];n}) \\
    & t_{M_2} = ( q_i , 0 , 0 , 0 )
    = (t_i,t^{m[2]},t_i{}^{m[4]},t^{m[5];n}) \\
\hline
\text{D3}
    & t_{M_1} = (P_m, F_i \wedge \rd Y^{m} , \rd Y^{m_1} \wedge \rd Y^{m_2} \wedge \rd Y^{m_3} , 0 , 0 ) \\
    & t_{M_2} = ( F_i , \rd Y^{m_1} \wedge \rd Y^{m_2} , 0 , 0 ) \\
    & t_{M_3} = ( \rd Y^{m} , 0 , 0 )
    = (t^{m},t_i{}^{m[3]},t^{m[4];n}) \\
    & t_{M_4} = ( 1 , 0 , 0 )
    = (t,t_i{}^{m[2]},t^{m[3];n}) \\
\hline
\text{NS5/D5}
    & t_{M_1} = ( P_m, G_i \wedge \rd Y^m , F \wedge \rd Y^{m_1} \wedge \rd Y^{m_2} \wedge \rd Y^{m_3} , q_i \, \rd Y^{m_1} \wedge \dots \wedge \rd Y^{m_5} , 0 ) \\
    & t_{M_2} = ( G_i , F \wedge \rd Y^{m_1} \wedge \rd Y^{m_2} , q_i \, \rd Y^{m_1} \wedge \dots \wedge \rd Y^{m_4} , 0 ) \\
    & t_{M_3} = ( F \wedge \rd Y^m , q_i \, \rd Y^{m_1} \wedge \rd Y^{m_2} \wedge \rd Y^{m_3} , 0 ) \\
    & t_{M_4} = ( F,q_i \, \rd Y^{m_1} \wedge \rd Y^{m_2} , 0 ) \\
    & t_{M_5}=(q_i \, \rd Y^m,0)=(t_i{}^m,t^{m[2];n}) \\ 
    & t_{M_6}=(q_i,0)=(t_i,t^{m;n})
\end{array}
\end{align*}
    \caption{Solutions to the charge constraints in the IIB theory section.
    Here, $i=1,2$ is an $\mathrm{SL}(2,\mathbb{R})$ (S-duality) index, while $m,\ldots=1,\dots,n-1$ are indices for the internal part of the IIB section.
    The vector $q_i$ characterises the F1/D1 doublet, and $F_i=(F,\star F)$, with $F = \rd A$, is the electric-magnetic doublet of the coupling of the worldvolume gauge field $A$ to the D3 worldvolume.
    Lastly, in the third row, $G_i=(q_1\,{\star}F,q_2F\wedge F)$ with $F=\rd A$ for the internal D5- or NS5-brane gauge field $A$, where $q_i$ now characterises the NS5/D5 doublet.}
    \label{tab:solutions_IIB}
\end{table}
Note that the KK monopole in type IIB theory has not been discussed, nor, for that matter, has it been considered in non-maximal solutions to the section constraint such as that corresponding to type IIA theory.
Nevertheless, it was shown in \cite{Osten:2021fil} that the M2-brane in the M-theory section reduces to the IIA F1-string after double dimensional reduction.
Another related approach was followed in \cite{Arvanitakis:2023dud} for massive IIA theory.
The brane charge constraints for $p>1$, i.e.~beyond \cite{Blair:2017gwn,Arvanitakis:2017hwb}, have so far only been discussed for the maximal M-theory and type IIB sections.
We leave an analysis of solutions of the brane charge constraints in the IIA section and other non-maximal sections to future work.

\subsubsection*{Embedding into supergravity in ten and eleven dimensions}

Even when working with $E_n$ for $n<8$, it is possible to embed the worldvolume into the supergravity theory in either ten or eleven dimensions.
As described in \cite{Osten:2024mjt}, the coupling of the worldvolume to the external metric and tensor hierarchy gauge fields is encoded in the kinematic momentum 
\begin{equation}
    \pi_\mu = P_\mu - A_\mu{}^M t_M - B_{\mu\nu}{}^{M_2} t_{M_2} - \dots \,,
\end{equation}
where $P_\mu$ is the canonical momentum to the embedding fields $X^\mu(\sigma)$.
The total Hamiltonian takes the form
\begin{equation}
    H = \frac12 g^{\mu\nu} \pi_\mu \pi_\nu + \frac12 \cM^{MN} t_M t_N \,.
    \label{eq:HamiltonianFull}
\end{equation}
The momentum $\pi_\mu$ is not canonical.
Its non-trivial Poisson bracket is 
\begin{equation}
    \{ \pi_\mu(\sigma),\pi_\nu(\sigma^\prime) \}
    = \big( F_{\mu\nu}{}^M t_M(\sigma) + H_{\mu\nu\rho}{}^{M_2} t_{M_2} \wedge \rd X^\rho (\sigma) + \dots \, \big) \delta(\sigma-\sigma^\prime) \,,
\end{equation}
where $F_{\mu\nu}{}^M$, $H_{\mu\nu\rho}{}^{M_2}$, etc.~are the field strengths for $A_\mu{}^M$, $B_{\mu\nu}{}^{M_2}$, and so on.
Importantly, the spatial directions along the brane are restricted to lie in the internal directions of the target space split.

\subsection{Hamiltonians for exotic branes}
\label{subsec:Hamiltonians_exotic_branes}

Consider the $\cR_1$ representation of $E_n$ in its M-theory or $\mathrm{GL}(n)$ decomposition\footnote{We emphasise that the first few components are universal for all exceptional groups. Those given here are found in the $E_8$ case and beyond. For clarity, we present only the $n$-dimensional M-theory section here. The $(n-1)$-dimensional IIA and IIB cases are completely analogous, and we leave details for follow-up work.}
\begin{equation}
    t_M = (t_m, t^{m[2]}, t^{m[5]}, t^{m[7],n}, t^{m[8]}, t^{m[8],n[3]}, t^{m[8],n[6]}, t^{m[8],n[8],p}, \dots) \,.
    \label{eq:R1Decomp}
\end{equation}
We have already seen that $t^{m[2]}$ and $t^{m[5]}$ naturally characterise the M2- and M5-branes, respectively, and that $t^{m[7],n}$ corresponds to the Kaluza--Klein monopole KK$6^1$.
In general, we observe the following pattern that was described already in the literature, for example in \cite{Fernandez-Melgarejo:2019mgd}:
\begin{align*}
\begin{array}{c|cccccccc}
\text{GL}(n)\text{ component}
    & t_m & t^{m[2]} & t^{m[5]} & t^{m[7];n} & t^{m[8],n[3]} & t^{m[8],n[6]} & t^{m[8],n[8],p} & \dots \\ \hline
\text{M-theory brane}
    & \text{M0} & \text{M2} & \text{M5} & \text{KK}6^1 & 5^3 & 2^6 & 0^{(1,7)} & \dots 
\end{array}
\end{align*}
Our notation is that $m[k]$ is one set of $k$ antisymmetric $\mathrm{GL}(n)$ indices $m_1\dots m_k$ and different sets of antisymmetric indices that are separated by a comma belong to different columns inside an irreducible Young tableau.
For example, $t^{m[7],n}$ is short for $t^{m_1\dots m_7,n}$ which satisfies the over-antisymmetrisation constraint $t^{[m_1\dots m_7,n]}=0$\,.
Antisymmetrising any set of antisymmetric indices with a single index to the right of it causes the tensor to vanish identically.
We can also denote \emph{reducible} tensors using semicolons to separate the indices corresponding to different irreducible components.
For example, $t^{m[7],n}$ and $t^{m[8]}$ can be packaged in a single reducible tensor $t^{m[7];n}$ that amounts to a tensor product between a seven-form and a one-form.
This gives a mixed-symmetry component and a totally antisymmetric eight-form component.

In the following, we shall formalise this pattern.
The formalism presented here will in principle allow us to do this analysis for arbitrary $n$.
This gives a constructive and compact way to describe the dynamics of objects that are predicted to lie in the $E_n$ multiplet \eqref{eq:R1Decomp}.
In fact, our analysis will show a generic behaviour for any $t^{m[A],n[B],p[C],q[D],\dots}$ in the $\cR_1$ representation.
So, crucially, each $\mathrm{GL}(n)$ component of $\cR_1$ will imply the existence of a corresponding brane.
This means that the structure of $E_n$ will correspond to a \emph{brane scan}.
Exceptional groups and their representations have been well-studied in the pursuit of such a brane scan.
Very interesting work in this direction concerning infinite-dimensional Kac--Moody symmetries $E_9$, $E_{10}$, and $E_{11}$ was carried out in \cite{Brown:2004jb,West:2004kb,Englert:2007qb,Cook:2008bi,Cook:2009ri,Houart:2009ya,Houart:2011sk,West:2018lfn}.

In order to observe how the components $t^{m[A],n[B],p[C],q[D],\dots}$ correspond to solutions of the brane charge constraints, we need to understand the tensor hierarchy in the $\mathrm{GL}(n)$ decomposition.
It is the case -- see e.g.~\cite{Bossard:2017wxl} -- that the tensor hierarchy in this decomposition is arranged into tensor products of a $p$\hspace{0.4mm}-form and an irreducible generator.
We have already seen this with $t^{7;1}=t^{7,1}\oplus t^8$, where we use a shorthand for the (ir)reducible symmetry types of each generator.
At the following level there is in principle a $t^{8;3}=t^{8,3}\oplus t^{9,2}\oplus t^{10,1}\oplus t^{11}$, only the first of which is visible in the $E_8$ case.
A brief look at the $E_{11}$ tensor hierarchy reveals a $t^{9;1,1}=t^{9,1,1}\oplus t^{10,1}$ in $\cR_1$ at the same level, of course coupling to some non-propagating gauge field $B_{m[10],n,p}$ in $E_{11}$ that is thought to be the top-down origin of Romans massive IIA theory.
For $E_8$ this is only visible in $\cR_2$ where we find a $t^{8;1,1}=t^{8,1,1}$.
Does this correspond to a brane?
It has been called the $8^{(1,0)}$-brane using our standard notation for branes with specific scalings with respect to different radii, and sometimes it has been called the M9-brane.
Since it only begins to appear in $\cR_1$ from $E_9$ onwards, we do not consider it or any other such branes in this paper, but they remain a very interesting aspect of the pursuit of $\frac12$-BPS multiplets.
To be concise, the tensor hierarchy corresponds to this tensor product structure $t^{A;B,C,D,\dots}=t^A\otimes t^{B,C,D,\dots}$ with the first component an $A$-form generator.
For our purposes we take the `shortest' irreducible component in each such tensor product as the source of the brane, e.g.~the $t^{8,3}$ in $t^{8;3}=t^{8,3}\oplus t^{9,2}\oplus t^{10,1}\oplus t^{11}$ has the shortest columns in its Young tableau compared to the other components.\footnote{For $t^{m[8];n[3]}$, $t^{m[8];n[6],p}$, and $t^{m[8];n[8],p}$, the semicolon is redundant since we only consider $n\leq8$, they only appear for $n\geq8$, and in $n=8$ a tensor product between an eight-form and any $\mathrm{GL}(8)$ tensor gives a tensor with an additional column of height eight glued to the left-hand side of its Young tableau. Note that there are extra generators that start to appear from $\cR_2$ onwards, including a $t^{m[8];n,p}$ and a $t^{m[8];n[4],p}$ in $\cR_2$, that we do not write explicitly. The $t^{m[8];n,p}$ can be thought of as descending from the $t^{m[9];n,p}$ in the $\cR_1$ of $E_n$ for $n\geq9$. For more details on the generators in the tensor hierarchy formulated as a differential graded superalgebra, see \cite{Bossard:2017wxl}.}
\begin{align*}
\begin{array}{c||c|c|c|c|c|c|c|c}
\cR_1
    & t_m & t^{m[2]} & t^{m[5]} & t^{m[7];n} & t^{m[8];n[3]} & t^{m[8];n[6]} & t^{m[8];n[8],p} & \dots \\ 
\cR_2
    & & t^m & t^{m[4]} & t^{m[6];n} & t^{m[7];n[3]} \; t^{m[8];n,p} & t^{m[7];n[6]} \; t^{m[8];n[4],p} & t^{m[7];n[8],p} \; \dots & \dots \\
\cR_3
    & & t & t^{m[3]} & t^{m[5];n} & t^{m[6];n[3]} \; t^{m[7];n,p} & t^{m[6];n[6]} \; t^{m[7];n[4],p} \; \dots & t^{m[6];n[8],p} \; \dots & \dots \\
\cR_4
    & & & t^{m[2]} & t^{m[4];n} & t^{m[5];n[3]} \; t^{m[6];n,p} & t^{m[5];n[6]} \; t^{m[6];n[4],p} \; \dots & t^{m[5];n[8],p} \; \dots & \dots \\
\vdots
    & & & \vdots & \vdots & \vdots & \vdots & \vdots & \ddots 
\end{array}
\end{align*}
We only consider finite-dimensional exceptional groups, but if we were to include groups at least as large as $E_9$ then we would need to include components with blocks of at least nine antisymmetric indices, which are omitted.
The ``\,$\dots$'' denote generators that are present up to and including $E_8$ but here we do not write them all.
A component $t^{m[A];n[B],p[C],\dots}$ in $\cR_1$ descends to a $t^{m[A-(k-1)];n[B],p[C],\dots}$ in $\cR_k$.
In the $\mathrm{GL}(n)$ decomposition, the differential $\hat{\partial}:\cR_k\to\cR_{k-1}$ is just the de Rham differential, treating $t^{m[A];n[B],p[C],\dots}$ as an $A$-form.

One observes \cite{Fernandez-Melgarejo:2019mgd} that $t^{m[A],n[B],p[C],\dots}$ in the decomposition characterises a brane $b^{(\dots,d,c)}$ with
\begin{equation}
    b = A - B \,,\quad
    c = B-C \,, \quad
    d = C-D \,, \quad
    \dots
    \label{eq:b,c,d,...}
\end{equation}
One can understand $A$ as some `total' (non-transverse) dimensionality of the worldvolume, including genuine brane degrees of freedom plus smearing, and $B$ as the total number of isometry (smearing) directions, and so on.
Consequently, there are $B$ isometry vectors
\begin{equation}
    V_I^m = (V_1^m,\dots,V_B^m)
    \equiv ( v_1^m, \dots, v_c^m, w_1^m, \dots, w_d^m, \dots) \,.
\end{equation}

As above, the final characterising brane charge for any $b$\hspace{0.4mm}-brane is expected to be valued in $\cR_{b+1}$ and the corresponding brane charge constraint at $\cR_{b+2}$ is
\begin{equation}
    t_{(b+1)}^{m[B],n[B],p[C],q[D],\dots} \otimes \partial \, \big|_{\cR_{b+2}} = 0 \,.
\end{equation}
This is because the $t^{m[A],n[B],p[C],q[D],\dots}\in\cR_1$ corresponds to $t^{m[A-b],n[B],p[C],q[D],\dots}$ in the $\cR_{b+1}$, which becomes $t^{m[B],n[B],p[C],q[D],\dots}$ using \eqref{eq:b,c,d,...}.

In general, this characterising brane charge can be parameterised by the $B$ vectors $V_I$ as
\begin{equation}
    t^{m[B],n[B],p[C],\dots}
    \;\sim\;
    V_{1}^{[m_1} \dots V_B^{m_B]} \;
    V_{1}^{[n_1} \dots V_B^{n_B]} \;
    V_{1}^{[p_1} \dots V_C^{p_C]} \dots
\end{equation}
with the brane charge constraint amounting to
\begin{equation}
     t^{m[B],n[B],p[C],\dots} \otimes \partial \, \big|_{\cR_{b+2}} = 0
     \;\;\sim\;\;
     V_{1}^{[m_1} \dots \, V_B^{m_B]} \;
     V_{1}^{[n_1} \dots \, V_C^{n_C]}
     \dots \, \partial_{m_1} = 0 \,.
\end{equation}
Consequently, the brane charges and the branes themselves can only exist when these $B$ vectors $V_I$ correspond to isometry directions, i.e.~all functions $f$ on the worldvolume satisfy
\begin{equation}
    V_I^m \partial_m f =0 \, .
\end{equation}
Hence, the only branes $b^{(\dots,d,c)}$ corresponding to objects in the $\frac{1}{2}$-BPS brane multiplet in supergravity are the objects that appear in the $\mathrm{GL}(n)$ or $\mathrm{GL}(n-1)$ decomposition of $E_{n}$ exceptional field theory, and which possess $B=c+d+\dots$ isometry directions.

The complete $\cR_1$ current as a solution\footnote{Of course, in principle, this is only a solution as a complete tensor hierarchy of currents, but the construction of this is straightforward and does not reveal any new structure here.} to the constraints \eqref{eq:BraneChargeSmall} for an exotic brane $b^{(\dots,d,c)}$ is
\begin{align}
\begin{aligned}
    t_M^{b^{(\dots,d,c)}} &= (P_m,\, 0\,,\, \dots\,,\, 0\,,\, t^{m[A],n[B],p[C],\dots},\, 0\,,\,\dots\,) \\
    &\hspace{-10mm}= (P_m,\, 0\,,\, \dots\,,\, 0\,,\, \rd Y^{[ m_1} \dots \rd Y^{m_b} V_1^{m_{b+1}} \dots V_B^{m_A]} V_1^{[n_1} \dots V_B^{n_B]} V_1^{[p_1} \dots V_C^{p_C]}\,\dots\,,\, 0\,,\,\dots\,) \,.
    \label{eq:braneChargeR1exoticbrane}
\end{aligned}
\end{align}
Let us assume that the relevant components of the target space metric $\cM_{MN}$ are parameterised in terms of the on-section metric $g_{mn}$ as
\begin{equation}
    \cM_{m[A],n[B],\dots|a[A],b[B],\dots}
    = A! \, B! \, \dots \, g_{m_1[a_1|} g_{m_2|a_2|} \dots \, g_{m_A|a_A]} \, g_{n_1[b_1|} \dots \, g_{n_B|b_B]} \, \dots
\end{equation}
The Hamiltonian \eqref{eq:exoticBraneHamiltonianConjecture} that was proposed in the previous section for the conjectured action \eqref{eq:exoticNGgauged} of the worldvolume dynamics of exotic branes $b^{(\dots,d,c)}$ is naturally reproduced from the duality-covariant Hamiltonian density is
\begin{equation}
    H = \cM^{MN} t_M t_N   
\end{equation}
for the brane charge solution \eqref{eq:braneChargeR1exoticbrane}.
As mentioned above, there is a natural coupling of an exotic brane to the associated mixed-symmetry field $A_{m[A+1],n[B],p[C],\dots}=A_{m[1+b+c+\dots],n[c+\dots],\dots}$.
In the generalised metric, and hence the Hamiltonian, this is included in the component
\begin{equation}
    \cM^{k}{}_{m[A],n[B],\dots} \sim \cM^{k}{}_{m_1\dots m_A,n_1\dots n_B,\dots}
    = g^{km_{A+1}} A_{m_1\dots m_{A+1},n_1\dots n_B,\dots} \,.
\end{equation}
The associated term in the Hamiltonian is
\begin{equation}
    H = \dots + \cM^{k}{}_{m[A],n[B],\dots} \, t_{k} \, t^{m[A],n[B],\dots} \,,
\end{equation}
and the contribution of this term to the worldvolume action is
\begin{equation}
    S_A \sim \int\!\rd Y^{m_1} \wedge \dots \wedge \rd Y^{m_{b+1}} \, V_{1}^{m_{b+2}} \dots V_B^{m_{A+1}} \, V_{1}^{n_{1}} \dots V_B^{n_B} \, V_{1}^{p_{1}} \dots V_C^{p_C} \, A_{m[A+1],n[B],p[C],\dots} \,.
\end{equation}
From this, it is clear that the worldvolume has to be transverse to these isometry directions.

To summarise this procedure, the choice of a brane charge solution breaks the $E_n$-covariance down to that of $\mathrm{GL}(n)$ or $\mathrm{GL}(n-1)$, and specifies one of the possible $\frac{1}{2}$-BPS branes.
As the example discussed here, the brane charge solution \eqref{eq:braneChargeR1exoticbrane} characterises a possibly exotic brane $b^{(\dots,d,c)}$ together with the generic Hamiltonian \eqref{eq:PhaseSpaceFormalismHamiltonian} and Poisson structure \eqref{eq:CurrentAlgebraSmall}.

\subsection{Duality-covariant phase space formalism for zero-branes}
\label{subsec:duality-covariant_phase_space_zero-branes}

The analysis in Section~\ref{subsec:Hamiltonians_exotic_branes} describes the dynamics of arbitrary $\frac{1}{2}$-BPS branes in supergravity and ExFT backgrounds.
Now we specify to the $p=0$ case.
This will provide an opportunity to compare with the well-established duality-covariant $p$\hspace{0.4mm}-brane actions for $p=0$ up to $E_7$ \cite{Blair:2017gwn} and for $p=1$ up to $E_6$ \cite{Arvanitakis:2018hfn}.
We shall also discuss the worldline theory of the exotic zero-brane $0^{(1,7)}$ here.
For this, we need to extend the setup in the previous section to include ancillary transformations of $E_8$ ExFT in terms of the constrained gauge parameter $\Sigma_M$.

\subsubsection*{Equivalence to the $E_n$ exceptional particle action for $n<8$}

Let us first compare Blair's particle action reviewed earlier in Section~\ref{subsec:exceptional-particle} with the duality-covariant phase space formalism introduced above.
The latter takes a particularly simple form for zero-branes, but has not been discussed in the literature explicitly.

Again, we consider the embedding $(X^\mu(\tau),Y^M(\tau))$ of the particle inside an ExFT background.
Here we will describe the dynamics in the Hamiltonian formalism.
The external phase space then has coordinates $X^\mu$ and their canonical momentum $P_\mu$.
For the internal dynamics, the hierarchy of currents for a zero-brane reduces to something that we can consider to be the generalised momentum $P_M\in\cR_1$.
There are no currents $t_{M_k}\in\cR_k$ with $k>1$.
The current algebra \eqref{eq:CurrentAlgebraSmall} is trivial and can be interpreted as commuting (equal time) momenta
\begin{equation}
    \{ P_M,P_N \} = 0 \,.
\end{equation}
Subject to constraints, the `internal phase space' is described by functions of the extended coordinate fields $Y^M$ and their canonical momentum $P_M$.
The charge constraints \eqref{eq:BraneChargeConstraintZero} and \eqref{eq:BraneChargeSmall} can be given as
\begin{align}
    P \otimes \partial \, \big|_{\cR_2} &= 0 \,,&
    P \otimes P \, \big|_{\cR_2} = 0 \,,
    \label{eq:BraneChargeConditionZeroBrane}
\end{align}
in addition to the standard section constraint $\partial\otimes\partial\,|_{\cR_2}=0$ that restricts the dependence of the fields and parameters on the internal coordinates $Y^M$.
From the ExFT point of view, these constraints will ensure that the generalised Lie derivative is realised on the phase space of the zero-brane for sections $\Phi=\Phi^MP_M$.
Of course, on a solution to the charge constraint such as the standard M-theory solution $P_M=(P_m,0,\dots,0)$, these sections of generalised vector fields reduce to sections of regular vector fields, and the generalised Lie derivative becomes the standard Lie derivative.
Another way to view these constraints is the following.
One expects a particle in $n$-dimensional supergravity to have only $n$ degrees of freedom.
The constraints \eqref{eq:BraneChargeConditionZeroBrane} remove the unphysical components of $P_M$ that are introduced to achieve duality-covariance.

The coupling of this worldline to an $E_n$ ExFT background $(g_{\mu\nu},\cM_{MN},A_\mu{}^M)$ for $n<8$ is given by the Hamiltonian $H=g^{\mu\nu}\pi_\mu\pi_\nu+\cM^{MN}t_{M}t_{N}$ in \eqref{eq:HamiltonianFull} where $\pi_\mu=P_\mu-A_\mu{}^M P_M$.
The Legendre transform\footnote{Note that a Legendre transform of this current algebra approach for higher-dimensional branes is in principle possible but much more complicated since the duality-covariant currents do not satisfy trivial canonical Poisson brackets, cf.~\eqref{eq:CurrentAlgebraSmall}.}
\begin{equation}
    L = P_M \dot{Y}^M + P_\mu \dot{X}^\mu - H
\end{equation}
of this Hamiltonian is the Lagrangian of Blair's exceptional action in \eqref{eq:S_Blair}.
To be precise, we obtain a version without covariant derivatives but with constraints that correspond to potential isometries.
In particular, the brane charge constraint $P\otimes\partial\,|_{\cR_2}$ is equivalent to the equation of motion \eqref{eq:Blair_eomV} for the worldline field $V^M$ in the exceptional particle model.

\subsubsection*{Introducing ancillary transformations on the worldline}

As reviewed in Section~\ref{subsec:E8_ExFT}, when trying to generalise this to $E_8$ and beyond, the algebra of generalised Lie derivatives no longer closes.
This can be resolved by the introduction of ancillary transformations.
In addition to the gauge parameter $\Lambda^M\in\cR_1$, the new covariantly constrained parameter $\Sigma_M\in\overline{\cR}_1$ is introduced.
More generally, this problem always occurs when the generalised Lie derivative acts on sections of the $\cR_{9-n}$-bundle in $E_n$ ExFT.
For example, $E_7$ ExFT has a constrained two-form $B_{\mu\nu M}$ that transforms with a constrained gauge parameter.
This article focuses mainly on point particles and zero branes, so only the $\cR_1$ is relevant.
We shall investigate only the $E_8$ case here.

Let us present a simple realisation of the ancillary transformations on the zero-brane phase space.
One way to solve this is to introduce a new generator, i.e.~fields $Q^M$, with equal-time Poisson brackets
\begin{align}
    \{ Q^M , Q^N \} &= f^{MN}{}_P \, Q^P \,,&
    \{ P_M , Q^N \} &= f_M{}^N{}_P \, Q^P \,,
    \label{eq:PoissonQ}
\end{align}
where $f^{MN}{}_P$ are the $\mathfrak{e}_8$ structure constants.
The aim would be that
\begin{equation}
    \big\{ \Lambda^M t_M + \Sigma_M Q^M , \Phi^N P_N \big\}
    = \cL_{(\Lambda,\Sigma)} \Phi^N t_N \,,
    \label{eq:Poisson_gen_Lie_deriv}
\end{equation}
with the generalised Lie derivative \eqref{eq:gen_Lie_deriv} acting on a vector $\Phi$ of generalised density weight $\lambda_{(\Phi)}=0$\,.
Again, for this we emphasise that $P_M$ is subject to the charge constraints \eqref{eq:BraneChargeConditionZeroBrane}.
When comparing with \eqref{eq:gen_Lie_deriv}, one notices that \eqref{eq:Poisson_gen_Lie_deriv} is true on the worldline if we assume
\begin{equation}
    Q^M P_M = 0
    \qquad\Longrightarrow\qquad
    Q^M\partial_M = 0 \label{eq:BraneChargeQ}
\end{equation}
as a brane charge constraint for the new fields $Q^M$.
The ancillary part then drops out consistently for the algebra \eqref{eq:PoissonQ}.

This condition does \textit{not} make the term $\Sigma_MQ^M$ trivial in general.
The presence of isometries leads to non-trivial solutions to this constraint.
The constraints \eqref{eq:BraneChargeQ} on $Q^M$ are analogous to those of the gauge field $V^M$ in Blair's exceptional particle action -- see Section~\ref{subsec:exceptional-particle}.
However, $V^M$ does not appear there as a physical field with non-trivial Poisson brackets.

On the $E_8$ ExFT target space, the constrained one-form $B_{\mu M}$ is part of the tensor hierarchy.
This naturally couples to the worldvolume via the minimal coupling in terms of
\begin{equation}
    \pi_\mu = P_\mu - P_M A_\mu{}^M - Q^M B_{\mu M} \,.
    \label{eq:pi_mu_worldvolume}
\end{equation}
The new worldline fields $Q^M(\tau)$ are introduced in this top-down approach as generators that enlarge the phase space and which satisfy the Poisson algebra \eqref{eq:PoissonQ}.
In Section~\ref{sec:E8_worldline_action} we will attempt to realise and interpret this new worldline degree of freedom $Q^M$ in an action.
The answer will be that it arises as an internal degree of freedom in a new coadjoint orbit term.

\subsubsection*{Zero branes in the $E_8$ case}

In the M-theory section, the brane charge constraints $P\otimes P\,|_{\cR_2}=0$ and $P\otimes\partial\,|_{\cR_2}=0$ for zero-branes \eqref{eq:BraneChargeConstraintZero} have two distinct solutions.\footnote{Similarly, one could obtain the F0/D0 doublet in IIA string theory, but this goes beyond the scope of this paper.}
We will repeat this analysis in Section~\ref{sec:E8_worldline_action} from the point of view of worldline actions.

The first solution is the standard point particle.
The corresponding solution is $P_M=(P_m,0,\dots,0)$ if we work on the non-trivial section $\partial_M=(\partial_m,0,\dots,0)$.
Consequently, the constraint \eqref{eq:BraneChargeQ} implies that $Q^M=(0,Q_{\widetilde{m}})$, where $Q_{\widetilde{m}}$ lies along the unphysical directions that are transverse to the physical section.
The generalised Lie derivative for sections $\Phi=\Phi^MP_M$ reduces to the usual Lie derivative, as expected.
Since the ancillary parameter is covariantly constrained, i.e.~$\Sigma_M=(\Sigma_m,0,\dots,0)$, the fate of the ancillary transformations in \eqref{eq:gen_Lie_deriv} for the particle is that $\Sigma_MQ^M$ vanishes.

More interesting is the exotic $0^{(1,7)}$ solution.
Exotic branes rely on the presence of isometries in the target space.
The associated solutions to the brane charge constraints become possible only due to these isometries.
Indeed, only in the presence of eight isometries is a more general solution possible for $E_n$ with $n\geq8$.
Considering the trivial solution to the section constraint where $\partial_Mf$ vanishes for any physical (worldline) function $f$, the brane charge constraint
\begin{equation}
    P \otimes \partial \, \big|_{\mathbf{1}\,\oplus\,\mathbf{248}\,\oplus\,\mathbf{3875}} = 0
\end{equation}
is satisfied trivially and does not restrict the form of $P_M$.
Instead, all the restriction comes from the second condition in \eqref{eq:BraneChargeConditionZeroBrane},
\begin{equation}
    P \otimes P \, \big|_{\mathbf{1}\,\oplus\,\mathbf{248}\,\oplus\,\mathbf{3875}} = 0 \,.
\end{equation}
Given the M-theory decomposition $E_8\to\mathrm{GL}(8)$ in equation \eqref{eq:R1Decomp}, there are two inequivalent possible choices for maximal, i.e.~eight-dimensional, solutions:
\begin{itemize}[leftmargin=*]
\item
    $P_M= (P_m,0,\dots,0)$ in the case of the standard point particle subject to a dimensional reduction along the eight isometry directions.
\item
    $P_M=(0,\dots,0,P^{m[8],n[8],p})$.
    Without the isometry condition $\partial_Mf=0$ for worldline functions $f$, this solution would be forbidden by the charge constraint $P\otimes\partial\,|_{\cR_2}$.
    This is the solution we want to discuss here since it corresponds to the exotic zero-brane $0^{(1,7)}$.
\end{itemize}
One way to understand this solution is the following.
Consider the parameterisation
\begin{equation}
    P_{M} = (P_m , 0 , \dots , 0 , P^{m[8],n[8],p})
    = \big( P_m , 0 , \dots , 0 , v_1^{[m_1} \dots v_7^{m_7} w^{m_8]} v_1^{[n_1} \dots v_7^{n_7} w^{n_8]} w^p \big) \,,
\end{equation}
where $P_m$ is the canonical momentum, and where $v^m_i$ and $w^m$ are scalar fields on the worldline that are target space vectors.
This is analogous to the higher-dimensional exotic branes \eqref{eq:braneChargeR1exoticbrane}.
Of course, this violates the charge constraint $P\otimes P\,|_{\cR_2}$.
However, in fact, we obtain the isometry constraints on $v$ and $w$ as a consequence of this charge constraint.
The charge constraint $P\otimes P\,|_{\mathbf{1}\,\in\,\cR_1}\equiv\kappa^{MN}P_MP_N=0$ corresponds to $w$ being an isometry in the sense that
\begin{equation}
    \kappa^{MN} P_M P_N
    = 2\,\varepsilon_{m_1\dots m_8} \varepsilon_{n_1\dots n_8} v_1^{m_1} \dots v_7^{m_7} w^{m_8} v_1^{n_1} \dots v_7^{n_7} w^{n_8} w^p p_p
    \;\sim\; w^m p_m = 0 \,.
\end{equation}
Similarly,\footnote{Our shorthand notation $P\otimes P\,|_{\cR_2} = 0$ for this part of the charge constraint also leads to the simple fact that $P\otimes P\,|_{\mathbf{248}}\equiv P_M P_N f^{MN}{}_K \equiv 0$ is satisfied due to the skewsymmetry of $f^{MN}{}_K$.}
\begin{equation}
    P\otimes P\,|_{\mathbf{3875}\,\in\,\cR_2}
    \equiv \mathbb{P}_{(\mathbf{3875})}{}^{MN}{}_{PQ} P_M P_N = 0
    \qquad\Longrightarrow\qquad
    v^m_i p_m = 0 \,.
\end{equation}
Alternatively, we could have set $P_m$ to zero from the very beginning.
As discussed in Section~\ref{subsec:overview_of_branes}, this is a consequence of an exotic brane being delocalised or smeared.
The fact that we need $\partial_N=0$ is a coincidence.
We work in $E_8$ ExFT and the $0^{(1,7)}$ is the only non-standard zero-brane requiring at most eight transverse directions, and which requires eight isometry directions.
A different way to interpret the $P_m=0$ solution is to consider the $0^{(1,7)}$-brane as the counterpart to the non-geometric section $\tilde{\partial}^p:=\varepsilon_{m[8]}\varepsilon_{n[8]}\partial^{m[8],n[8],p}\neq0$ as a solution of the $E_8$ section constraint.
The natural solution to $P\otimes\partial\,|_{\cR_2}=0$ is then $P_M=(0,\dots,0,P^{m[8],n[8],p})$.

The dynamics along the external direction $x^\mu$ is unrestricted.
We can write the Hamiltonian in a pure metric background as
\begin{equation}
    H = g^{\mu\nu} \pi_\mu \pi_\nu + \cM^{MN} P_M P_N
    = g^{\mu\nu} \pi_\mu \pi_\nu + v_1^4 \dots v_7^4 w^6 \approx 0 \,.
\end{equation}
This agrees with the Hamiltonian system of the $0^{(1,7)}$-brane action proposed in Section~\ref{subsec:exotic_brane_dynamics}.

Now let us return to the ancillary transformation.
Assume that the target space parameter $\Sigma_M$ is still constrained, and choose the 11D solution to the section constraint on the target space.
In fact, the only non-trivial contribution from the generalised Lie derivative pulled back to the worldline is
\begin{equation}
    \cL_{(\Lambda,\Sigma)} \Phi \, \big|_{0^{(1,7)}}
    = f^{MN}{}_K \, \Sigma_M P_N \Phi^K \,.
\end{equation}
This does not vanish trivially if $P_M$ and $\Sigma_M$ are not constrained with respect to each other.
Hence, the ancillary transformation describes a non-trivial shift of the $\Phi_m$ component.\footnote{In our earlier notation this is the $\Phi_{m[8],n[8],p}$ component of the $\cR_1$ representation.}

\section{An $E_8$-covariant worldline action}
\label{sec:E8_worldline_action}

In this section we shall present an $E_8$-covariant worldline action pulled back from the $E_8$ ExFT target space.
This model generalises Blair's exceptional particle \cite{Blair:2017gwn} and is invariant under all the $E_8$ gauge transformations, including those involving the constrained gauge parameter.
The constrained gauge field $B_{\mu M}$ is incorporated in the most minimal way via a coadjoint orbit term.

\subsection{A worldline action for $E_8$ particles}
\label{subsec:E8_particle}

We consider a worldline embedding into the $E_8$ ExFT target space.
The background fields are the external metric $g_{\mu\nu}$, internal metric $\cM_{MN}$, and the external one-forms $A_\mu{}^M$ and $B_{\mu M}$.
The one-forms appear via their worldline pullbacks $\dot{X}^\mu A_\mu{}^{M}$ and $\dot{X}^\mu B_{\mu M}$ that do not transform as gauge connections when the gauge parameters $(\Lambda^M,\Sigma_M)$ depend on $Y^M$ along the worldline.
This makes it necessary to introduce some additional worldline fields, the first of which is $V^M$ that had been introduced in \cite{Blair:2017gwn} to define the covariantised internal velocity $\cD_\tau Y^M$ in \eqref{eq:DY_Blair}.
It continues to satisfy a constraint \eqref{eq:V_partial_constraint} that should not be confused with the section constraint.
We also introduce a new worldline one-form $W_M$ carrying a constrained index so that the completed pullback
\begin{equation}
    \widetilde{B}_M := \dot{X}^\mu B_{\mu M} + W_M
    \label{eq:hatB=dotX_B+W}
\end{equation}
transforms as a worldline gauge connection.

Another new ingredient is the unconstrained object $Q^M$ that couples to $\widetilde{B}_M$.
This `duality charge' works differently to the usual momentum $P_M$ and in order to include it in a gauge-covariant way we introduce a standard coadjoint orbit term.
Let $g\in E_8$ be a group-valued worldline variable and let $\Gamma=\Gamma^Mt^*_M\in\mathfrak{e}_8^*$ be a fixed element of the dual algebra, where $t^M$ forms an $\mathfrak{e}_8$ basis and $t^*_M$ forms a basis of $\mathfrak{e}_8^*$\,.
The duality charge is defined by the coadjoint action of $g$ on the fixed element, i.e.
\begin{align}
    Q(\tau) = Q^M(\tau) t^*_M := \Ad^*_{g(\tau)} \Gamma \,,
    \label{eq:Q^M}
\end{align}
where the coadjoint action $\Ad^*$ is defined by $(\Ad^*_gt^*_M)X:=t^*_M(\Ad_{g^{-1}}X)$ for any $X\in\mathfrak{e}_8$.
Then, using \eqref{eq:hatB=dotX_B+W}, define $\widetilde{B}:=\widetilde{B}_M\,t^M$.
Note that $\mathfrak{e}_8$ is semisimple, so we can identify $\mathfrak{e}_8$ with $\mathfrak{e}_8^*$ and rewrite the coadjoint-adjoint pairing using the isomorphism $\flat:\mathfrak{e}_8\to\mathfrak{e}_8^*:X\mapsto\kappa(X,\cdot\,)$ and its inverse $\sharp:\mathfrak{e}_8^*\to\mathfrak{e}_8$.

The $E_8$ particle action is
\begin{equation}
    S_{E_8} = S_\mathrm{kin} + S_\mathrm{orb} \,,
    \label{eq:S_total}
\end{equation}
where the kinetic sector $S_\mathrm{kin}$ is Blair's action \eqref{eq:S_Blair}, and the orbit sector is given by
\begin{equation}
    S_\mathrm{orb}
    = \int\!\rd\tau\, \big\langle \Gamma , J \big\rangle
    = \int\!\rd\tau\, \Gamma^M J_M \,,
    \label{eq:S_orbit}
\end{equation}
with the natural pairing $\langle t^*_M,t^N\rangle:=t^*_M(t_N)=\delta_M^N$.
We defined currents
\begin{align}
    J &= J_M t^M := g^{-1} \big( \partial_\tau + \widetilde{B} \big) g \,,&
    j &= j_M t^M := g^{-1}\dot{g} \,,
\end{align}
so that we have
\begin{equation}
    J_M = j_M + g^N{}_M \widetilde{B}_N \,,
\end{equation}
where $g^M{}_N$ is the adjoint representation matrix, i.e.~$g^{-1}t^Mg=g^M{}_Nt^N$ and can write $Q^M=g^M{}_N\Gamma^N$ for the orbit charge.

It is helpful to understand orbit degrees of freedom as `internal' in the same sense as spin or colour.
The internal orbit mechanics should not be confused with the internal part of the external-internal split in the target space ExFT.

The stabiliser of $\Gamma$ and the orbit through $\Gamma$ are defined by
\begin{align}
    G_\Gamma &:= \{g\in E_8\,|\,\Ad^*_g\Gamma=\Gamma\} \,,&
    \cO_\Gamma &:= \{\Ad^*_g\Gamma\,|\,g\in E_8\} \cong E_8/G_\Gamma \,.
\end{align}
A decomposition $\mathfrak{e}_8=\mathfrak{g}_\Gamma\oplus\mathfrak{m}_\Gamma$ is chosen with
\begin{equation}
    \mathfrak{m}_\Gamma
    \cong \rT_\Gamma\cO_\Gamma
    = \big\{ \ad^*_X\Gamma \,\big|\, X\in\mathfrak{e}_8 \big\}
    \cong \mathfrak{e}_8/\mathfrak{g}_\Gamma \,,
\end{equation}
where $\rT_\Gamma\cO_\Gamma$ denotes the tangent space to the coadjoint orbit at the point $\Gamma$.
We have chosen $\mathfrak{m}_\Gamma$ to be the $\kappa$-orthogonal complement since $E_8$ is semisimple.
This choice may not be available for a generic orbit, especially a nilpotent one, but such a complement can be chosen locally.

The orbit sector includes a Kirillov--Kostant--Souriau (KKS) term \cite{Kirillov:1962KKS,Kostant:1970KKS,Souriau:1970KKS} and a Wong term \cite{Wong:1970fu}
\begin{align}
    S_\mathrm{orb}^\mathrm{(KKS)}
    &= \int\!\rd\tau\, \big\langle \Gamma , g^{-1}\dot{g} \big\rangle
    = \int\!\rd\tau\, \Gamma^M j_M \,,&
    S_\mathrm{orb}^\mathrm{(Wong)}
    &= \int\!\rd\tau\, \big\langle \Gamma , g^{-1}\widetilde{B}g \big\rangle
    = \int\!\rd\tau\, Q^M \widetilde{B}_M \,,&
    \label{eq:S_orbit_KKS_Wong}
\end{align}
using \eqref{eq:Q^M} and
\begin{equation}
    \big\langle \Gamma , g^{-1} \widetilde{B} g \big\rangle
    = \big\langle \Ad^*_g \Gamma , \widetilde{B} \big\rangle
    = \big\langle Q , \widetilde{B} \big\rangle
    = Q^M \widetilde{B}_M \,.
\end{equation}
The Euler--Lagrange equation of $W_M$ implies that the Wong term is trivial.
The reason is that $W_M$ is constrained to lie along the physical directions, and it appears only in $Q^MW_M$.
Its Euler--Lagrange equation thus kills the only components of $Q^M$ that enter the action in the first place.
This leaves only the KKS term which is `topological' in the sense that it is a metric-independent first-order orbit action.
This is consistent with the role of $B_{\mu M}$ in $E_8$ ExFT as part of the Chern--Simons term responsible for enforcing the on-shell duality equation between the vectors and scalars.
In the 11D section, the $\Sigma_M$ symmetry becomes a St\"uckelberg shift symmetry that removes dual graviton components, and in the compactification to 3D the dual scalars can propagate while the $B$ field is pure gauge.
So, the absence of $B_{\mu M}$ from the local geodesic action is not a problem needing a solution.
That said, including a term quadratic in $J$ is a way to allow $W_M$ to be non-trivial and the orbit to carry dynamics.
We shall explore this possibility in Appendix~\ref{app:deformation}.

Some comments about the kinetic and orbit terms are needed.
The kinetic term $S_{\mathrm{kin}}$ is separately invariant under the target space gauge transformations, and the orbit term $S_{\mathrm{orb}}$ is not introduced to cancel its variation.
Instead, its main purpose is to allow the worldline to probe the part of the theory with the constrained field $B_{\mu M}$ and gauge parameter $\Sigma_M$ that distinguish $E_8$ ExFT from the ExFTs with smaller exceptional groups.
Moreover, the phase space analysis of Section~\ref{subsec:duality-covariant_phase_space_zero-branes} proposes to introduce new generators $Q^M$ obeying the $\mathfrak{e}_8$ Poisson algebra.
From the worldline perspective, we are using the orbit term to supply a minimal symplectic realisation of this algebra, with $Q$ the moment map that is minimally coupled to the worldline connection $\widetilde{B}$.

\subsection{Symmetries and consistency checks}

The $E_8$ model has the following worldline variables:~(i) embedding fields $X^\mu$ and $Y^M$, (ii) einbein $e$, (iii) internal conjugate momenta $P_M$ that satisfy the brane charge constraints, (iv) an orbit variable $g\in E_8$ and a fixed element $\Gamma\in\mathfrak{e}_8^*$ with dynamical charge $Q=\Ad^*_g\Gamma$, and (v) worldline compensator fields $V^M$ and $W_M$.
The background fields are those of $E_8$ exceptional field theory, namely the external and internal metrics and the pair of gauge fields $A_\mu{}^M$ and $B_{\mu M}$.
We will also consider the momenta $P_\mu$ conjugate to $X^\mu$ in Section~\ref{subsec:E8_Hamiltonian} when we perform a Hamiltonian analysis of the $E_8$ worldline model.

\subsubsection*{Worldline reparametrisations}

The $E_8$ action is invariant under reparametrisations.
This is the usual local diffeomorphism symmetry on the worldline, so infinitesimally the parameter is an arbitrary smooth function.
Worldline scalars $\Phi$ (i.e.~$X^\mu$, $Y^M$, $P_M$, $g$) transform as $\delta_\epsilon\Phi=\epsilon\dot{\Phi}$.
Worldline one-forms $A$ (i.e.~$e$, $V^M$, $W_M$) transform as $\delta_\epsilon A=\epsilon\dot{A}+\dot{\epsilon}A=\frac{\rd}{\rd\tau}(\epsilon A)$.
This is just the Lie derivative by $\epsilon\hspace{0.4mm}\partial_\tau$.

This ensures that $\delta_\epsilon(\cD_\tau Y^M)=\frac{\rd}{\rd\tau}(\epsilon \cD_\tau Y^M)$ and, similarly, $\delta_\epsilon J=\frac{\rd}{\rd\tau}(\epsilon J)$.
One can check that the different sectors of the Lagrangian transform as
\begin{align}
    \delta_\epsilon L_\mathrm{kin} &= \frac{\rd}{\rd\tau}(\epsilon L_\mathrm{kin}) \,,&
    \delta_\epsilon L_\mathrm{orb} &= \frac{\rd}{\rd\tau}(\epsilon L_\mathrm{orb}) \,,
\end{align}
so the action is reparametrisation invariant up to a boundary term.

\subsubsection*{Orbit symmetry and stabiliser redundancy}

As is standard \cite{Wong:1970fu,Alekseev:2015hda,Alekseev:2018pbv}, the orbit sector is invariant under two important symmetries.
The first of these has the orbit variable transform by group multiplication while $\widetilde{B}$ transforms as a connection:
\begin{align}
    g &\;\longmapsto\; h g \,,&
    \widetilde{B} &\;\longmapsto\; h \widetilde{B} h^{-1} - \dot{h} h^{-1} \,.
    \label{eq:g_transformation}
\end{align}
This leaves $J$ and hence $\langle\Gamma,J\rangle$ invariant.
The duality charge $Q$ transforms covariantly as $Q\mapsto\Ad^*_hQ$.
The infinitesimal form of this symmetry with parameter $\alpha\in\mathfrak{e}_8$ is
\begin{align}
    \delta_\alpha g &= -\alpha g \,,&
    \delta_\alpha \widetilde{B} = \dot{\alpha} + [\widetilde{B},\alpha] \,.
\end{align}

The second of the two symmetries is a redundancy of the orbit parametrisation, i.e.~the orbit point depends only on the coset $g\hspace{0.5mm}G_\Gamma\in E_8/G_\Gamma$ rather than $g$ itself.
Under $g\mapsto gk$ with $k\in G_\Gamma$ we have
\begin{align}
    J &\;\longmapsto\; k^{-1} J k + k^{-1} \dot{k} \,,&
    S_\mathrm{orb} &\;\longmapsto\;
    S_\mathrm{orb} + \int\!\rd\tau\, \big\langle \Gamma , k^{-1} \dot{k} \big\rangle \,.
\end{align}
This leaves $Q$ invariant since $Q=\Ad^*_g\Gamma\mapsto\Ad^*_{gk}\Gamma=\Ad^*_g\Ad^*_k\Gamma=\Ad^*_g\Gamma=Q$.
The infinitesimal form of this symmetry is $\delta_\beta g=g\beta$ and $\delta_\beta\widetilde{B}=0$ with parameter $\beta\in\mathfrak{g}_\Gamma$\,.
This leads to a boundary term that is zero for closed worldlines or when $\beta$ vanishes at the boundary, with $\tau\in[0,1]$\,:
\begin{equation}
    \delta_\beta S_\mathrm{orb}
    = \int\!\rd\tau\, \Big( \big\langle \Gamma , \dot{\beta} \big\rangle + \big\langle \Gamma,[\beta,J] \big\rangle \Big)
    = \int\!\rd\tau\, \frac{\rd}{\rd\tau} \big\langle \Gamma,\beta \big\rangle
    = \big\langle \Gamma,\beta \big\rangle \Big|_{\tau\hspace{0.2mm}=\hspace{0.2mm}0}^{\tau\hspace{0.2mm}=\hspace{0.2mm}1} \,.
\end{equation}

\subsubsection*{Induced $\Lambda$ and $\Sigma$ transformations}

At this stage, we need to work out the gauge transformations of $V^M$ and $W_M$ so that $\cD_\tau Y^M$ and $\widetilde{B}_M$ transform covariantly under the $E_8$ ExFT target space symmetries.
Under these transformations, $S_\mathrm{kin}$ must be invariant since it is built from covariant objects and the worldline scalar density $e^{-1}$, while $S_\mathrm{orb}$ is invariant up to a boundary term since it is a gauged coadjoint orbit action with connection $\widetilde{B}$.
Target space generalised diffeomorphisms act on the background fields and induce transformations on the pullbacks.
The constrained field $B_{\mu M}$ transforms under a $\Sigma$ transformation as $\delta_\Sigma B_{\mu M}=D_\mu\Sigma_M$ with $\Sigma_M$ also covariantly constrained.
If we had chosen the naive pullback $\dot{X}^\mu B_{\mu M}$ as the connection $\widetilde{B}_M$ then this would have given $\delta_\Sigma\widetilde{B}_M=\dot{X}^\mu D_\mu\Sigma_M$.
However, $\Sigma$ depends on $X$ and $Y$, so
\begin{equation}
    \dot{\Sigma}_M = \dot{X}^\mu \partial_\mu \Sigma_M + \dot{Y}^N \partial_N \Sigma_M \,.
\end{equation}
Thus the naive pullback does not transform as a connection.

One way around this could have been to restrict parameters or trajectories via a constraint like $\dot{Y}^N\partial_N\Sigma_M=0$ along the worldline, so that $Q^M\dot{X}^\mu B_{\mu M}$ would be invariant up to a boundary term.
Our approach is instead to introduce a constrained worldline one-form $W_M$ and define the completed pullback \eqref{eq:hatB=dotX_B+W} with $\widetilde{B}:=\widetilde{B}_Mt^M\in\mathfrak{e}_8$.
We require that $\widetilde{B}$ transforms as a connection:
\begin{align}
    \delta_\Sigma \widetilde{B} &= \dot{\Sigma} + [\widetilde{B},\Sigma] \,,&
    \delta_\Sigma \widetilde{B}_M &= \dot{\Sigma}_M + f^{NP}{}_M \widetilde{B}_N \Sigma_P \,.
\end{align}
Using $\delta_\Sigma(\dot{X}^\mu B_{\mu M})=\dot{X}^\mu D_\mu\Sigma_M$ fixes $\delta_\Sigma W_M$ uniquely as
\begin{align}
    \delta_\Sigma W &= \dot{\Sigma} + [\widetilde{B},\Sigma] - \dot{X}^\mu D_\mu \Sigma \,,&
    \delta_\Sigma W_M &= \dot{\Sigma}_M + f^{NP}{}_M \widetilde{B}_N \Sigma_P - \dot{X}^\mu D_\mu \Sigma_M \,.
\end{align}

The orbit sector is invariant under the $\Sigma$ transformation
\begin{align}
    \delta_\Sigma g &= - \Sigma g \,,&
    \delta_\Sigma \widetilde{B} &= \dot{\Sigma} + [\widetilde{B},\Sigma] \,,&
    \delta_\Sigma Q &= - \ad^*_\Sigma Q \,,
\end{align}
where $\ad^*_\Sigma Q$ is the coadjoint action of $\Sigma\in\mathfrak{e}_8$ on $Q\in\mathfrak{e}_8^*$\,.
This symmetry is of the type discussed just below equation \eqref{eq:g_transformation}.
The $\Sigma$ transformation of $Q$ can be expressed in terms of the $\mathfrak{e}_8$ adjoint action as $\delta_\Sigma Q^\sharp=-\ad_\Sigma Q^\sharp=-[\Sigma,Q^\sharp]$, where we used the inverse $\sharp:\mathfrak{e}_8^*\to\mathfrak{e}_8$ of $\flat:\mathfrak{e}_8\to\mathfrak{e}_8^*:X\mapsto\kappa(X,\cdot\,)$.

Consider the action of generalised diffeomorphisms with parameter $\Lambda^M$ on the pulled-back objects of the theory.
We choose the duality charge $Q^M$ to be inert under $\Lambda$ transformations:
\begin{align}
    \delta_\Lambda g &= 0 \,,&
    \delta_\Lambda \Gamma^M &= 0 \,,&
    \delta_\Lambda Q^M &= 0 \,.
    \label{eq:delta_Lambda_g,Gamma,Q=0}
\end{align}
This is natural since none of these objects are target space gauge fields.
The KKS term is invariant, i.e.~$\delta_\Lambda\langle\Gamma,g^{-1}\dot g\rangle=0$\,.
The Wong term transforms as
\begin{equation}
    \delta_\Lambda S_\mathrm{orb}^\mathrm{(Wong)} = \int\!\rd\tau\, \langle Q , \delta_\Lambda \widetilde{B} \rangle
    = \int\!\rd\tau\, Q^M \delta_\Lambda \widetilde{B}_M \,.
\end{equation}
A sufficient condition for a $\Lambda$-invariant Wong term is $\delta_\Lambda \widetilde{B}_M=0$\,.
A weaker condition $Q^M\delta_\Lambda\widetilde{B}_M=0$ could have been used, but this would be less natural since the symmetry would depend on the orbit.
The stronger option fixes the $\Lambda$ transformation of $W_M$ to be\footnote{Following \cite{Blair:2017gwn}, we distinguish the fixed-$Y$ variation $\delta_\Lambda$ from the total variation $\tilde{\delta}_\Lambda$ which includes transport in the extended coordinates.
If one works with $\tilde{\delta}_\Lambda$ instead, then $\tilde{\delta}_\Lambda W_M=-\tilde{\delta}_\Lambda(\dot{X}^\mu B_{\mu M})$\,. This is equivalent to the fixed-$Y$ variation once the same convention is used consistently throughout.}
\begin{equation}
    \delta_\Lambda W_M
    = - \dot{X}^\mu \delta_\Lambda B_{\mu M}
    = \dot{X}^\mu \Lambda^N \partial_M B_{\mu N} - \dot{X}^\mu f^N{}_{PQ}\,\Lambda^P \partial_M \partial_N A_\mu{}^Q \,.
    \label{eq:deltaLambdaW-fixedY}
\end{equation}

In the $E_8$ model there is a new transformation that is not present for smaller groups.
Recall that a vector $T^M$ transforms as
\begin{equation}
    \delta_{\Sigma} T^M
    = f^{MN}{}_P \Sigma_N T^P \,.
    \label{eq:delta_Sigma_vector}
\end{equation}
Correspondingly, the generalised metric transforms as
\begin{equation}
    \delta_{\Sigma} \cM_{MN}
    = f^{PQ}{}_{M} \Sigma_P \cM_{QN}
    + f^{PQ}{}_{N} \Sigma_P \cM_{MQ} \,.
    \label{eq:delta_Sigma_internal_metric}
\end{equation}
We require that $\cD_\tau Y^M$ transforms as a vector, i.e.~as in \eqref{eq:delta_Sigma_vector}.
If we take $Y^M$ and $A_\mu{}^M$ to be inert, then this condition fixes the transformation of $V^M$ to be
\begin{equation}
    \delta_{\Sigma}V^M
    = f^{MN}{}_{P} \Sigma_N \cD_\tau Y^P \,.
    \label{eq:delta_Sigma_V^M}
\end{equation}
One finds that $\cM_{MN} \cD_\tau Y^M \cD_\tau Y^N$ is invariant using \eqref{eq:delta_Sigma_internal_metric} and \eqref{eq:delta_Sigma_V^M}.
Note that \eqref{eq:delta_Sigma_V^M} is compatible with $V^M\partial_M=0$ in \eqref{eq:V_partial_constraint} due to the section constraint since $\Sigma$ carries a constrained index.

\subsection{Equations of motion in the orbit sector}

Identify $\mathfrak{e}_8\cong\mathfrak{e}_8^*$ using the Killing form.
Let $\Gamma^\sharp\in\mathfrak{e}_8$ be the generator corresponding to $\Gamma\in\mathfrak{e}_8^*$ so that $\langle\Gamma,J\rangle=\kappa(\Gamma^\sharp,J)$.
Varying the $E_8$ particle action \eqref{eq:S_total} with respect to $g(\tau)$ gives $\delta J=\dot{\eta}+[J,\eta]$ with $\eta:=g^{-1}\delta g$.
As a result, after integrating by parts and using invariance of the Killing form, we find
\begin{equation}
    \delta S_\mathrm{orb}
    = \int\!\rd\tau\, \kappa \big( \Gamma^\sharp , \dot{\eta} + [J,\eta] \big)
    = \Big[ \kappa(\Gamma^\sharp,\eta) \Big]_{\tau\hspace{0.2mm}=\hspace{0.2mm}0}^{\tau\hspace{0.2mm}=\hspace{0.2mm}1} - \int\!\rd\tau\, \kappa \big( \dot{\Gamma}{}^\sharp + [J,\Gamma^\sharp] , \eta \big) \,,
\end{equation}
and so the equation for $g$ is
\begin{equation}
    \dot{\Gamma}^\sharp + [J,\Gamma^\sharp] \approx 0 \,.
    \label{eq:EOM_g}
\end{equation}
Now recall the definition
\begin{equation}
    Q{}^\sharp := \Ad_g \Gamma^\sharp \,,
\end{equation}
so that \eqref{eq:EOM_g} is equivalent to
\begin{equation}
    \dot{Q}^\sharp + [\widetilde{B},Q{}^\sharp] \approx 0 \,.
    \label{eq:EOM_g_Q}
\end{equation}
In other words, we have a covariant transport equation for $Q$, namely the standard Wong equation.

Now vary the $E_8$ action with respect to $W$.
Since $\delta J=g^{-1}\delta Wg$, we find
\begin{equation}
    \delta S_\mathrm{orb}
    = \int\!\rd\tau\, \kappa(\Gamma^\sharp,g^{-1}\delta Wg)
    = \int\!\rd\tau\, \kappa(Q^\sharp,\delta W ) \,.
\end{equation}
Recall that $W$ is constrained, so the allowed variations $\delta W$ can only belong to the physical subspace of $\cR_1=\mathbf{248}$.
The Euler--Lagrange equation tells us that the duality charge $Q^\sharp$ must be orthogonal to the physical directions with respect to the Killing form.
In the 11D section, the dependence on the internal $y^M$ is restricted to the $y^m\in\mathbf{8}$ ($m=1,\dots,8$), i.e.
\begin{align}
    \partial_M &= (\partial_m,0,\dots,0) \,,&
    B_{\mu M} &= (B_{\mu m},0,\dots,0) \,,&
    \Sigma_M &= (\Sigma_m,0,\dots,0) \,.
\end{align}
The equation of motion for $W$ then becomes $Q_m^\sharp=0$\,.

\subsection{Hamiltonian analysis of the $E_8$ particle}
\label{subsec:E8_Hamiltonian}

Here we give a Hamiltonian description of the $E_8$ worldline model and we discuss its Dirac constraints.
For the canonical analysis it is convenient to use the second-order form where the external velocity is quadratic while the internal momentum $P_M$ is first-order.

Take the full $E_8$ action with kinetic sector \eqref{eq:S_kinetic_P} and orbit sector \eqref{eq:S_orbit}.
The conjugate momenta for the einbein and one-forms $V^M$ and $W_M$ vanish.
In other words we have the primary constraints
\begin{align}
    \pi_{(e)} &\approx 0 \,,& \pi_M^{(V)} &\approx 0 \,,&
    \pi^M_{(W)} &\approx 0 \,.
\end{align}
Let us compute the conjugate momentum for $X^\mu$.
Varying with respect to $\dot{X}^\mu$ gives
\begin{equation}
    P_\mu = \frac{\partial L}{\partial \dot{X}^\mu}
    = \frac{1}{e}\,g_{\mu\nu} \dot{X}^\nu + P_M A_\mu{}^M + Q^M B_{\mu M} \,.
    \label{eq:p_mu_def}
\end{equation}
This motivates the definition of gauge covariant external momentum
\begin{equation}
    \pi_\mu := P_\mu - P_M A_\mu{}^M - Q^M B_{\mu M} \,.
    \label{eq:pi_mu_def}
\end{equation}
Then \eqref{eq:p_mu_def} is $\pi_\mu=e^{-1}g_{\mu\nu}\dot{X}^\nu$, i.e.~$\dot{X}^\mu=e\,g^{\mu\nu}\pi_\nu$\,.
This is precisely the same as \eqref{eq:pi_mu_worldvolume} in Section~\ref{subsec:duality-covariant_phase_space_zero-branes}, and the $Q^MB_{\mu M}$ term in the $E_8$ action could have been reverse-engineered from this.
Similarly, the momentum conjugate to $Y^M$ is
\begin{equation}
    \Pi_M := \frac{\partial L}{\partial \dot{Y}^M} = P_M \,,
    \label{eq:piY_equals_P}
\end{equation}
subject to the section constraint.

\subsubsection*{The symplectic structure of the coadjoint orbit}

In the following, we show that the symplectic structure on the coadjoint orbit leads to an associated Poisson bracket that realises the $E_8$ Lie algebra.
The reader might prefer to skip directly to equation \eqref{eq:Lie_Q} rather than to read through the technical details, since we will not use the details again in the rest of the paper.
However, an exception to this is Appendix~\ref{app:deformation} in which we consider an alternative approach based on an extension of the following discussion.

For a fixed $\Gamma\in\mathfrak{e}_8^*$ the coadjoint orbit is $\cO_\Gamma=\{\Ad_g^*\Gamma\,|\,g\in E_8\}\cong E_8/G_\Gamma$ and we consider the bundle $\pi:E_8\to E_8/G_\Gamma\cong\cO_\Gamma:g\mapsto\Ad_g^*\Gamma$.
The canonical KKS symplectic form $\omega_\mathrm{KKS}$ is defined instrinsically at $Q:=\Ad_g^*\Gamma\in\cO_\Gamma$ by
\begin{equation}
    \omega_\mathrm{KKS}(\ad^*_XQ,\ad^*_YQ) = - \langle Q,[X,Y] \rangle
    \label{eq:KKS_def}
\end{equation}
for all $X,Y\in\mathfrak{e}_8$ up to an overall sign convention chosen here to be negative.

In order to relate the orbit geometry to the worldline action, first choose a group element $g\in E_8$ and consider the Maurer--Cartan form $g^{-1}\rd g$ with $\rd$ the exterior derivative on the group.
Define the symplectic potential
\begin{equation}
    \vartheta_\Gamma := \langle \Gamma,g^{-1}\rd g \rangle \,,
    \label{eq:vartheta_gamma}
\end{equation}
whose derivative is
\begin{equation}
    \rd\vartheta_\Gamma
    =-\frac12\big\langle \Gamma,[g^{-1}\rd g,g^{-1}\rd g]\big\rangle \,,
    \label{eq:d_vartheta_gamma}
\end{equation}
where we have used the Maurer--Cartan equation.
The potential is naturally defined on $E_8$ but it does not in general descend globally to a one-form on the orbit since it transforms as $\vartheta_\Gamma\mapsto\vartheta_\Gamma+\langle\Gamma,k^{-1}\rd k\rangle$ under $g\mapsto gk$ for $k\in G_\Gamma$\,.
However, its derivative $\rd\vartheta_\Gamma$ does, and $\omega_\mathrm{KKS}$ is the unique globally-defined two-form on $\cO_\Gamma$ satisfying
\begin{equation}
    \pi_\Gamma^* \, \omega_\mathrm{KKS} = \rd \vartheta_\Gamma \,.
\end{equation}

To obtain an honest symplectic potential on the orbit, choose a local section $\sigma:U\subset\cO_\Gamma\to E_8$ of $\pi:E_8\to\cO_\Gamma$ and then define on the patch $U$ the local symplectic potential
\begin{equation}
    \theta_\Gamma^{(U)}
    := \sigma^* \vartheta_\Gamma
    = \langle \Gamma,\sigma^{-1}\rd\sigma \rangle \,.
    \label{eq:local_theta_KKS}
\end{equation}
By construction,
\begin{equation}
    \omega_\mathrm{KKS}\big|_U
    = \rd \theta_\Gamma^{(U)}
    = -\frac12 \big\langle \Gamma,[\sigma^{-1}\rd\sigma,\sigma^{-1}\rd\sigma] \big\rangle \,.
\end{equation}
Now let $\tau\mapsto Q(\tau)\in U\subset\cO_\Gamma$ be a worldline in the orbit patch.
Lifting to the group gives a map $g(\tau)$ satisfying $\pi(g(\tau))=Q(\tau)$.
Having chosen a local section, every such lift can be written locally as
\begin{equation}
    g(\tau) = \sigma(\xi(\tau))\,k(\tau) \,,
\end{equation}
where $k(\tau)$ is an element of $G_\Gamma$ and $\xi^a(\tau)$ are local coordinates on the orbit.
Pulling back \eqref{eq:local_theta_KKS} to the worldline then gives $\big\langle \Gamma,g^{-1}\dot g\big\rangle\,\rd\tau$.
Thus the KKS term in  \eqref{eq:S_orbit_KKS_Wong} is the pullback of the local symplectic potential.

For each $X\in\mathfrak{e}_8$ the action on the orbit $\cO_\Gamma$ induces a fundamental vector field $V_X$ that is given by
$V_X(Q):=\frac{\rd}{\rd t}\big(\Ad_{e^{tX}}^*Q\big)\big|_{t=0}=\ad_X^*Q$.
One has that the KKS form \eqref{eq:KKS_def} satisfies
\begin{equation}
    \omega_\mathrm{KKS}(V_X,V_Y) = -\langle Q,[X,Y] \rangle \,.
    \label{eq:KKS_VX_VY}
\end{equation}
A vector field $V_X$ on $(\cO_\Gamma,\omega_\mathrm{KKS})$ is Hamiltonian if there exists a Hamiltonian function $H_X:\cO_\Gamma\to\mathbb{R}$ such that
\begin{equation}
    \iota_{V_X} \omega_\mathrm{KKS} = -\rd H_X \,.
    \label{eq:iVX_omega_dHX}
\end{equation}
With this convention for Hamiltonian vector fields, the associated Poisson bracket is
\begin{equation}
    \{f,g\}_\omega = -\omega_\mathrm{KKS} (V_f,V_g) \,,
\end{equation}
with $\iota_{V_f}\omega_\mathrm{KKS}=-\rd f$ and $\iota_{V_g}\omega_\mathrm{KKS}=-\rd g$.
These functions $H_X$ are often called charges because they are the Noether generators of the corresponding symmetries, and their Poisson brackets reproduce the symmetry algebra.

Equation \eqref{eq:KKS_VX_VY} already suggests that the Hamiltonian functions will be defined by
\begin{equation}
    H_X(Q) := \langle Q,X \rangle \,.
    \label{eq:HX_def}
\end{equation}
We now verify that these are the Hamiltonians generating the infinitesimal action.
One computes that
\begin{equation}
    \rd H_X(V_Y)
    = \frac{\rd}{\rd t} \big\langle \Ad_{e^{tY}}^*Q,X \big\rangle \Big|_{t=0}
    = \langle \ad_Y^*Q,X \rangle
    = \langle Q,[X,Y] \rangle
    = -\omega_\mathrm{KKS}(V_X,V_Y) \,.
    \label{eq:dHX_on_VY}
\end{equation}
for all $Y\in\mathfrak{e}_8$\,.

The same conclusion follows directly from the worldline action.
Consider the pure KKS Lagrangian
$L_\mathrm{KKS}=\langle\Gamma,j\rangle$ with $j=g^{-1}\dot{g}$ and consider an infinitesimal transformation $\delta_\epsilon g=\epsilon X g$.
This gives
\begin{equation}
    \delta_\epsilon L_{\mathrm{KKS}}
    = \langle \Gamma, \delta_\epsilon(g^{-1}\dot{g}) \rangle
    = \dot{\epsilon}\,\langle \Gamma,g^{-1}Xg \rangle
    = \dot{\epsilon}\,\langle Q,X\rangle \,.
    \label{eq:delta_LKKS}
\end{equation}
The coefficient of $\dot{\epsilon}$ is the Noether generator $H_X(Q)=\langle Q,X\rangle$.
This explains why the moment map on the orbit is simply the inclusion
\begin{equation}
    \mu : \cO_\Gamma \hookrightarrow \mathfrak{e}_8^* : Q \mapsto Q \,,
    \label{eq:moment_map_inclusion}
\end{equation}
in the sense that $(\mu\circ\pi)(g)=\mu(\Ad_g^*\Gamma)=\Ad_g^*\Gamma$.
By definition, the moment map is the unique $\mathfrak{e}_8^*$-valued map such that
\begin{equation}
    H_X(\cdot) = \langle \mu(\cdot),X \rangle
\end{equation}
for all $X\in\mathfrak{e}_8$.
Equation \eqref{eq:HX_def} shows that this is achieved by $\mu(Q)=Q$.
This is also the reason why the Wong term is $\langle Q,\widetilde{B}\rangle$ since a gauge connection always couples to the moment map of the symmetry that it gauges.

We can now compute the Poisson brackets of these Hamiltonians.
We obtain
\begin{equation}
    \{H_X,H_Y\}_\omega
    := - \omega_\mathrm{KKS}(V_X,V_Y)
    = \langle Q,[X,Y] \rangle
    = H_{[X,Y]} \,.
    \label{eq:HX_HY_bracket}
\end{equation}
Choose a basis $t^M$ of $\mathfrak{e}_8$ with $[t^M,t^N]=f^{MN}{}_P\,t^P$ and define the moment map components
\begin{equation}
    Q^M := \langle Q,t^M \rangle = H_{t^M} \,.
\end{equation}
Then \eqref{eq:HX_HY_bracket} gives that the Poisson brackets realise the Lie algebra on the orbit, i.e.
\begin{equation}
    \{Q^M,Q^N\}_\omega
    = f^{MN}{}_P\,Q^P \,.
    \label{eq:Lie_Q}
\end{equation}

\subsubsection*{Hamiltonian equations and secondary constraints}

Now we return to the full $E_8$ worldline model.
Up to the term $\langle Q,W\rangle$ that is not velocity-dependent, the canonical Hamiltonian is
\begin{equation}
    H_{E_8}
    = \frac{e}{2} \big( g^{\mu\nu} \pi_\mu \pi_\nu + \cM^{MN} P_M P_N \big)
    - P_M V^M - Q^M W_M \,,
    \label{eq:H_canonical}
\end{equation}
Preserving the primary constraint $\pi_{(e)}\approx0$ under time evolution gives the secondary constraint
\begin{equation}
    \cH := \frac12 \big( g^{\mu\nu} \pi_\mu \pi_\nu + \cM^{MN} P_M P_N \big) \approx 0 \,.
\end{equation}
This is the worldline Hamiltonian constraint generating reparametrisations.
In one dimension, $\cH$ is first class since $\{\cH,\cH\}\approx0$.
Together, $\pi_{(e)}\approx0$ and $\cH\approx0$ form the usual pair of first class constraints for a reparametrisation-invariant relativistic particle with extra contribution from the orbit sector.

Preserving $\pi_M^{(V)}\approx0$ and $\pi^M_{(W)}\approx0$ in time leads to the constraints on $P_M$ and $Q^M$ that we found previously.
The internal momentum $P_M$ is still constrained to lie along the physical directions, and for the duality charge $Q^M$ we have a constraint forcing it to lie along the unphysical directions.

Let us show that $Q$ is parallel transported by the worldline connection $\widetilde{B}$, giving the Hamiltonian version of the Wong equation.
First we use \eqref{eq:pi_mu_def} and \eqref{eq:Lie_Q} to compute
\begin{align}
    \{Q^M,\pi_\mu\}_\omega
    &= - f^{MN}{}_P B_{\mu N} Q^P \,,&
    \{Q^M,\langle Q,W\rangle\}_\omega
    &= f^{MN}{}_P\,W_N Q^P \,.
    \label{eq:Q_Poisson_computation}
\end{align}
Crucially, $\{Q^M,P_N\}=0$ since $Q$ depends only on $g$, i.e.~on $\xi^a$ in local coordinates, and $P_M$ belongs entirely to the $Y^M$ sector of the theory.
The symplectic structure is block-diagonal between these two sectors, so there are no mixed brackets.
From \eqref{eq:Lie_Q} and \eqref{eq:H_canonical} we obtain
\begin{equation}
    \dot{Q}^M = \{Q^M,H_{E_8}\}_\omega
    = \frac{e}{2} \, g^{\mu\nu} \{Q^M,\pi_\mu \pi_\nu\}_\omega
    - \{Q^M,\langle Q,W\rangle\}_\omega \,.
    \label{eq:Qdot_step1}
\end{equation}
Substituting \eqref{eq:Q_Poisson_computation} back into \eqref{eq:Qdot_step1} leads to
\begin{equation}
    \dot{Q}^M = - e\,g^{\mu\nu} \pi_\nu f^{MN}{}_P \, B_{\mu N} Q^P - f^{MN}{}_P \, W_N Q^P \,.
    \label{eq:Qdot_step2}
\end{equation}
Finally, from \eqref{eq:p_mu_def} and \eqref{eq:pi_mu_def} we have $\dot{X}^\mu=e\,g^{\mu\nu}\pi_\nu$ which gives the generalised Wong equation
\begin{equation}
    \dot{Q}^M + f^{MN}{}_P \, \widetilde{B}_N Q^P = 0 \,,
    \label{eq:Wong_equation_final}
\end{equation}
with $\widetilde{B}_M=\dot{X}^\mu B_{\mu M}+W_M$\,.
This is nothing other than the Euler--Lagrange equation \eqref{eq:EOM_g_Q}.

The evolution of $\pi_\mu$ gives the expected equation
\begin{align}
\begin{aligned}
    \dot{\pi}_\mu + \frac{e}{2} \big( \tilde{D}_\mu g^{\nu\rho} \pi_\nu \pi_\rho + \tilde{D}_\mu \cM^{MN} P_M P_N \big) - \big( P_M F_{\mu\nu}{}^M + Q^M G_{\mu\nu M} \big) \dot{X}^\nu \hspace{10mm} \\
    {}+ \cD_\tau Y^M \big( P_N \partial_M A_\mu{}^N + Q^N \partial_M B_{\mu N} \big) = 0 \,,
\end{aligned}
\label{eq:pi_evolution}
\end{align}
where $\tilde{D}_\mu:=\partial_\mu-A_\mu{}^M\partial_M$ is based on how $P_\mu$ and $P_M$ occur in \eqref{eq:pi_mu_def}, and it is not the full $E_8$ ExFT covariant derivative.
We defined $F_{\mu\nu}{}^M:=2\,\tilde{D}_{[\mu}A_{\nu]}{}^M$ and $G_{\mu\nu M}:=2\,\tilde{D}_{[\mu}B_{\nu]M}$\,.
These are contained in the covariant field strengths $\cF_{\mu\nu}{}^M$ and $\cG_{\mu\nu M}$ in $E_8$ ExFT.
The point is that we have a generalisation of the Lorentz force law.

Lastly, the evolution of $\Pi_M=P_M$ gives
\begin{equation}
    \dot{P}_M + \frac{e}{2} \big( \partial_M g^{\mu\nu} \pi_\mu \pi_\nu + \partial_M \cM^{NP} P_N P_P \big) - \big( P_N \partial_M A_\mu{}^N + Q^N \partial_M B_{\mu N} \big) \dot{X}^\mu = 0 \,.
    \label{eq:P_evolution}
\end{equation}
Unsurprisingly, the internal force is given in terms of internal derivatives of background couplings.
No dependence on $Y^M$ means that $P_M$ is constant.

\section{From $E_8$ to zero-branes in 11D}
\label{sec:from_E8_to_11D}

In this section we specialise our $E_8$ worldline action to the eleven-dimensional solution to the section constraint, where it reduces either to the known geodesic action for the M0-brane, i.e.~the M-theory point particle, or to that of the exotic zero-brane.
This reduction is entirely local, while the exotic zero-brane $0^{(1,7)}$ also has global monodromy that is not seen by the orbit sector.
We impose the $E_8$ ExFT truncation to the bosonic fields relevant for the $(3+8)$-dimensional split of 11D supergravity.

\subsection{Recovering known backgrounds}

Decomposing $E_8$ with respect to its $\mathrm{GL}(8)$ subgroup, the adjoint representation $\mathbf{248}$ decomposes as
\begin{equation}
    \mathbf{248} \;\cong\;
    \mathbf{8}_{+3} \oplus \mathbf{28}_{+2} \oplus \mathbf{56}_{+1}
    \oplus ( \mathbf{63} \oplus \mathbf{1} )_{0}
    \oplus \mathbf{56}_{-1} \oplus \mathbf{28}_{-2} \oplus \mathbf{8}_{-3} \,,
    \label{eq:248-GL8}
\end{equation}
where the subscripts denote the $\mathrm{GL}(1)$ level grading.
Accordingly, the adjoint generator $t^M$ splits as
\begin{equation}
    t^M = \big( t^{m[8],n}, t^{m[6]}, t^{m[3]}, t^m{}_n, t_{m[3]}, t_{m[6]}, t_{m[8],n} \big) ,
    \label{eq:t-GL8}
\end{equation}
and the extended coordinates $y^M$ split as
\begin{equation}
    y^M = \big( y^m , y_{m[2]} , y_{m[5]} , y_{m[7],n} , y_{m[8]} , y_{m[8],n[3]} , y_{m[8],n[6]} , y_{m[8],n[8],p} \big) ,
    \label{eq:Y-GL8-coords}
\end{equation}
where indices $m,n,\dots$ run from $1$ to $8$.
As in Section~\ref{sec:Hamiltonian}, our notation is such that $m[k]$ is one set of $k$ antisymmetric indices $m_1\dots m_k$ and different sets of antisymmetric indices that are separated by a comma correspond to different columns in an irreducible Young tableau.
Semicolons separate indices corresponding to different irreducible components of a reducible tensor.
The internal 8D Levi--Civita symbol can be used to dualise any set of indices, or to `pull out' any eight antisymmetric indices:
\begin{align}
    \tilde{y}^{m[3]} &:= \varepsilon^{m[3]n[5]} y_{n[5]} \,,&
    \tilde{y}^m{}_n &:= \varepsilon^{mp[7]} y_{p[7];n} \,,&
    \tilde{y}_m &:= \varepsilon^{n[8]} \varepsilon^{p[8]} y_{n[8],p[8],m} \,.
\end{align}
Thus the internal spacetime can also be written as
\begin{equation}
    y^M = \big( y^m , y_{m[2]} , \tilde{y}^{m[3]} , \tilde{y}^m{}_n , \tilde{y}_{m[3]} , \tilde{y}^{m[2]} , \tilde{y}_m \big) ,
\end{equation}
just as in \cite{Hohm:2014fxa}, and the decompositions \eqref{eq:t-GL8} and \eqref{eq:Y-GL8-coords} are exactly the same up to a set of eight indices that has been pulled out with a Levi--Civita symbol.

The eleven-dimensional solution to the section constraint is obtained by restricting the coordinate dependence of all fields and parameters to the external $x^\mu$ and internal $y^m\in\mathbf{8}_{+3}$\,.
This is expressed in terms of the internal derivatives:
\begin{equation}
    \partial_M=(\partial_m,0,\dots,0) \,.
\end{equation}
The constrained gauge field $B_{\mu M}$ and parameter $\Sigma_M$ must be truncated in the same way:
\begin{align}
    B_{\mu M}& = (B_{\mu m},0,\ldots,0) \,,&
    \Sigma_M& = (\Sigma_m,0,\ldots,0) \,.
    \label{eq:B-truncation}
\end{align}

In addition to the scalars coming from the internal components of the 11D metric, three-form, and six-form fields, the internal metric $\cM_{MN}=(\cV\cV^\rT)_{MN}$ provides eight more scalars $\varphi_m$ \cite{Hohm:2014fxa}.
We write
\begin{equation}
    \cV = \exp(h_m{}^n t^m{}_n)
    \exp(A_{m[3]} t^{m[3]})
    \exp(\tilde{A}_{m[6]} t^{m[6]})
    \exp(\tilde{h}_{m[8],n} t^{m[8],n}) \,,
    \label{eq:V_triangular_with_varphi}
\end{equation}
where external scalars $h_m{}^n$ come from the internal metric, $A_{m[3]}$ from the three-form, and $\tilde{A}_{m[6]}$ from the magnetic dual six-form.
The dual gravity scalars $\varphi_m$ come from the Curtright field, i.e.~the dual graviton $\tilde{h}_{m_1\dots m_8,n}:=\varepsilon_{m_1\dots m_8}\varphi_n$ \cite{Curtright:1980yk,Hull:2000zn,Hull:2001iu,Boulanger:2003vs}.
This mixed-symmetry field satisfies $\tilde{h}_{[m_1\dots m_8,n]}=0$\,.

Importantly, the scalars $\varphi_m$ enter the $E_8$ ExFT Lagrangian only through the covariant derivative $D_\mu\varphi_m\supset\partial_\mu\varphi_m+B_{\mu m}$ and they transform such that the combination $D_\mu\varphi_m$ is gauge invariant:
\begin{align}
    \delta_\Sigma \varphi_m &= - \Sigma_m \,,&
    \delta_\Sigma B_{\mu m} &= \partial_\mu \Sigma_m \,.
    \label{eq:017_sigma_varphi_B}
\end{align}
This resolves the dual graviton problem as explained in \cite{Hohm:2014fxa,Hohm:2018qhd}.
One gauges away $\varphi_m$ using $\Sigma_m$ and correspondingly the $E_8$ scalar potential does not depend on $\varphi_m$.
Thus the scalars $\varphi_m$ are locally pure gauge, and $B_{\mu m}$ is auxiliary rather than a propagating gauge field.

\subsubsection*{Identifying the Kaluza--Klein variables for the M0-brane}

We now choose explicit $E_8$ ExFT fields that reproduce the M0-brane in 11D.
First assemble the three external coordinates $x^\mu$ and the eight internal coordinates $y^m$ into the usual eleven coordinates
\begin{align}
    (\hat{x}^0,\dots,\hat{x}^{10}) = \hat{x}^{\hat\mu} := (x^\mu,y^m) = (x^0,x^1,x^2,y^1,\dots,y^8) \,.
\end{align}
A generic line element in the 11D section takes the form
\begin{equation}
    \rd s^2 = \hat{g}_{\hat\mu\hat\nu} \, \rd \hat{x}^{\hat\mu} \rd \hat{x}^{\hat\nu}
    = g_{\mu\nu} \, \rd x^\mu \rd x^\nu
    + \cM_{mn} \big( \rd y^m + A_\mu{}^m \rd x^\mu \big) \big( \rd y^n + A_\nu{}^n \rd x^\nu \big) \,,
    \label{eq:ds11}
\end{equation}
where $g_{\mu\nu}$ is the external metric (scalar under generalised diffeomorphisms), $A_\mu{}^m$ is the Kaluza--Klein vector that arises from $A_\mu{}^M$, and $\cM_{mn}$ is the physical internal metric obtained from the generalised internal metric $\cM_{MN}$\,.
All other components of $A_\mu{}^M$ are retained in the full theory, and they couple to embedding fields for the other extended coordinates in \eqref{eq:Y-GL8-coords}.
The worldline one-form $V^M$ is used to gauge away the velocities of these unphysical extended coordinates.

Recall the coordinate split $x^\mu=(t,r,\theta)$ and $y^m=(z^1,\ldots,z^8)$ from Section~\ref{subsec:overview_of_branes}, so that $z=z_1$ is the wave direction and the remaining seven internal coordinates $\vec{z}_{(7)}=(z_2,\dots,z_8)$ are spectator.
Thus we have
\begin{equation}
    \hat{x}^{\hat\mu} = (x^\mu,y^m)
    = (t,r,\theta,z_1,\dots,z_8) = (t,r,\theta,z,\vec{z}_{(7)}) \,.
\end{equation}
We shall use the codimension-two
smeared harmonic function $H(r)$ in \eqref{eq:H_K_def} for comparison with the exotic zero-brane \cite{Berman:2018okd}.
For convenience we reproduce here the metric \eqref{eq:metric_M0_smeared} for the M0-brane\footnote{One can also shift $z\mapsto z-t$ in order to write the M0 with the asymptotically vanishing vector $A_t{}^{z}=1-H^{-1}$.}
\begin{equation}
    \rd s^2_{\mathrm{M0}}
    = -2\,\rd t\,\rd z + H\,\rd z^2 + \rd r^2+r^2\rd\theta^2 + \rd \vec{z}_{(7)}^{\,2} \,.
\end{equation}
We want to write this in Kaluza--Klein form \eqref{eq:ds11}, for which we have
\begin{align}
    \hat{g}_{mn} &= \cM_{mn} \,,&
    \hat{g}_{\mu m} &= \cM_{mn} A_\mu{}^n \,,&
    \hat g_{\mu\nu} &= g_{\mu\nu} + \cM_{mn} A_\mu{}^m A_\nu{}^n \,,
\end{align}
with inverse relations $A_\mu{}^m=\cM^{mn}\hat{g}_{\mu n}$ and $g_{\mu\nu}=\hat{g}_{\mu\nu}-\hat{g}_{\mu m}\cM^{mn}\hat{g}_{\nu n}$\,.

For the M0 metric, we have $\hat{g}_{tz}=-1$ and $\hat{g}_{zz}=H$.
Therefore the physical internal 8D metric is
\begin{align}
    \cM &= \mathrm{diag}(H,1,\dots,1) \,,&
    \cM_{mn} \rd y^m \rd y^n
    &= H\,\rd z^2 + \rd \vec{z}_{(7)}^{\,2} \,.
\end{align}
The only non-zero vector component is $A_t{}^{z}=\cM^{zz}\hat{g}_{tz}=-H^{-1}$ and the external 3D metric is
\begin{equation}
    g_{\mu\nu} \rd x^\mu \rd x^\nu
    = - H^{-1} \rd t^2 + \rd r^2 + r^2 \rd \theta^2 \,.
\end{equation}
In particular,
\begin{equation}
    \hat{g}_{tt} = g_{tt} + \cM_{zz} (A_t{}^{z})^2
    = - H^{-1} + H (H^{-1})^2
    = 0 \,,
    \label{eq:hat_g_tt=0}
\end{equation}
while
\begin{equation}
    \hat{g}_{tz} = \cM_{zz} A_t{}^{z}
    = H(-H^{-1})
    = -1 \,.
\end{equation}
Combining all this, one reconstructs the metric representative \eqref{eq:metric_M0_smeared} for the smeared M0-brane.\footnote{Note that $\hat{g}_{tt}$ is a component of the 11D metric, and its vanishing does not leave the ExFT fields underdetermined.
As soon as the full 11D metric and the coordinate $z$ are fixed, \eqref{eq:ds11} uniquely determines $\cM_{zz}=\hat{g}_{zz}$, $A_t{}^z=\cM^{zz}\hat{g}_{tz}$, and $g_{tt}=\hat{g}_{tt}-\hat{g}_{tz}\cM^{zz}\hat{g}_{zt}$.
A coordinate transformation of $z$ is embedded in the physical internal generalised diffeomorphisms and induces the gauge transformation of $A_\mu{}^z$.
However, it does not provide extra freedom in the relative definitions of the ExFT field components once the 11D metric representative has been fixed.}

In terms of $E_8$ ExFT fields, this means that we take $\cM_{mn}$ as above, and
\begin{align}
    A_t{}^m &= -H^{-1} \delta^{\hspace{0.1mm}m}_{z} \,,&
    A_r{}^m &= A_\theta{}^m = 0 \,,
\end{align}
in the physical $\mathbf 8_{+3}$ block.
All three-form, six-form, and dual graviton scalar components are set to zero.
The remaining components of $\cM_{MN}$ are then fixed by the $E_8$ generalised metric, i.e.~they are not independent fields.

Let us check that the components given above define a local solution of the $E_8$ ExFT equations, at least locally away from the source.
The 11D section of the theory is equivalent to 11D supergravity, so it is enough to verify the 11D equations of motion.
For the M0 representative where all form fields are set to zero, the inverse metric in the $(t,z)$ block is
\begin{align}
    \hat{g}^{tt} &= -H \,,&
    \hat{g}^{tz} &= -1 \,,&
    \hat{g}^{zz} &= 0 \,,&
    \hat{g}^{ij} &= \delta^{ij} \,,
\end{align}
where $i,j,\dots$ denote all other directions.
The only non-constant metric component is $\hat{g}_{zz}=H$.
One computes the Ricci tensor
\begin{equation}
    R_{zz} = -\frac12 \delta^{ij} \partial_i \partial_j H \,,
    \label{eq:Ricci_Rzz}
\end{equation}
so the 11D equations are solved if this vanishes away from the source.
This has long been understood for usual M0-branes/pp-waves \cite{Blau:2001ne}.
In the codimension-two smeared representative, there are only two transverse directions $(r,\theta)$ and the Laplacian of $H$ vanishes for $r>0$ as required.

The fields have non-trivial components along the physical directions, e.g.~$\cM_{zz}$ and $A_t{}^{z}$, but for the M0 and the $0^{(1,7)}$ their coefficients are are given in terms of $H$ and $K$ which depend only on $(r,\theta)$ in the smeared case.
As such, $\partial_M\Phi=0$ for every ExFT field $\Phi$ in the backgrounds that are smeared along all eight physical internal directions.
For a fully localised M0, the harmonic function may depend on the internal $y^m$ coordinates.
Then $\partial_M\Phi\neq0$ and the $E_8$ potential does not vanish.
The verification of the field equations is then that for full 11D supergravity, i.e.~the $\mathrm{GL}(8)$ decomposition of the $E_8$ ExFT equations.
This gives a solution whenever the M0 is Ricci-flat, i.e.~whenever $H$ is harmonic away from the source, as implied by \eqref{eq:Ricci_Rzz}.

\subsubsection*{Identifying the Kaluza--Klein variables for the exotic zero-brane}

In order to obtain the worldline action for the exotic zero-brane, we stay inside the 11D section but with a different local $E_8$ frame.
The line element was given in \eqref{eq:metric_0^(1,7)} and is reproduced here:
\begin{equation}
    \rd s^2_{0^{(1,7)}}
    = -2 \, \rd t \, \rd z
    + HK^{-1} \rd z^2
    + K \big( \rd r^2 + r^2 \rd \theta^2 \big)
    + \rd \vec{z}_{(7)}^{\,2} \,.
    \label{eq:metric_0^(1,7)_Brinkmann}
\end{equation}
Note that this is the same as the smeared M0-brane \eqref{eq:metric_M0}, but with $H$ replaced by $HK^{-1}$ and with
the $(r,\theta)$ part multiplied by $K=H^2+\sigma^2\theta^2$ \cite{Berman:2018okd}.

Comparing the $0^{(1,7)}$ metric with the Kaluza--Klein ansatz, we can read off the internal metric
\begin{align}
    \cM &= \mathrm{diag}(HK^{-1},1,\dots,1) \,,&
    \cM_{mn} \rd y^m \rd y^n
    &= HK^{-1} \rd z^2 + \rd \vec{z}_{(7)}^{\,2} \,.
    \label{eq:internal_metric_0^(1,7)}
\end{align}
The vector is $A_t{}^{z}=-H^{-1}K$ with all other components zero, and the external metric is
\begin{equation}
    g_{\mu\nu} \rd x^\mu\rd x^\nu
    = - H^{-1}K\,\rd t^2 + K \left(\rd r^2+r^2\rd\theta^2\right).
    \label{eq:017_external_metric}
\end{equation}
Compute the components of the 11D metric as
\begin{align}
    \hat{g}_{tt}
    &= g_{tt} + \cM_{zz} (A_t{}^{z})^2
    = 0 \,,&
    \hat{g}_{tz}
    &=\cM_{zz} A_t{}^{z}
    = -1 \,.
\end{align}
Altogether, this reconstructs the $0^{(1,7)}$ representative \eqref{eq:metric_0^(1,7)_Brinkmann}.

As we did for the M0-brane, we will check that the field components define a local ExFT solution.
One subtlety is that the external metric $g_{\mu\nu}$ is in the 3D Einstein frame, so if $\bar{g}_{\mu\nu}$ denotes the raw external part of the ansatz, then $\bar{g}_{\mu\nu}=\Delta^{-1}g_{\mu\nu}$ with $\Delta=\det\cM_{mn}$\,.
For the $0^{(1,7)}$ we found that the internal metric is \eqref{eq:internal_metric_0^(1,7)}.
Hence,
\begin{equation}
    \Delta_{0^{(1,7)}} = \det \cM_{mn}
    = HK^{-1} \,.
\end{equation}
The raw external metric is
\begin{equation}
    \bar{g}_{\mu\nu}\,\rd x^\mu \rd x^\nu
    = -H^{-1}K \rd t^2 + K ( \rd r^2 + r^2 \rd \theta^2 ) \,,
\end{equation}
and so the external metric in the Einstein frame is
\begin{equation}
    g_{\mu\nu}\,\rd x^\mu \rd x^\nu
    = \Delta_{0^{(1,7)}}
    \big( {-}H^{-1}K \rd t^2 + K ( \rd r^2 + r^2 \rd \theta^2 ) \big)
    = - \rd t^2 + H ( \rd r^2 + r^2 \rd \theta^2 ) \,.
    \label{eq:external_metric_Einstein_frame}
\end{equation}
This is the same as that of the smeared M0 representative \eqref{eq:metric_M0}.

As is usual from 4D gravity reduced to 3D, the Kaluza--Klein vector can be dualised into a second scalar, giving 3D gravity coupled to the $\mathrm{SL}(2,\mathbb{R})/\mathrm{SO}(2)$ sigma model \cite{Breitenlohner:1987dg,Galtsov:2008zz,Hohm:2013jma}.
We introduce
\begin{align}
    f &:= \cM_{zz} = HK^{-1} \,,&
    \chi &:= -K^{-1} \sigma \theta \,.
    \label{eq:0^(1,7)_f_chi_def}
\end{align}
The vector $A^{z}=\rd x^\mu A_\mu{}^{z}$ in the Kaluza--Klein metric is $A^{z}=-H^{-1}K\rd t=-f^{-1}\rd t$ and it is dual to the scalar $\chi$ with respect to the Einstein frame metric.
Choosing
\begin{equation}
    \mathrm{vol}_3 = \sqrt{g}\,\rd t \wedge \rd r \wedge \rd \theta
    = Hr\,\rd t \wedge \rd r \wedge \rd \theta \,,
\end{equation}
one obtains, as is well-understood, a first-order duality equation
\begin{equation}
    \rd\chi
    = - f^2 \star_3 \rd A^{z} \,.
\end{equation}
This is just $\rd\chi={-}\star_2\rd f$ on the transverse $(r,\theta)$ plane, where $\star_2$ is the Hodge star for the flat metric $\rd r^2+r^2\rd\theta^2$.
One can check directly that the Cauchy--Riemann equations hold, i.e.~that
\begin{align}
    \partial_\rho f
    &= \frac{\sigma(H^2-\sigma^2\theta^2)}{K^2} = \partial_\theta \chi \,,&
    \partial_\theta f
    &= -\frac{2\sigma^2\theta H}{K^2} = -\partial_\rho \chi \,,
\end{align}
where $\rho=\log(\mu r^{-1})$ so $\partial_\rho=-r\partial_r$.
In particular, $f$ and $\chi$ are harmonic on the
punctured $(r,\theta)$ plane.

The relevant $\mathrm{SL}(2,\mathbb{R})/\mathrm{SO}(2)$ scalar equations are those of the 3D sigma model \cite{Breitenlohner:1987dg}, namely
\begin{equation}
    R_{\mu\nu} = \frac{1}{2f^2} \big( \partial_\mu f \partial_\nu f + \partial_\mu\chi\,\partial_\nu\chi \big) \,,\label{eq:SL(2,R)/SO(2)_eq_Einstein}
\end{equation}
and two other equations.
For \eqref{eq:external_metric_Einstein_frame} with $H$ harmonic away from $r=0$, the Ricci components are
\begin{align}
    R_{rr} &= \frac{\sigma^2}{2r^2H^2} \,,&
    R_{\theta\theta} &= \frac{\sigma^2}{2H^2} \,,&
    R_{tt} &= R_{r\theta} = 0 \,.
\end{align}
These match the components of the scalar stress tensor in \eqref{eq:SL(2,R)/SO(2)_eq_Einstein}, verifying the Einstein equation.

\subsection{Local geodesic actions for zero-branes}

We have a universal kinetic sector $S_\mathrm{kin}$ reproducing the usual massless particle dynamics in 11D \cite{Blair:2017gwn}.
Now let us obtain the worldline geodesic action for the M0-brane.
Start from \eqref{eq:S_Blair} and restrict to the 11D section.
The physical coordinates are $y^m\in\mathbf{8}_{+3}$ and the auxiliary worldline field $V^M$ gauges the unphysical directions.
Only the physical components of $A_\mu{}^M$ survive, so it is consistent to choose the unphysical coordinates to be constant, i.e.~their covariant velocities vanish
\begin{align}
    \cD_\tau Y^M
    &= \big( \cD_\tau Y^m , 0 , \ldots , 0 \big) \,,&
    \cD_\tau Y^m
    &= \dot{Y}^m + \dot{X}^\mu A_\mu{}^m \,.
\end{align}
The kinetic term then becomes
\begin{equation}
    S_\mathrm{kin}^{\mathrm{M0}}
    = \frac12 \int\!\rd\tau \, e^{-1} \Big( g_{\mu\nu} \dot{X}^\mu \dot{X}^\nu + \cM_{mn} \big( \dot{Y}^m + \dot{X}^\mu A_\mu{}^m \big) \big( \dot{Y}^n + \dot{X}^\nu A_\nu{}^n \big) \Bigr)
    = \frac12 \int\!\rd\tau \, e^{-1} \hat{g}_{\hat\mu\hat\nu}^{\mathrm{M0}} \dot{X}^{\hat\mu} \dot{X}^{\hat\nu} \,,
    \label{eq:S-kin-11D}
\end{equation}
where $X^{\hat\mu}$ are the embedding fields for $\hat{x}^{\hat\mu}=(x^\mu,y^m)$ and $\hat{g}_{\hat\mu\hat\nu}^{\mathrm{M0}}$ is the M0 metric \eqref{eq:metric_M0}.
One obtains
\begin{align}
\begin{aligned}
    S_{\mathrm{kin}}^{\mathrm{M0}}
    &= \frac12\int\!\rd\tau\, e^{-1}
    \Big( {-}H^{-1} \dot{t}^{\,2} + H ( \dot{z} - H^{-1} \dot{t}\hspace{0.4mm})^2 + \dot{r}^2 + r^2 \dot{\theta}^2 + \dot{\vec{z}}_{(7)}^{\;2} \Big) \\
    &= \frac12\int\!\rd\tau\, e^{-1}
    \Big( {-}2\,\dot{t} \hspace{0.2mm} \dot{z} + H \dot{z}^2 + \dot{r}^2 + r^2 \dot{\theta}^2 + \dot{\vec{z}}_{(7)}^{\;2} \Big) \,.
\end{aligned}
\label{eq:geodesic_M0}
\end{align}
This is the massless M0-brane action, and it can be expressed in first-order form as
\begin{equation}
    S_\mathrm{kin}^{\mathrm{M0}(1)}
    = \int\!\rd\tau \Big( \hat{p}_{\hat\mu} \dot{X}^{\hat\mu} - \frac12 e \hspace{0.4mm} \hat{g}^{\hat\mu\hat\nu}_\mathrm{M0} \, \hat{p}_{\hat\mu} \hat{p}_{\hat\nu} \Big) \,,
\end{equation}
where the mass shell constraint $\hat{g}^{\hat\mu\hat\nu}\hat{p}_{\hat\mu}\hat{p}_{\hat\nu}\approx0$ is enforced by the einbein.
The orbit sector is not needed for this local reduction.

For the exotic zero-brane, one proceeds in the same way as for the M0.
The kinetic sector is
\begin{align}
\begin{aligned}
    S_{\mathrm{kin}}^{0^{(1,7)}}
    &= \frac12\int\!\rd\tau\, e^{-1}
    \Big( {-}H^{-1}K \dot{t}^{\,2} + HK^{-1} ( \dot{z} - H^{-1}K \dot{t}\hspace{0.4mm})^2 + K \big( \dot{r}^2 + r^2 \dot{\theta}^2 \big) + \dot{\vec{z}}_{(7)}^{\;2} \Big) \\
    &= \frac12\int\!\rd\tau\, e^{-1}
    \Big( {-}2\,\dot{t} \hspace{0.2mm} \dot{z} + HK^{-1} \dot{z}^2 + K \big( \dot{r}^2 + r^2 \dot{\theta}^2 \big) + \dot{\vec{z}}_{(7)}^{\;2} \Big) \,.
\end{aligned}
\label{eq:geodesic_0^(1,7)}
\end{align}
We should compare this with the M0 action \eqref{eq:geodesic_M0}.
Since $H$ is independent of $\theta$, a loop $\gamma$ wrapping $n$ times around $r=0$ sends $\theta$ to $\theta+2\pi n$, and the Lagrangian is invariant: $L_{\mathrm{kin}}^{\mathrm{M0}}(\theta+2\pi n)=L_{\mathrm{kin}}^{\mathrm{M0}}(\theta)$.
For the exotic zero-brane, travelling along $\gamma$ gives
\begin{equation}
    K(r,\theta) \;\longmapsto\;
    K_n(r,\theta) := H(r)^2 + \sigma^2 (\theta + 2 \pi n)^2 \,.
\end{equation}
Therefore, the Lagrangian is not single-valued since
\begin{equation}
    L_{\mathrm{kin}}^{0^{(1,7)}}(\theta + 2 \pi n)
    = \frac{1}{2e}
    \Big( {-}2\,\dot{t} \hspace{0.2mm} \dot{z} + HK_n^{-1} \dot{z}^2 + K_n \big( \dot{r}^2 + r^2 \dot{\theta}^2 \big) + \dot{\vec{z}}_{(7)}^{\;2} \Big)
    \neq L_{\mathrm{kin}}^{0^{(1,7)}}(\theta) \,.
    \label{eq:L_017_shifted}
\end{equation}
The background is globally patched by an $E_8$ transformation.

\subsubsection*{Action of monodromy elements}

The exotic zero-brane is not supported by the three-form or its magnetic dual six-form.
We therefore set all their associated scalar components to zero.
Non-geometric information is contained in the dual gravity part of the $E_8/\mathrm{SO}(16)$ scalar matrix.
Earlier we wrote the dual graviton component $-K^{-1}\sigma\theta$.
This is best understood by isolating the subgroup $\mathrm{SL}(2,\mathbb{R})\subset E_8$ that mixes the physical wave direction $z$ in $\mathbf{8}_{+3}$ with its `dual gravity partner' direction in $\mathbf{8}_{-3}$ that we denote by $\tilde{z}$.

We should not overstate what this subgroup does.
The unipotent element $g_z$ introduced below does not by itself generate the full 11D background.
Instead, it generates part of the relevant scalar block.
The background also contains the external metric and the Kaluza--Klein vector, both of which must be specified separately.
In the local comparison between the M0 and the $0^{(1,7)}$, the $2\times2$ scalar block inside $\cM_{MN}$ is
\begin{equation}
    \cM^{\mathrm{M0}}_{(\tilde{z},z)} =
    \begin{pmatrix}
        H&0\\
        0&H^{-1}
    \end{pmatrix}.
    \label{eq:H_M0_SL2_block_rewrite}
\end{equation}
This is the diagonal $\mathrm{SL}(2,\mathbb{R})/\mathrm{SO}(2)$ representative with no axion.

Now let $E_z$ denote the nilpotent generator in this subgroup.
In the $(\tilde{z},z)$ block, we choose $E_z$ and the unipotent element $g_z$ as
\begin{align}
    E_z &=
    \begin{pmatrix}
        0&1\\
        0&0
    \end{pmatrix} ,&
    g_z &= \exp(-\sigma\theta E_z) = \begin{pmatrix}
        1&-\sigma\theta\\
        0&1
    \end{pmatrix} .
    \label{eq:unipotent_frame_g_u}
\end{align}
Here $g_z$ is a factor in the coset representative $\cV_M{}^A\in E_8/\mathrm{SO}(16)$.
This and $\cM=\cV\cV^\rT$ transform as
\begin{align}
    \cV_M{}^A &\;\longmapsto\; (g_z)_M{}^N \cV_N{}^A \,,&
    \cM_{MN} &\;\longmapsto\; (g_z)_M{}^P \cM_{PQ} (g_z)^Q{}_N \,.
\end{align}
Note that the representation matrix $(g_z)$ is the matrix in \eqref{eq:unipotent_frame_g_u}.
As a result, we obtain
\begin{equation}
    \cM^{\mathrm{M0}}_{(\tilde{z},z)}
    \;\longmapsto\; \cM^{0^{(1,7)}}_{(\tilde{z},z)}
    = (g_z) \, \cM^{\mathrm{M0}}_{(\tilde{z},z)} \, (g_z)^\rT
    = \frac1H
    \begin{pmatrix}
        K&-\sigma\theta\\
        -\sigma\theta&1
    \end{pmatrix} .
\end{equation}

Recall that the worldline field $V^{\tilde{z}}$ eliminates the unphysical direction $\tilde{z}$.
The $(\tilde{z},z)$ contribution to the kinetic sector is
\begin{equation}
    L_{(\tilde{z},z)} = \frac{1}{2e} \big( \cM_{zz} (\cD_\tau z)^2 + 2 \cM_{z\tilde{z}} (\cD_\tau z) (\cD_\tau \tilde{z}) + \cM_{\tilde{z}\tilde{z}} (\cD_\tau \tilde{z})^2 \big)
\end{equation}
with $\cD_\tau z=\dot{z}+\dot{X}^\mu A_\mu{}^z$ and $\cD_\tau\tilde{z}=\dot{\tilde{z}}+\dot{X}^\mu A_\mu{}^{\tilde{z}}+V^{\tilde{z}}$.
The Euler--Lagrange equation for $V^{\tilde{z}}$ gives
\begin{equation}
    \cD_\tau \tilde{z} = - \frac{\cM_{z\tilde{z}}}{\cM_{\tilde{z}\tilde{z}}} \cD_\tau z \,.
\end{equation}
Substituting back gives the Schur complement
\begin{equation}
    L_{(\tilde{z},z)}
    =\frac{1}{2e}
    \bigg( \cM_{zz} - \frac{\cM_{z\tilde{z}}\cM_{\tilde{z}z}}{\cM_{\tilde{z}\tilde{z}}} \bigg) (\cD_\tau z)^2.
\end{equation}
In the exotic case, the coefficient in parentheses is
\begin{equation}
    H^{-1} - \frac{(H^{-1}\sigma\theta)^2}{H^{-1}K}
    = H K^{-1} \,.
\end{equation}
This is just the coefficient of $\rd z^2$ in the local $0^{(1,7)}$ metric \eqref{eq:metric_0^(1,7)_Brinkmann}.
The M0 is obtained separately by choosing the usual Kaluza--Klein fields $\cM_{zz}=H$, $A_t{}^z=-H^{-1}$, and $g_{tt}=-H^{-1}$.

We can equivalently package $\cM^{0^{(1,7)}}_{(z,\tilde{z})}$ in the upper half-plane coordinate
\begin{equation}
    \tau_z = - \sigma \theta + iH \,.
\end{equation}
This is often written as the Ernst potential \cite{Ernst:1967wx}, and it gives $\mathrm{Im}\,\tau_z=H$, $\mathrm{Re}\,\tau_z=-\sigma\theta$, and $|\tau_z|^2=K$, leading to the well-known $\mathrm{SL}(2,\mathbb{R})/\mathrm{SO}(2)$ Ehlers scalar matrix\footnote{This notation is well-known, but here it may be confusing since $\tau$ is also used for the worldline parameter.}
\begin{equation}
    \cM(\tau_z)
    = \frac1{\mathrm{Im}\,\tau_z}
    \begin{pmatrix}
        |\tau_z|^2&\mathrm{Re}\,\tau_z\\
        \mathrm{Re}\,\tau_z&1
    \end{pmatrix} .
\end{equation}
The fields appearing in the $0^{(1,7)}$ metric are often written in terms of the inverse complex function
\begin{equation}
    \tau_z \;\longmapsto\; -\frac{1}{\tau_z}
    = -\frac{1}{-\sigma\theta+iH}
    = \frac{\sigma\theta+iH}{K} \,.
\end{equation}

We now compute the monodromy.
This is the transition function relating two local frames after travelling along a loop $\gamma$ around the defect.
Let $\gamma$ have winding number $n$ in the transverse $(r,\theta)$ plane.
The monodromy $\Omega_z$ that acts on the old frame and sends it to the new frame is\footnote{
Strictly speaking, this is true up to a local $\mathrm{SO}(16)$ element acting from the right.
Since such an element preserves the invariant $\delta_{AB}$ in $\cM_{MN}=\cV_M{}^A\cV_N{}^B\delta_{AB}$, it drops out of the internal metric.}
\begin{equation}
    \Omega_z \, g_z(\theta)
    := g_z(\theta+2\pi n) \,.
    \label{eq:monodromy_definition}
\end{equation}
Of course, the idea that exotic branes are patched by monodromy transformations is not new -- some useful references are \cite{deBoer:2010ud,deBoer:2012ma,Berman:2018okd,Kimura:2023nvt}.
Using $[E_z,E_z]=0$ one computes
\begin{equation}
    \Omega_z
    = \exp ( {-}\sigma (\theta + 2 \pi n) E_z ) \exp ( \sigma \theta E_z )
    = \exp( -2 \pi n \sigma E_z ) \,.
\end{equation}
Treating the particle as a localised source, vector--scalar duality suggests a relation between this charge and the scalar monodromy, although this is not derived here.
One could formally choose
\begin{equation}
    \Gamma^\sharp
    = \frac{1}{2\pi n} \log \Omega_z
    = -\sigma E_z \,,
\end{equation}
such that the coadjoint orbit exponentiates to the conjugacy class of $\Omega_z$.
This was not imposed by the action.
Our worldline variable $g(\tau)$ is \emph{not} the same as $g_z(\theta)$, and the orbit construction does not impose the quantisation for $E_8(\mathbb{Z})$.

For the M0-brane, there is no such nilpotent factor $\Omega_z$ and we find instead that the monodromy is trivial.
The same result can be obtained using the Maurer--Cartan form
\begin{equation}
    g_z^{-1} \rd g_z
    = - \sigma \, \rd \theta \, E_z \,.
\end{equation}
Since this connection is abelian along the chosen nilpotent subgroup, no path ordering is needed, and
\begin{equation}
    \Omega_z
    = \exp \bigg( \oint_{\gamma} g_z^{-1} \rd g_z \bigg)
    = \exp \bigg( {-}\sigma E_z \oint_{\gamma} \rd \theta \bigg)
    = \exp( - 2 \pi n \sigma E_z) \,.
\end{equation}
Under this transition function, the internal metric and generalised velocity satisfy
\begin{align}
    \cM_{MN}(\theta+2\pi n)
    &= (\Omega_z)_M{}^P \cM_{PQ}(\theta) (\Omega_z)^Q{}_N \,,&
    \cD_\tau Y^M(\theta+2\pi n)
    &= (\Omega_z^{-1})^M{}_N \cD_\tau Y^N (\theta) \,.
\end{align}
Using index-free notation, $\cM'=\Omega\cM\Omega^\rT$ and $(\cD_\tau Y)'=\Omega^{-\rT}\cD_\tau Y$.
Consequently, the internal part of the line element $\cM_{MN}\cD_\tau Y^M\cD_\tau Y^N$ is invariant and globally well-defined.

Importantly, an $E_8$ group element acts on the generalised frame, not directly on some solution or Lagrangian already in the 11D section.
Only after transforming $\cM_{MN}$ and $\cD_\tau Y^M$ do we reduce to 11D and eliminate the unphysical velocities using $V^M$ to obtain a local geodesic Lagrangian.

\subsubsection*{A brief look at the Kaluza--Klein monopole}

The smeared KK$6^1$ \eqref{eq:metric_KK6^1} has a smeared base direction $z=z_1$ and Taub--NUT fibre $z_2$.
Recall that travelling along a loop of winding number $n$ around the $(r,\theta)$ plane amounts to a shift $\theta\mapsto\theta+2\pi n$ that causes $\rd z_2-\sigma\theta\,\rd z_1$ to transform non-trivially unless we also take $z_2\mapsto z_2+2\pi\sigma z_1$.
Hence the $E_8$ monodromy transforms the KK$6^1$ to itself up to a large diffeomorphism in the $(z_1,z_2)$ torus.

Note that the non-zero Taub--NUT connection $A:=-\sigma\theta\,\rd z_1$ is not to be confused with the ExFT field $A_\mu{}^M$.
Both $z_1$ and $z_2$ are internal coordinates, so the term $H^{-1}\big(\rd z_2+A\,\rd z_1\big)^2$ contributes to the internal 11D metric $\cM_{mn}$ only.
Since $\hat{g}_{\mu m}=0$, the Kaluza--Klein vectors vanish, i.e.~$A_\mu{}^m=0$, and the non-trivial monodromy is instead encoded in $\cM_{z_1z_2}=-\sigma\theta H^{-1}$.

The local geodesic Lagrangian for the monopole is
\begin{equation}
    L_{\mathrm{KK}6^1}
    = \frac{1}{2e}
    \Big( {-}\dot{t}^{\,2} + H \big( \dot{r}^2 + r^2 \dot{\theta}^2 + \dot{z}_1^2 \big)
    + H^{-1} \big( \dot{z}_2 - \sigma \theta\,\dot{z}_1 \big)^2
    + \dot{z}_3^2 + \ldots + \dot{z}_8^2 \Big) \,.
\end{equation}
This Lagrangian is single-valued up to a large diffeomorphism.
In this sense, the KK$6^1$ has geometric monodromy while that of the exotic zero-brane $0^{(1,7)}$ is non-geometric \cite{Bergshoeff:1997gy,deBoer:2010ud,deBoer:2012ma}.

\section{Conclusion}
\label{sec:conclusion}

In this paper, we constructed worldvolume descriptions of particles and exotic branes from the point of view of DFT and ExFT.
The idea is that a duality-covariant target space should have corresponding probes that are also duality-covariant.
For particles, strings, and also wrapped branes, this is already realised in the double and exceptional worldline and worldsheet models \cite{Blair:2017gwn,Arvanitakis:2017hwb,Arvanitakis:2018hfn}.
Our aim was to extend this in two directions.
Firstly, we proposed dynamics for generic exotic branes where special isometry directions are gauged on the worldvolume itself.
We showed that this dynamics is natural from the duality-covariant Hamiltonian point of view.
Secondly, we investigated a possible $E_8$ extension of the exceptional worldline model, where the target space ExFT contains the constrained gauge field $B_{\mu M}$ and parameter $\Sigma_M$.
The construction that we developed is not supersymmetric, and it is an effective classical description of worldvolume dynamics, i.e.~not derived from some complete quantum theory.
As such, reductions to the M0-brane and exotic zero-brane are to be understood as consistency checks of the duality-covariant formalism.

The first part concerned the bosonic worldvolume dynamics of exotic branes in generality.
These objects are not characterised only by their longitudinal directions, they also have transverse isometry directions along which they are smeared in specific ways.
This differentiates exotic from usual branes.
After reduction to lower dimensions, one finds particle-like objects with masses that `remember' not only how the brane was wrapped but also the sizes of the special directions that had been gauged away on the worldvolume.
This led to our universal proposal for a Hamiltonian description of worldvolume dynamics.
We showed that this generalises the known treatment of the Kaluza--Klein monopole.

There are several ways in which this proposal could be tested beyond our Hamiltonian derivation.
One could couple the action to supergravity/ExFT fields and verify that its source terms reproduce the corresponding exotic brane background.
This would generalise the calculation already performed for the monopole.
Secondly, actions for specific exotic branes could be derived by applying sequences of (T- and U-)dualities to known brane actions.
For this one would need to keep track of the gauged isometries and Wess--Zumino couplings.
Another test would be to wrap the proposed actions, i.e.~perform double dimensional reduction, to see if we obtain actions for descendant branes.
This would be particularly straightforward.
Lastly, the strongest test would be a supersymmetric completion, and it would tell us if these actions describe the expected half BPS objects.

As has already been observed in the case of string dynamics \cite{Tseytlin:1990nb,Siegel:1993bj,Alekseev:2004np}, going to the phase space makes duality-covariance more transparent, also for higher dimensional $p$\hspace{0.4mm}-forms.
As suggested in \cite{Hatsuda:2012uk,Hatsuda:2012vm,Hatsuda:2013dya}, it is possible to arrange the worldvolume degrees of freedom into representations $\cR_1,\cR_2,\ldots$ of the ExFT tensor hierarchy in the form of so-called currents.
Here, we employ the formalism worked out in \cite{Osten:2021fil,Osten:2024mjt}, that starts with manifest duality-covariant current.
Realising generalised diffeomorphisms on phase space requires the brane charge constraints \eqref{eq:BraneChargeSmall} to be imposed.
Moreover, this formalism offers a very natural choice of Hamiltonian and, hence, brane dynamics.
We demonstrated that this formalism reproduces the Blair's point particle action \cite{Blair:2017gwn} for $p=0$.
The gauging, corresponding to the special target space directions, is imposed via first order constraints that can be constructed from the brane charge constraints.

For $E_n$ with $n\leq7$, the exceptional particle couples to the one-form field with Wess--Zumino term $p_MA_\mu{}^M\dot{X}^\mu$, where $p_M$ is some fixed momentum.
However, for $E_8$ the algebra of local transformations closes only when ancillary transformations are introduced, and the corresponding gauge field $B_{\mu M}$ can be seen by the worldline.
In the 11D section this provides a St\"uckelberg mechanism that removes the dual gravity part of the scalar sector.
A worldline action probing $E_8$ geometry \cite{Cederwall:2015ica} should be capable of coupling to $B_{\mu M}$.
We proposed such a coupling:~our coadjoint orbit term $Q^M(B_{\mu M}\dot{X}^\mu+W_M)$ with a `duality charge' $Q^M$ whose Kirillov--Kostant--Souriau brackets reproduce the $E_8$ Lie algebra.
This is the minimal coupling to (the pullback of) $B_{\mu M}$, supplemented by a new constrained worldline field $W_M$ so that everything transforms as needed.

We then applied the above to the M0-brane, i.e.~the 11D pp-wave, and the exotic zero-brane $0^{(1,7)}$.
On the 11D section, the universal kinetic sector reduces locally to the massless geodesic action for the M0-brane, as expected.
Locally, after choosing a different frame, the same kinetic sector reduces to the geodesic action for the exotic zero-brane.
Globally, travelling around the defect acts on the fields by an $E_8$ transition function.
The reduced fields in 11D are not single-valued in a regular supergravity frame.
One should not think of $E_8$ as acting directly on some already-reduced 11D Lagrangian, and sending it to another Lagrangian.
It acts instead on the exceptional frame, and only afterwards does one make a choice of section and eliminate any pesky unphysical velocities.

One might wish to couple the worldline action directly to the bulk $E_8$ ExFT action, and to treat the particle as a localised source.
The coupling $P_M A_\mu{}^M\dot X^\mu$ will source the $A_\mu{}^M$
equation and, after making use of vector--scalar duality, it may relate Kaluza--Klein momentum $P_z$ to scalar monodromy.
It seems that the coupling $Q^M B_{\mu M}\dot X^\mu$ would instead source the $B_{\mu M}$ equation, and hence the first-order duality relation directly.
In our minimal model, the relevant components of $Q^M$ are eliminated by the $W_M$ equation of motion, so a non-trivial realisation of the duality between gravity and dual gravity may need the quadratic deformation that is discussed in Appendix~A.

There are several limitations of our construction that we should emphasise.
Firstly, everything is bosonic and we have not taken supersymmetry or kappa symmetry into account.
The coadjoint orbit term is, as said above, the minimal way to incorporate the $B_{\mu M}$ field, but there are of course other possible couplings that could and should be considered.
Moreover, our comparison with zero-branes in 11D does not by itself prove the uniqueness of the worldline action.
For this one needs to consider worldline gauge invariance including possible higher-order terms or topological terms.

It would also be quite useful to understand better the mixed-symmetry gauge potentials that are used to classify exotic branes.
The exotic zero-brane $0^{(1,7)}$ is understood to couple to the higher dual gravity field $h_{9,8,1}$ in eleven dimensions, but this only starts to show up in $E_n$ ExFT from $n=9$ onwards \cite{Bossard:2018utw,Bossard:2021jix,Bossard:2019ksx,Bossard:2021ebg}.
Understanding which mixed-symmetry fields propagate 11D supergravity degrees of freedom therefore continues to be one of the most direct methods of initiating a brane scan that does not require the use of supersymmetry \cite{Brown:2004jb,West:2004kb,Englert:2007qb,Cook:2008bi,Cook:2009ri,Houart:2009ya,Houart:2011sk,West:2018lfn}.

An interesting direction for future work is to try to understand what a duality-covariant worldline action would look like for larger Kac--Moody symmetries $E_9$, $E_{10}$, and $E_{11}$.
The associated ExFTs have many more fields carrying a constrained index compared to just $B_{\mu M}$ for $E_8$ \cite{Bossard:2018utw,Bossard:2021jix,Bossard:2019ksx,Bossard:2021ebg}.
One may try to introduce a corresponding set of $Q^M$ objects, but it is not clear what any kind of orbit sector might look like.
A direct analogue of our coadjoint orbit coupling may be insufficient.
It may be that a hierarchy of new variables is needed, or a current algebra valued in a larger module.

The kinetic sector is more straightforward.
For $E_9$ the external spacetime $(x^0,x^1)$ is now only two-dimensional, but the nine internal dimensions $(y^1,\dots,y^9)$ are extended to $\cR_1$ of $E_9$, which is infinite-dimensional.
In the $E_{10}$ case, the only external dimension $(x^0)$ is an evolution parameter that is interpreted unsurprisingly as time, and all ten spatial dimensions $(y^1,\dots,y^{10})$ are internal.
It would be illuminating to understand if or how an $E_{10}$-covariant worldline \textit{\`a la} Blair \cite{Blair:2017gwn} would have a relation to the $E_{10}$ sigma model Lagrangian \cite{Damour:2002cu,Damour:2004zy,Henneaux:2007ej}.
It would especially be important to understand the role of constraints that are known to be needed in the $E_{10}$ sigma model in order to reproduce the dynamics of 11D supergravity \cite{Damour:2007dt}.
See also the thought-provoking proposal in Appendix~E of \cite{Bossard:2021ebg}.

The case of $E_{11}$ is particular in that there is no external spacetime, only an internal coordinate $y^M$ in the $\cR_1$ of $E_{11}$ \cite{West:2001as,West:2003fc}.
It is known what (a lot of) the target space theory is \cite{Tumanov:2015yjd,Tumanov:2016abm,Bossard:2017wxl,Bossard:2019ksx,Bossard:2021ebg}, and the kinetic sector of its worldline pullback would essentially be $\frac12e^{-1}\cM_{MN}\cD_\tau Y^M\cD_\tau Y^N$ in terms of the covariant velocity $\cD_\tau Y^M=\dot{Y}^M+V^M$.
The tricky part is to work out how its many constrained fields would or could enter the worldline action.
There is also a separate line of work attempting to formulate M-theory in
terms of current algebras based on $E_{11}$ \cite{Sugawara:2017fds,Shiba:2017oaa,Glennon:2026tuk}.
Perhaps the current algebra appearing in our construction is in some way related to these proposals.

Another important direction is to consider duality-covariant worldsheets.
The exceptional string model \cite{Arvanitakis:2018hfn} provides an $E_6$-covariant action, where an $\cR_2$-valued two-form $B_{\mu\nu}{}^{M_2}$ and fixed charge $q_{M_2}$ in the dual representation $\overline{\cR}_2$ appear in the Wess--Zumino term.
For $E_7$, however, the tensor hierarchy also contains an additional constrained two-form $B_{\mu\nu M}$ just as the $E_8$ tensor hierarchy contains $B_{\mu M}$.
This suggests that an $E_7$-covariant worldsheet action may require an extra top-form coupling beyond the Wess--Zumino term that worked for $E_6$.
A natural first guess is to supplement the usual term
\begin{equation}
    q_{M_2} \varepsilon^{\alpha\beta}
    \Big( B_{\mu\nu}{}^{M_2}
    \partial_\alpha X^\mu \partial_\beta X^\nu
    + \eta^{M_2}{}_{MN}
    \big( \partial_\alpha X^\mu A_\mu{}^M \cD_\beta Y^N
    + \partial_\alpha Y^M V_\beta^N \big) \Big)
\end{equation}
by some additional coadjoint orbit coupling of the form
\begin{equation}
    Q^M \varepsilon^{\alpha\beta} B_{\alpha\beta M}
    = Q^M \varepsilon^{\alpha\beta} \partial_\alpha X^\mu \partial_\beta X^\nu B_{\mu\nu M} \,.
\end{equation}
This is obviously not the unique possibility, but it is the most direct generalisation of what we proposed for an $E_8$-covariant particle.
Moreover, a new constrained worldsheet field analogous to $W_M$ in our $E_8$ worldline action would need to be present.
One would then need to determine if such a term can be made invariant under the tensor hierarchy gauge transformations and external diffeomorphisms, and whether or not its coefficient is fixed relative to the standard Wess--Zumino term.

It would be even more subtle to construct an $E_8$-covariant worldsheet.
The tensor hierarchy for $E_8$ contains at the first level two external one-forms $A_\mu{}^M$ and $B_{\mu M}$, the latter of which is constrained.
At the next level it contains a two-form $C_{\mu\nu}{}^{M_2}$ valued in $\cR_2=\mathbf{1}\oplus\mathbf{3875}$ and a constrained two-form $C_{\mu\nu M}{}^N$ valued in $\cR_1=\mathbf{248}$ \cite{Hohm:2014fxa}.
This is tied to the $E_8$ section constraint \eqref{eq:SC_generic} which projects onto the $\mathbf{1}\oplus\mathbf{248}\oplus\mathbf{3875}$.
An $E_8$-covariant worldsheet clearly requires a lot more thought.
The constrained one-form $B_{\mu M}$ and two-form $C_{\mu\nu M}{}^N$ should presumably enter together, with relative coefficients fixed by gauge covariance, but the correct structure remains to be found.

Taking a step back, the pattern seems to be as follows.
The duality-covariant worldvolume story for $E_n$ with $n\leq 6$ is controlled by the tensor hierarchy and all the corresponding charge constraints.
Starting at $E_7$ and becoming unavoidable at $E_8$, the constrained fields associated with dual gravity enter directly and are visible to our probes.
These fields are invisible in simpler cases, or they appear only through duality equations, but they are essential if the worldvolume action is going to couple to the full ExFT background.
In more maximal cases such as that of $E_{11}$, one would need to work out how a probe behaves in the presence of the infinite set of propagating higher dual potentials and the infinite set of non-propagating potentials.

It would also be useful to revisit the doubled string model \cite{Hull:2004in,Hull:2006va,Arvanitakis:2018hfn} in which the worldsheet gauge field $V_\alpha^M$ removes the unphysical dual coordinates and the Wess--Zumino term is compatible with the $\mathrm{O}(d,d)$ tensor hierarchy.
This is a worldsheet theory pulled back from the target space DFT, and it is the simplest example of the mechanism that becomes increasingly elaborate for ExFT target spaces.
Comparing the doubled string, the exceptional string, and the $E_8$ particle may help to isolate which aspects of the construction are specific to dual gravity.

Yet another important direction would be to understand the $8^{(1,0)}$-brane, also called the M9-brane, in 11D from the perspective of duality-covariant worldvolumes and Hamiltonians \cite{Bergshoeff:1998bs,Sato:1999bu,Sato:2000mw,Bergshoeff:2006qw}.
This is not a space-filling brane, it has eight standard spatial directions and one special isometry direction.
It is more accurately viewed as an exotic brane, and its worldvolume theory is a gauged sigma model in which translations along the special isometry direction are gauged.
Reducing the $8^{(1,0)}$-brane gives the D8-brane in IIA theory when reduced along its special isometry direction.
The other inequivalent reductions are the exotic brane $8^{(1,0)}_3$ (KK8A) and a space-filling brane $9_2$ (NS9A).
In contrast to the usual M2, M5, and KK$6^1$, the $8^{(1,0)}$ appears to require top-forms and mixed-symmetry potentials.
Such fields are found in $E_{11}$ theory \cite{West:2001as} where they can be interpreted as 11D ancestors of these 10D top-forms \cite{Tumanov:2016dxc}.
For the $8^{(1,0)}$-brane, one could try to understand the duality-covariant phase space variables, currents, and so on.
Extending to the $E_9$, $E_{10}$, and $E_{11}$ cases may provide answers.

Starting from $E_9$, another question arises.
Compactifying 4D gravity or 11D supergravity to 2D gives a theory with affine Kac--Moody symmetry given by the Geroch group, i.e.~$A_1^+=\mathrm{SL}(2,\mathbb{R})^+$ or $E_8^+=E_9$.
It was shown in \cite{Cesaro:2024ipq,Cesaro:2025msv} that these theories in 2D possess a unique higher-derivative deformation that keeps most underlying structures intact.
Existing work towards an $E_9$ brane scan associates a $\tfrac12$-BPS solution of 11D supergravity or an exotic dual formulation to each real root of $E_9$, recovering M0, M2, M5, and KK$6^1$ as basic members \cite{Englert:2007qb}.
The fate of worldvolume dynamics in this direction is not yet understood, nor is it known if the unique higher-derivative deformation has counterparts in other $E_n$-covariant theories.

Lastly, it is important to mention that everything we have investigated for the $E_8$ group and 11D supergravity can also be understood for $\mathrm{SL}(2,\mathbb{R})$ and 4D gravity.
There is the good old pp-wave, and a Kaluza--Klein monopole that is essentially a (Euclidean) $(-1)$-brane, so it seems sensible to label it as KK$(-1)^1$.
The exotic zero-brane $0^{(1,7)}$ has seven spectator indices, and the four-dimensional analogue should be labelled as $0^{(1,0)}$, where its metric representative $\rd s_{0^{(1,0)}}^2$ is simply that of equation \eqref{eq:metric_0^(1,7)} without the $\rd \vec{z}_{(7)}^{\,2}$ term.
This is non-geometric, of course, and travelling in a loop around this defect involves an $\mathrm{SL}(2,\mathbb{R})$ duality transformation.
The exponents appearing in \eqref{eq:exoticPolyakovgauged} imply that the $b=-1$ case of the monopole could be tricky to work out.
It would be interesting to study the Hamiltonians for the pp-wave and the exotic zero-brane in four dimensions, and to formulate Ehlers-covariant worldline and worldvolume models.

In the present paper, we hope to have made a step towards a more systematic duality-covariant theory of particles, strings, and branes.
The exceptional worldline and worldsheet models show that momentum, winding, and wrapped brane charges can be treated in the same way.
Exotic branes force us to include special isometry directions, monodromy transformations, and constrained ExFT fields associated with dual gravity degrees of freedom.
The $E_8$ worldline action proposed here is a (minimal) attempt to include all these features.
Extending this to larger exceptional symmetries, worldsheets and higher-dimensional worldvolumes, and to supersymmetric branes, all remain open problems.

\subsection*{Acknowledgements}

We wish to thank Isma\"el Ahlouche Lahlali, Alex Arvanitakis, Mattia Cesaro, and Axel Kleinschmidt for useful comments and discussions, and also the organisers of the workshop \emph{FDA meets Generalised Geometry} in Mons, Belgium.
J.O.~would like to thank Falk Hassler and D.O.~for hospitality at the University of Wroc{\l}aw, Poland, and Axel Kleinschmidt at the Max Planck Institute for Gravitational Physics, Germany, where parts of this work were completed.
The work of J.O.~was supported in November 2025 by the SONATA BIS grant 2021/42/E/ST2/00304 of the National Science Center (NCN), Poland, and in December 2025 by the Max Planck Institute for Gravitational Physics, Germany.
The work of J.O.~is supported by the Croatian Science Foundation project IP-2022-10-5980 ``Non-relativistic supergravity and applications''.
The research of D.O.~was part of the SONATA grant No.~2024/55/D/ST2/01205 funded by NCN.

\appendix

\section{Deformation of the orbit sector}
\label{app:deformation}

In this appendix we explore a possible modification of the $E_8$ particle action of Section~\ref{sec:E8_worldline_action} to include a term quadratic in the orbit current $J$.
This guarantees that imposing the equation of motion for $W_M$ does not remove or trivialise the dynamics in the orbit sector.

\subsection{Structure and symmetries of the deformed model}

The orbit sector \eqref{eq:S_orbit} of the deformed model \eqref{eq:S_total} is given by
\begin{equation}
    S_\mathrm{orb}
    = \int\!\rd\tau\, \Big( \big\langle \Gamma , J \big\rangle + \frac{\ell}{2e} \kappa \big( \Pi J , \Pi J \big) \Big)
    = \int\!\rd\tau\, \Big( \Gamma^M J_M + \frac{\ell}{2e} \kappa^{MN} \Pi^P{}_M \Pi^Q{}_N J_P J_Q \Big) \,,
    \label{eq:S_orbit_quadratic}
\end{equation}
where $\ell$ is a new coupling, and where $\Pi$ is a projector onto the directions transverse to the stabiliser algebra $\mathfrak{g}_\Gamma=\mathrm{Lie}(G_\Gamma)\subset\mathfrak{e}_8$\,.
Recall that $\mathfrak{e}_8=\mathfrak{g}_\Gamma\oplus\mathfrak{m}_\Gamma$ such that $\Pi:\mathfrak{e}_8\to\mathfrak{m}_\Gamma$ is the projector.
We need $\Pi(\Ad_gX)=\Ad_g(\Pi X)$ for all $X\in\mathfrak{e}_8$ and $g\in G_\Gamma$ so that the action depends only on the point inside the orbit and not on the representative $g$.
Let $\Pi^\dagger$ denote the adjoint of $\Pi$ with respect to the Killing form, i.e.~$\kappa(\Pi X,Y)=\kappa(X,\Pi^\dagger Y)$.
If the complement is $\kappa$-orthogonal to the stabiliser, then $\Pi^\dagger=\Pi$.
We use $\Pi^\dagger=\Pi$ in this paper.

The orbit sector \eqref{eq:S_orbit_quadratic} is
\begin{equation}
    S_\mathrm{orb}
    = S_\mathrm{orb}^{(J)} + S_\mathrm{orb}^{(JJ)} \,,
    \label{eq:S_orbit_terms}
\end{equation}
where the term linear in $J$ is \eqref{eq:S_orbit} and the term quadratic in $J$ is given by
\begin{equation}
    S_\mathrm{orb}^{(JJ)}
    = \int\!\rd\tau\, \frac{\ell}{2e} \kappa \big( \Pi J , \Pi J \big)
    = \int\!\rd\tau\, \frac{\ell}{2e} \kappa^{MN} \Pi_M{}^P \Pi_N{}^Q J_P J_Q \,.
    \label{eq:S_orbit_JJ}
\end{equation}
Without the quadratic term, the Euler--Lagrange equation of $W_M$ would trivialise the internal orbit dynamics as described in Section~\ref{sec:E8_worldline_action}.

Note that $S_\mathrm{orb}^{(JJ)}$ does not vanish due to the section constraint.
Of course we have $\kappa^{MN}C_M\otimes\tilde{C}_N=0$ for any pair of constrained objects, but this does not guarantee that $\kappa(\Pi J,\Pi J)$ vanishes.
Firstly, $J_M$ is not covariantly constrained.
It contains $j_M$ which has nothing to do with the constrained worldline connection $\widetilde{B}$, so already the term quadratic in $j$ is generally non-zero.
Moreover, conjugation $g^{-1}\widetilde{B}g$ does not need to preserve the constrained subspace.
The singlet part of the section constraint \eqref{eq:SC_generic} is $\kappa(C,\tilde{C})=0$ for all constrained objects, and the unprojected form of the quadratic term would include $\kappa(g^{-1}\widetilde{B}g,g^{-1}\widetilde{B}g)=\kappa(\widetilde{B},\widetilde{B})=0$ since $\kappa$ is $E_8$-invariant, but $\kappa(\Pi J,\Pi J)$ does not need to vanish.
Even if some $X\in\mathfrak{e}_8$ is null in the sense that $\kappa(X,X)=0$, its projection $\Pi X$ need not be null in general.

As for symmetries, the model is invariant under worldline reparametrisations.
We also have that $J$ is invariant under \eqref{eq:g_transformation}, so $\langle\Gamma,J\rangle$ and $\kappa(\Pi J,\Pi J)$ are each invariant.
Moreover, under $g\mapsto gk$ with $k\in G_\Gamma$ we have that the quadratic term is invariant since $k^{-1}\dot{k}\in\mathfrak{g}_\Gamma$ and since $\Pi J$ projects $J$ onto $\mathfrak{g}_\Gamma$.
Lastly, $J$ is inert under $E_8$ generalised diffeomorphisms, so $S_\mathrm{orb}^{(J)}$ and $S_\mathrm{orb}^{(JJ)}$ are separately invariant.

Define the adjoint element
\begin{equation}
    \widetilde{\Gamma}^\sharp := \Gamma^\sharp + \frac{\ell}{e} \, \Pi^\dagger \Pi J \in E_8 \,,
\end{equation}
which simplifies to
\begin{equation}
    \widetilde{\Gamma}^\sharp := \Gamma^\sharp + \frac{\ell}{e} \, \Pi J
    \label{eq:Gamma^sharp=Gamma+PiJ}
\end{equation}
if $\Pi$ is self-adjoint with respect to the Killing form.
The Euler--Lagrange equation for $g$ is found to be
\begin{equation}
    \dot{\widetilde{\Gamma}}{}^\sharp + [J,\widetilde{\Gamma}^\sharp] \approx 0
    \qquad\Longleftrightarrow\qquad
    \dot{\!\widetilde{Q}}{}^\sharp + [\widetilde{B},\widetilde{Q}{}^\sharp] \approx 0 \,,
    \label{eq:EOM_g_Q_tilde}
\end{equation}
where we defined the dressed duality charge $\widetilde{Q}{}^\sharp := \Ad_g \widetilde{\Gamma}^\sharp$.
This covariant transport equation reduces in the $\ell\to0$ limit to $\dot{Q}^\sharp+[\widetilde{B},Q^\sharp]\approx0$ with $Q^\sharp=\Ad_g\Gamma^\sharp$.
Separately, the Euler--Lagrange equation of $W$ tells us that, just as for $Q^\sharp$ in the undeformed model, $\widetilde{Q}{}^\sharp$ must lie along the unphysical directions.

Since the einbein appears in \eqref{eq:S_orbit_JJ}, the variation of the orbit sector with respect to $e$ is
\begin{equation}
    \delta S_\mathrm{orb} = - \int\!\rd\tau\, \frac{\ell}{2e^2} \, \kappa(\Pi J,\Pi J) \, \delta e \,.
\end{equation}
The full Euler--Lagrange equation of $e$ is
\begin{equation}
    g_{\mu\nu} \dot{X}^\mu \dot{X}^\nu + \cM_{MN} \cD_\tau Y^M \cD_\tau Y^M + \ell \, \kappa(\Pi J, \Pi J) \approx 0 \,.
\end{equation}
As such, the orbit current $J$ contributes to the mass-shell constraint of the particle.

\subsection{First-order formulation of the orbit sector}

The first-order form of the deformed orbit sector action \eqref{eq:S_orbit_quadratic} is
\begin{equation}
    S_\mathrm{orb}^{(1)} = \int\!\rd\tau\, \Big( \big\langle\Gamma,J\big\rangle + \kappa(\varpi,\Pi J) - \frac{e}{2\ell} \, \kappa(\varpi,\varpi) \Big) \,,
    \label{eq:S_orbit_1st_order}
\end{equation}
where $\varpi(\tau)\in\mathfrak{m}_\Gamma:=\mathrm{im}\hspace{0.5mm}\Pi$ is the auxiliary orbit momentum.
Varying with respect to $\varpi$ gives
\begin{equation}
    \varpi = \frac{\ell}{e} \, \Pi J \,,
    \label{eq:varpi=PiJ}
\end{equation}
and substituting back recovers the second-order form \eqref{eq:S_orbit_quadratic}.
Using $j=g^{-1}\dot{g}$ we write the first-order orbit sector as
\begin{equation}
    S_\mathrm{orb}^{(1)} = \int\!\rd\tau\, \Big( \big\langle\Gamma,j\big\rangle + \kappa\big(\varpi,\Pi j\big) + \big\langle\widetilde{Q},\widetilde{B}\big\rangle - \frac{e}{2\ell} \, \kappa(\varpi,\varpi) \Big) \,.
    \label{eq:S_orbit_1st_order_jQB}
\end{equation}
The charge $\widetilde{Q}{}^\sharp=\Ad_g(\Gamma^\sharp+\varpi)$ is defined by the orbit representative $\Gamma$ and momentum $\varpi$.
Note also that the covariant momentum \eqref{eq:pi_mu_def} becomes $\pi_\mu=P_\mu-P_MA_\mu{}^M+\widetilde{Q}^MB_{\mu M}$ with $Q$ replaced by $\widetilde{Q}$.

\subsubsection*{Conservation of the quadratic orbit norm}

A useful consequence of the first-order formulation is that $\kappa(\varpi,\varpi)$ is conserved along the worldline.
Recall $\widetilde\Gamma^\sharp:=\Gamma^\sharp+\varpi$ from \eqref{eq:Gamma^sharp=Gamma+PiJ} and \eqref{eq:varpi=PiJ}.
Dependence of the Lagrangian on $J$ is entirely through its pairing with $\widetilde{\Gamma}^\sharp$.
The equation of motion for $g\in E_8$ is $\dot{\widetilde{\Gamma}}{}^\sharp+[J,\widetilde{\Gamma}^\sharp]=0$ or equivalently the transport equation $\,\dot{\!\widetilde{Q}}{}^\sharp+[\widetilde{B},\widetilde{Q}{}^\sharp]\approx0$ in \eqref{eq:EOM_g_Q_tilde}.

Now differentiate the quadratic invariant built from $\widetilde\Gamma^\sharp$ to find
\begin{equation}
    \frac{\rd}{\rd\tau} \kappa(\widetilde{\Gamma}^\sharp,\widetilde{\Gamma}^\sharp)
    = 2\,\kappa(\dot{\widetilde\Gamma}{}^\sharp,\widetilde{\Gamma}^\sharp)
    = -2\,\kappa([J,\widetilde{\Gamma}^\sharp],\widetilde{\Gamma}^\sharp)
    = -2\,\kappa(J,[\widetilde{\Gamma}^\sharp,\widetilde{\Gamma}^\sharp])
    = 0 \,.
    \label{eq:kappa_Gamma_tilde_conserved}
\end{equation}
We have used the equation of motion \eqref{eq:EOM_g} and invariance of the Killing form.
In order to relate this conserved quantity to $\kappa(\varpi,\varpi)$, use $\widetilde{\Gamma}^\sharp=\Gamma^\sharp+\varpi$ to obtain
\begin{equation}
    \kappa(\widetilde{\Gamma}^\sharp,\widetilde{\Gamma}^\sharp)
    = \kappa(\Gamma^\sharp,\Gamma^\sharp)
    + 2\,\kappa(\Gamma^\sharp,\varpi)
    + \kappa(\varpi,\varpi) \,.
    \label{eq:expand_tildeGamma_norm}
\end{equation}
Let us show that the mixed term vanishes.
Note that $\Gamma^\sharp$ itself lies in the stabiliser algebra $\mathfrak{g}_\Gamma$ since 
\begin{equation}
    \langle \ad_{\Gamma^\sharp}^*\Gamma,X \rangle
    = - \langle \Gamma,[\Gamma^\sharp,X] \rangle
    = - \kappa(\Gamma^\sharp,[\Gamma^\sharp,X])
    = -\kappa([\Gamma^\sharp,\Gamma^\sharp],X)
    = 0 \,,
\end{equation}
for all $X\in\mathfrak{e}_8$\,.
On the other hand, $\varpi$ belongs to the orthogonal complement $\mathfrak{m}_\Gamma$ of $\mathfrak{g}_\Gamma$\,.
Therefore we have $\kappa(\Gamma^\sharp,\varpi)=0$ and equation \eqref{eq:expand_tildeGamma_norm} simplifies to
\begin{equation}
    \kappa(\widetilde{\Gamma}^\sharp,\widetilde{\Gamma}^\sharp)
    = \kappa(\Gamma^\sharp,\Gamma^\sharp) + \kappa(\varpi,\varpi) \,.
\end{equation}
Moreover, $\kappa(\Gamma^\sharp,\Gamma^\sharp)$ is a constant number since $\Gamma$ is fixed.
As a result, \eqref{eq:kappa_Gamma_tilde_conserved} implies
\begin{equation}
    \frac{\rd}{\rd\tau} \kappa(\varpi,\varpi) = 0 \,,
    \label{eq:varpi_norm_conserved}
\end{equation}
and therefore the contribution of $\varpi$ to the Hamiltonian is constant along the worldline.

\subsection{Symplectic structure of the enlarged phase space}

The quadratic term in the orbit sector enlarges the phase space from the orbit to a twisted cotangent bundle over it.
Recall the splitting $\mathfrak{e}_8=\mathfrak{g}_\Gamma\oplus\mathfrak{m}_\Gamma$ with $\mathfrak{g}_\Gamma$ the stabiliser algebra and $\mathfrak{m}_\Gamma$ a complement identified with $\rT_\Gamma\cO_\Gamma$\,.
We denote by $\Pi:\mathfrak{e}_8\to\mathfrak{m}_\Gamma$ the projector that extracts the components tangent to the orbit.
On a patch $U\subset\cO_\Gamma$ the $\mathfrak{m}_\Gamma$ projection of the pulled-back Maurer--Cartan form is
\begin{equation}
    \Pi(\sigma^{-1}\rd\sigma)
    = \rd \xi^a E_a{}^A t_A
    =: \rd \xi^a E_a \,,
    \label{eq:orbit_frame_def}
\end{equation}
where $\{t_A\}$ is a basis of $\mathfrak{m}_\Gamma$ and we introduced $E_a:=E_a{}^At_A\in\mathfrak{m}_\Gamma$\,.
Thus $E_a$ is the $\mathfrak{m}_\Gamma$-valued frame induced by the section.
Equivalently, along a worldline $\xi^a(\tau)$, the projected Maurer--Cartan current is
\begin{equation}
    \Pi j
    = \Pi(g^{-1}\dot{g})
    = E_a\,\dot{\xi}^a \,.
    \label{eq:Pi_j_local}
\end{equation}

We now introduce the first-order auxiliary orbit momentum $\varpi=\varpi^At_A\in\mathfrak{m}_\Gamma$\,.
The term $\kappa(\varpi,\Pi j)$ in \eqref{eq:S_orbit_1st_order} is therefore $\kappa(\varpi,\Pi j)=\kappa(\varpi,E_a)\,\dot{\xi}^a$.
This motivates the definition of the local fibre momenta
\begin{equation}
    p_a := \kappa(\varpi,E_a) \,.
    \label{eq:pa_def}
\end{equation}
In terms of these variables,
\begin{equation}
    \kappa(\varpi,\Pi j) = p_a \dot{\xi}^a \,.
\end{equation}

Let $\rho:\rT^*\cO_\Gamma\to\cO_\Gamma$ denote the cotangent bundle projection.
On the local patch $\rT^*U\subset\rT^*\cO_\Gamma$ we have the canonical Liouville one-form $p_a\hspace{0.4mm}\rd \xi^a$.
Combining this with the local potential $\theta_{\Gamma}^{(U)}$ from \eqref{eq:local_theta_KKS}, we obtain the local symplectic potential on the enlarged phase space
\begin{equation}
    \Theta_\Gamma^{(U)}
    := \rho^*\theta_{\Gamma}^{(U)} + p_a \hspace{0.4mm} \rd \xi^a
    = \theta_a \hspace{0.4mm} \rd \xi^a + p_a \hspace{0.4mm} \rd \xi^a \,,
    \label{eq:Theta_orb_bundle_rewritten}
\end{equation}
where $\theta_\Gamma^{(U)}=\theta_a\hspace{0.4mm}\rd \xi^a$.
By construction, the worldline pullback of $\Theta_\Gamma^{(U)}$ gives the purely first-order part $\langle\Gamma,j\rangle+\kappa(\varpi,\Pi j)$ of the orbit action.

Taking the exterior derivative of \eqref{eq:Theta_orb_bundle_rewritten} gives the symplectic form on the enlarged phase space
\begin{equation}
    \Omega_\Gamma
    := \rd \Theta_\Gamma^{(U)}
    = \rho^* \omega_\mathrm{KKS} + \rd p_a \wedge \rd \xi^a
    = \frac12\,\omega_{ab}\,\rd \xi^a \wedge \rd \xi^b + \rd p_a \wedge \rd \xi^a \,,
    \label{eq:Omega_Gamma}
\end{equation}
where we have written
\begin{equation}
    \omega_\mathrm{KKS} \big|_U = \frac12\,\omega_{ab}\,\rd \xi^a\wedge \rd \xi^b \,.
\end{equation}
The canonical symplectic form $\rd p_a\wedge\rd\xi^a$ is shifted by the pullback of $\omega_\mathrm{KKS}$ on the base orbit.

It is useful to invert \eqref{eq:Omega_Gamma} explicitly.
In the local coordinates $(\xi^a,p_a)$ the symplectic matrix is
\begin{align}
    \Omega &=
    \begin{pmatrix}
        \omega_{ab} & -\delta_a^b \\
        \delta^a_b & 0
    \end{pmatrix},&
    \Omega^{-1} &=
    \begin{pmatrix}
        0 & \delta^a_b \\
        -\delta_a^b & \omega_{ab}(z)
    \end{pmatrix}.
\end{align}
Therefore the local Poisson brackets are
\begin{align}
    \{\xi^a,\xi^b\} &= 0 \,,&
    \{\xi^a,p_b\} &= \delta^a_b \,,&
    \{p_a,p_b\} &= \omega_{ab} \,.
    \label{eq:twisted_brackets}
\end{align}
In other words, the momentum brackets are deformed by the KKS curvature on the base space.

We next write the first-order orbit Lagrangian $L_\mathrm{orb}^{(1)}$ in local coordinates.
Writing
\begin{equation}
    \sigma^{-1} \widetilde{B} \sigma
    = -V_{\widetilde{B}}^a \, E_a + \beta_{\widetilde{B}}
    \label{eq:B_decomposition_local}
\end{equation}
with $\beta_{\widetilde{B}}\in\mathfrak{g}_\Gamma$ defines the orbit vector field $V_{\widetilde{B}}=V_{\widetilde{B}}^a\,\partial_a$ generated on the orbit by $\widetilde{B}\in\mathfrak{e}_8$\,.
The sign in \eqref{eq:B_decomposition_local} is chosen so that $V_{\widetilde{B}}^a$ obeys the standard Hamiltonian relation
\begin{equation}
    \partial_a \langle Q,\widetilde{B} \rangle
    = -\omega_{ab} \, V_{\widetilde{B}}^b \,,
    \label{eq:VB_moment_map_relation}
\end{equation}
where $Q(z)=\Ad_{\sigma(z)}^*\Gamma$ is the usual orbit moment map written in local coordinates.

The Lagrangian can now be written in terms of $(\xi^a,p_a)$.
Using \eqref{eq:orbit_frame_def}, \eqref{eq:pa_def}, and \eqref{eq:B_decomposition_local}, one finds
\begin{equation}
    \kappa \big( \varpi,\Pi(\sigma^{-1}\widetilde{B}\sigma) \big)
    = - p_a V_{\widetilde{B}}^a \,.
\end{equation}
Moreover, introducing the orbit metric $k_{ab}:=\kappa(E_a,E_b)$ with $k^{ab}k_{bc}=\delta^a{}_c$ leads to
\begin{equation}
    \kappa(\varpi,\varpi) = k^{ab} p_a p_b \,.
    \label{eq:varpi2_local}
\end{equation}
Hence the first-order orbit Lagrangian becomes
\begin{equation}
    L_{\mathrm{orb}}^{(1)}
    = \theta_a \dot{\xi}^a + p_a \dot{\xi}^a + \langle Q,\widetilde{B} \rangle - p_a V_{\widetilde{B}}^a - \frac{e}{2\ell} k^{ab} p_a p_b \,.
    \label{eq:L_orb_first_local}
\end{equation}

From \eqref{eq:L_orb_first_local}, the orbit Hamiltonian is
\begin{equation}
    H_{\mathrm{orb}}(\xi,p)
    = \frac{e}{2\ell} k^{ab} p_a p_b - \langle Q,\widetilde{B}\rangle + p_a V_{\widetilde{B}}^a \,.
    \label{eq:H_orb_local}
\end{equation}
One can use the twisted brackets \eqref{eq:twisted_brackets} to compute the Hamiltonian equations
\begin{align}
    \dot{\xi}^a
    &= \{\xi^a,H_{\mathrm{orb}}\}
    = \frac{\partial H_{\mathrm{orb}}}{\partial p_a}
    = \frac{e}{\ell} k^{ab} p_b + V_{\widetilde{B}}^a \,,
    \label{eq:zdot_local}\\
    \dot{p}_a
    &= \{p_a,H_{\mathrm{orb}}\}
    = -\frac{\partial H_{\mathrm{orb}}}{\partial \xi^a} + \omega_{ab} \frac{\partial H_{\mathrm{orb}}}{\partial p_b} \,.
\end{align}
Using \eqref{eq:H_orb_local} one finds that
\begin{align}
    \dot{p}_a
    = -\frac{e}{2\ell}\,\partial_a k^{bc} p_b p_c + \partial_a \langle Q,\widetilde{B} \rangle - p_b\,\partial_a V_{\widetilde{B}}^b + \omega_{ab} \left( \frac{e}{\ell} k^{bc} p_c + V_{\widetilde{B}}^b \right) ,
\end{align}
and \eqref{eq:VB_moment_map_relation} then gives
\begin{equation}
    \dot{p}_a
    = -\frac{e}{2\ell}\,\partial_a k^{bc} p_b p_c + \frac{e}{\ell}\,\omega_{ab} k^{bc} p_c - p_b\,\partial_a V_{\widetilde{B}}^b \,.
    \label{eq:pdot_local}
\end{equation}
Equations \eqref{eq:zdot_local} and \eqref{eq:pdot_local} are the explicit equations on the twisted cotangent bundle.

It is also useful to eliminate the $p_a$ and recover the second-order form.
Varying \eqref{eq:L_orb_first_local} with respect to $p_a$ gives
\begin{equation}
    \dot{\xi}^a - V_{\widetilde{B}}^a
    = \frac{e}{\ell} k^{ab} p_b \,,
\end{equation}
or equivalently
\begin{equation}
    p_a
    = \frac{\ell}{e} k_{ab} \big( \dot{\xi}^b - V_{\widetilde{B}}^b \big) \,.
\end{equation}
Substituting this back into \eqref{eq:L_orb_first_local}, one obtains the second-order action
\begin{equation}
    L_{\mathrm{orb}}
    = \theta_a \dot{\xi}^a
    + \langle Q,\widetilde{B} \rangle
    + \frac{\ell}{2e} k_{ab} \big( \dot{\xi}^a - V_{\widetilde{B}}^a \big) \big( \dot{\xi}^b - V_{\widetilde{B}}^b \big) \,.
    \label{eq:L_orb_second_local}
\end{equation}

The equations \eqref{eq:zdot_local} and \eqref{eq:pdot_local} have a straightforward interpretation.
Clearly, the first equation shows that the orbit velocity receives two distinct contributions.
The term $e\ell^{-1}k^{ab}p_b$ is the cotangent bundle velocity induced by the quadratic Hamiltonian $\frac{1}{2}e\ell^{-1}k^{ab}p_ap_b$ while $V_{\widetilde{B}}^a$ is the drift on the orbit generated by the coupling to $\widetilde{B}$.
Thus the quadratic term endows the orbit sector with its own local Hamiltonian motion along the fibres of $\rT^*\cO_\Gamma$\,.

The second equation \eqref{eq:pdot_local} encodes the evolution of the fibre momentum.
Position dependence of $k^{ab}$ affects the first term, while the second term is the force induced by the KKS curvature on the orbit, and the third describes how the orbit vector field generated by $\widetilde{B}$ acts on the fibre momentum.
Even in the absence of a background gauge field, the KKS term still affects the momentum evolution through the $e\ell^{-1}\omega_{ab}k^{bc}p_c$ term, and hence we refer to $\rT^*\cO_\Gamma$ as twisted.

Let us list some particular cases.
If $p_a=0$ then the momentum equation is automatically satisfied and the orbit motion reduces to
\begin{equation}
    \dot{\xi}^a = V_{\widetilde{B}}^a \,,
\end{equation}
so the trajectory along the orbit phase space is simply the Hamiltonian flow generated by the moment map coupling $\langle Q(\xi),\widetilde{B}\rangle$.
In contrast, if $\widetilde{B}=0$ then one obtains
\begin{align}
    \dot{\xi}^a 
    &= \frac{e}{\ell} k^{ab} p_b \,,&
    \dot{p}_a
    &= -\frac{e}{2\ell}\,\partial_a k^{bc} p_b p_c + \frac{e}{\ell}\,\omega_{ab} k^{bc} p_c \,,
\end{align}
which is essentially a geodesic flow on bundle.

It is also useful to compare the first-order and second-order formulations.
The relation
\begin{equation}
    p_a = \frac{\ell}{e} k_{ab} \big( \dot{\xi}^b - V_{\widetilde{B}}^b \big)
\end{equation}
shows that the fibre momentum measures the deviation of the orbit velocity from the drift generated by $\widetilde{B}$.
Substituting this into the first term of \eqref{eq:H_orb_local} gives the quadratic term in \eqref{eq:L_orb_second_local}.
The role of $\ell$ is now transparent, it controls the kinetic term.
In the $\ell\to0$ limit, the quadratic term disappears and the fibre momenta vanish.

\subsubsection*{The lifted group action and deformed moment map}

Recall that $V_X=V_X^a\,\partial_a$ is the fundamental vector field on the orbit for each $X\in\mathfrak{e}_8$\,.
In terms of the local section $\sigma:\cO_\Gamma\to\mathfrak{e}_8$ we have
\begin{equation}
    \Pi(\sigma^{-1}X\sigma) = V_X^a E_a \,,
    \label{eq:Pi_sXs}
\end{equation}
with $E_a$ the orbit frame in \eqref{eq:orbit_frame_def}.
The Hamiltonian function on $\cO_\Gamma$ is $H_X(Q(\xi))=\langle Q(\xi),X\rangle$ and it satisfies $\iota_{V_X}\omega_{\mathrm{KKS}}=-\rd H_X$.
In local coordinates this is $\partial_aH_X(z)=-\omega_{ab}\,V_X^b(z)$.

The action of $E_8$ on $\cO_\Gamma$ lifts canonically to an action on $\rT^*\cO_\Gamma$ where the lifted vector field is
\begin{equation}
    \widetilde{V}_X
    = V_X^a\,\partial_a - p_b\,\partial_a V_X^b \frac{\partial}{\partial p_a} \,.
    \label{eq:cotangent_lift_vector}
\end{equation}
The first term is the motion of the base point along the orbit, while the second term is the induced transformation of the fibre momentum.
The Hamiltonian generating $\widetilde{V}_X$ with respect to the symplectic form $\Omega_\Gamma$ is obtained by contracting $\widetilde{V}_X$ into $\Omega_\Gamma$\,.
A short computation gives
\begin{equation}
    \iota_{\widetilde V_X} \Omega_\Gamma
    = \iota_{V_X} \omega_{\mathrm{KKS}}
    + \iota_{\widetilde{V}_X} ( \rd p_a \wedge \rd \xi^a )
    = -\rd H_X - \rd \big( p_a V_X^a \big)
    = -\rd \widetilde{H}_X \,,
\end{equation}
where we defined
\begin{equation}
    \widetilde{H}_X := H_X + p_a V_X^a \,.
    \label{eq:Htilde_local_def}
\end{equation}

It is useful to match \eqref{eq:Htilde_local_def} with the first-order variables.
Using the auxiliary momentum $\varpi\in\mathfrak{m}_\Gamma$ and the relation $p_a=\kappa(\varpi,E_a)$, together with \eqref{eq:Pi_sXs}, one finds $p_aV_X^a=\kappa\big(\varpi,\Pi(\sigma^{-1}X\sigma)\big)$.
Therefore,
\begin{equation}
    \widetilde{H}_X
    = \langle Q,X\rangle
    + \kappa\big(\varpi,\Pi(\sigma^{-1}X\sigma)\big)
    = \big\langle \Ad_\sigma^*\Gamma,X \big\rangle
    + \big\langle \Ad_\sigma^*\varpi,X \big\rangle
    = \big\langle \Ad_\sigma^* \big( \Gamma + \varpi \big), X \big\rangle \,.
\end{equation}
In other words, if we define the deformed moment map $\widetilde{\mu}:\rT^*\cO_\Gamma\to\mathfrak{e}_8^*$ as
\begin{equation}
    \widetilde{\mu}(z,p)
    = \Ad_\sigma^* \big( \Gamma + \varpi \big)
    =: \widetilde{Q} \,,
    \label{eq:deformed_moment_map_def}
\end{equation}
then the lifted Hamiltonians are
\begin{equation}
    \widetilde{H}_X = \langle \widetilde{Q},X \rangle \,.
\end{equation}

This can also be seen from the first-order symplectic part of the action.
Under an infinitesimal transformation $\delta_\epsilon\hspace{0.4mm}g=\epsilon Xg$ with $\delta_\epsilon\varpi=0$ we find that
\begin{equation}
    \delta_\epsilon \big( \langle \Gamma,j \rangle + \kappa(\varpi,\Pi j) \big)
    = \dot{\epsilon} \big( \langle \Gamma,g^{-1}Xg \rangle
    + \kappa(\varpi,\Pi(g^{-1}Xg))
    \big)
    = \dot{\epsilon} \hspace{0.4mm} \langle \widetilde{Q},X \rangle
    = \dot{\epsilon} \hspace{0.4mm} \widetilde{H}_X \,.
\end{equation}
The coefficient of $\dot{\epsilon}$ is again the lifted Hamiltonian function $\widetilde{H}_X$ which confirms that $\widetilde{Q}$ is the correct deformed moment map for the enlarged phase space.

We can derive the Poisson algebra of these new Hamiltonian functions using the twisted symplectic form $\Omega_\Gamma$ in \eqref{eq:Omega_Gamma}, in analogy with \eqref{eq:HX_HY_bracket}.
The Poisson bracket is
\begin{equation}
    \{\widetilde{H}_X,\widetilde{H}_Y\}_\Omega
    := - \Omega_\Gamma(\widetilde{V}_X,\widetilde{V}_Y)
    = - \rho^*\omega_{\mathrm{KKS}}(\widetilde{V}_X,\widetilde{V}_Y)
    - (\rd p_a \wedge \rd \xi^a) (\widetilde{V}_X,\widetilde{V}_Y) \,.
    \label{eq:Htilde_bracket_split}
\end{equation}
Since $\rho_*\widetilde{V}_X=V_X$ one has
\begin{equation}
    \rho^*\omega_{\mathrm{KKS}}(\widetilde{V}_X,\widetilde{V}_Y)
    = \omega_{\mathrm{KKS}}(V_X,V_Y)
    = - \langle Q,[X,Y] \rangle
    = - H_{[X,Y]} \,.
    \label{eq:pullback_KKS_term}
\end{equation}
For the second term, use $\rd\xi^a(\widetilde{V}_X)=V_X^a$ and $\rd p_a(\widetilde{V}_X)=-p_b\,\partial_aV_X^b$ to compute
\begin{equation}
    (\rd p_a \wedge \rd \xi^a) (\widetilde{V}_X,\widetilde{V}_Y)
    = p_b \big( V_X^a \partial_a V_Y^b - V_Y^a \partial_a V_X^b \big)
    = p_b [V_X,V_Y]^b \,.
    \label{eq:canonical_term_bracket}
\end{equation}
The fundamental vector fields of the left $E_8$ action obey $[V_X,V_Y]=-V_{[X,Y]}$ so it follows that
\begin{equation}
    (\rd p_a \wedge \rd \xi^a) (\widetilde{V}_X,\widetilde{V}_Y)
    = -p_a V_{[X,Y]}^a \,.
    \label{eq:canonical_term_final}
\end{equation}

Substituting the objects above into \eqref{eq:Htilde_bracket_split} gives that the Poisson brackets are
\begin{equation}
    \{\widetilde H_X,\widetilde H_Y\}_\Omega
    = H_{[X,Y]} + p_a V_{[X,Y]}^a
    = \widetilde{H}_{[X,Y]} \,.
\end{equation}
Finally, with the basis $t^M$ of $\mathfrak{e}_8$ and defining the components
\begin{equation}
    \widetilde{Q}^M := \langle \widetilde{Q},t^M \rangle = \widetilde{H}_{t^M} \,,
\end{equation}
we obtain that the Poisson brackets realise the Lie algebra on the twisted cotangent bundle:
\begin{equation}
    \{\widetilde{Q}^M,\widetilde{Q}^N\}_{\Omega}
    = f^{MN}{}_P\,\widetilde{Q}^P \,.
    \label{eq:Lie_Q_tilde}
\end{equation}
In the $\ell\to0$ limit, this reduces to \eqref{eq:Lie_Q} in the undeformed model.

\subsubsection*{Hamiltonian equations and constraints}

Lastly, we shall present the Hamiltonian and constraints.
The canonical Hamiltonian is given by
\begin{equation}
    \widetilde{H}_{E_8}
    = \frac{e}{2} \big( g^{\mu\nu} \widetilde{\pi}_\mu \widetilde{\pi}_\nu + \cM^{MN} P_M P_N + \ell^{-1} \kappa(\varpi,\varpi) \big)
    - P_M V^M - \widetilde{Q}^M W_M \,,
    \label{eq:H_canonical_deformed}
\end{equation}
and we obtain the deformed Hamiltonian constraint
\begin{equation}
    \widetilde{\cH} := \frac12 \big( g^{\mu\nu} \widetilde{\pi}_\mu \widetilde{\pi}_\nu + \cM^{MN} P_M P_N +\ell^{-1} \kappa(\varpi,\varpi) \big) \approx 0 \,.
    \label{eq:mass_shell_constraint_ham_deformed}
\end{equation}
Once again, preserving $\pi_M^{(V)}\approx0$ and $\pi^M_{(W)}\approx0$ in time leads to the usual constraints on $P_M$ and $\widetilde{Q}^M$ to lie in the physical and unphysical directions, respectively.

In the same way that we computed \eqref{eq:Wong_equation_final} in Section~\ref{subsec:E8_Hamiltonian}, one finds that the Poisson brackets lead to the Wong equation
\begin{equation}
    \,\dot{\!\widetilde{Q}}{}^M + f^{MN}{}_P \, \widetilde{B}_N \widetilde{Q}^P = 0 \,.
    \label{eq:Wong_equation_final_deformed}
\end{equation}
This is the Euler--Lagrange equation \eqref{eq:EOM_g_Q_tilde} of $g$ in the deformed model.
The evolution equations \eqref{eq:pi_evolution} and \eqref{eq:P_evolution} for $\pi_\mu$ and $P_M$, respectively, are almost unchanged except that $Q$ becomes $\widetilde{Q}$.


\begin{thebibliography}{100}

\bibitem{Hull:2009mi}
C.~Hull and B.~Zwiebach, \emph{{Double Field Theory}}, \href{https://doi.org/10.1088/1126-6708/2009/09/099}{\emph{JHEP} {\bfseries 09} (2009) 099} [\href{https://arxiv.org/abs/0904.4664}{{\ttfamily 0904.4664}}].

\bibitem{Hohm:2010pp}
O.~Hohm, C.~Hull and B.~Zwiebach, \emph{{Generalized metric formulation of double field theory}}, \href{https://doi.org/10.1007/JHEP08(2010)008}{\emph{JHEP} {\bfseries 08} (2010) 008} [\href{https://arxiv.org/abs/1006.4823}{{\ttfamily 1006.4823}}].

\bibitem{Aldazabal:2013sca}
G.~Aldazabal, D.~Marqu{\'{e}}s and C.~Nu{\~{n}}ez, \emph{{Double Field Theory: A Pedagogical Review}}, \href{https://doi.org/10.1088/0264-9381/30/16/163001}{\emph{Class. Quant. Grav.} {\bfseries 30} (2013) 163001} [\href{https://arxiv.org/abs/1305.1907}{{\ttfamily 1305.1907}}].

\bibitem{Hohm:2013pua}
O.~Hohm and H.~Samtleben, \emph{{Exceptional Form of D=11 Supergravity}}, \href{https://doi.org/10.1103/PhysRevLett.111.231601}{\emph{Phys. Rev. Lett.} {\bfseries 111} (2013) 231601} [\href{https://arxiv.org/abs/1308.1673}{{\ttfamily 1308.1673}}].

\bibitem{Hohm:2013vpa}
O.~Hohm and H.~Samtleben, \emph{{Exceptional Field Theory I: $E_{6(6)}$ covariant Form of M-Theory and Type IIB}}, \href{https://doi.org/10.1103/PhysRevD.89.066016}{\emph{Phys. Rev. D} {\bfseries 89} (2014) 066016} [\href{https://arxiv.org/abs/1312.0614}{{\ttfamily 1312.0614}}].

\bibitem{Hohm:2013uia}
O.~Hohm and H.~Samtleben, \emph{{Exceptional field theory. II. E$_{7(7)}$}}, \href{https://doi.org/10.1103/PhysRevD.89.066017}{\emph{Phys. Rev. D} {\bfseries 89} (2014) 066017} [\href{https://arxiv.org/abs/1312.4542}{{\ttfamily 1312.4542}}].

\bibitem{Hohm:2014fxa}
O.~Hohm and H.~Samtleben, \emph{{Exceptional field theory. III. E$_{8(8)}$}}, \href{https://doi.org/10.1103/PhysRevD.90.066002}{\emph{Phys. Rev. D} {\bfseries 90} (2014) 066002} [\href{https://arxiv.org/abs/1406.3348}{{\ttfamily 1406.3348}}].

\bibitem{Berman:2020tqn}
D.S.~Berman and C.D.A.~Blair, \emph{{The Geometry, Branes and Applications of Exceptional Field Theory}}, \href{https://doi.org/10.1142/S0217751X20300148}{\emph{Int. J. Mod. Phys. A} {\bfseries 35} (2020) 2030014} [\href{https://arxiv.org/abs/2006.09777}{{\ttfamily 2006.09777}}].

\bibitem{Blair:2017gwn}
C.D.A.~Blair, \emph{{Particle actions and brane tensions from double and exceptional geometry}}, \href{https://doi.org/10.1007/JHEP10(2017)004}{\emph{JHEP} {\bfseries 10} (2017) 004} [\href{https://arxiv.org/abs/1707.07572}{{\ttfamily 1707.07572}}].

\bibitem{Arvanitakis:2017hwb}
A.S.~Arvanitakis and C.D.A.~Blair, \emph{{Unifying Type-II Strings by Exceptional Groups}}, \href{https://doi.org/10.1103/PhysRevLett.120.211601}{\emph{Phys. Rev. Lett.} {\bfseries 120} (2018) 211601} [\href{https://arxiv.org/abs/1712.07115}{{\ttfamily 1712.07115}}].

\bibitem{Arvanitakis:2018hfn}
A.S.~Arvanitakis and C.D.A.~Blair, \emph{{The Exceptional Sigma Model}}, \href{https://doi.org/10.1007/JHEP04(2018)064}{\emph{JHEP} {\bfseries 04} (2018) 064} [\href{https://arxiv.org/abs/1802.00442}{{\ttfamily 1802.00442}}].

\bibitem{deBoer:2012ma}
J.~de~Boer and M.~Shigemori, \emph{{Exotic Branes in String Theory}}, \href{https://doi.org/10.1016/j.physrep.2013.07.003}{\emph{Phys. Rept.} {\bfseries 532} (2013) 65} [\href{https://arxiv.org/abs/1209.6056}{{\ttfamily 1209.6056}}].

\bibitem{West:2003fc}
P.C.~West, \emph{{E(11), SL(32) and central charges}}, \href{https://doi.org/10.1016/j.physletb.2003.09.059}{\emph{Phys. Lett. B} {\bfseries 575} (2003) 333} [\href{https://arxiv.org/abs/hep-th/0307098}{{\ttfamily hep-th/0307098}}].

\bibitem{Hull:2007zu}
C.M.~Hull, \emph{{Generalised Geometry for M-Theory}}, \href{https://doi.org/10.1088/1126-6708/2007/07/079}{\emph{JHEP} {\bfseries 07} (2007) 079} [\href{https://arxiv.org/abs/hep-th/0701203}{{\ttfamily hep-th/0701203}}].

\bibitem{Hohm:2010jy}
O.~Hohm, C.~Hull and B.~Zwiebach, \emph{{Background independent action for double field theory}}, \href{https://doi.org/10.1007/JHEP07(2010)016}{\emph{JHEP} {\bfseries 07} (2010) 016} [\href{https://arxiv.org/abs/1003.5027}{{\ttfamily 1003.5027}}].

\bibitem{Coimbra:2011ky}
A.~Coimbra, C.~Strickland-Constable and D.~Waldram, \emph{{$E_{d(d)} \times \mathbb{R}^+$ generalised geometry, connections and M theory}}, \href{https://doi.org/10.1007/JHEP02(2014)054}{\emph{JHEP} {\bfseries 02} (2014) 054} [\href{https://arxiv.org/abs/1112.3989}{{\ttfamily 1112.3989}}].

\bibitem{Coimbra:2011nw}
A.~Coimbra, C.~Strickland-Constable and D.~Waldram, \emph{{Supergravity as Generalised Geometry I: Type II Theories}}, \href{https://doi.org/10.1007/JHEP11(2011)091}{\emph{JHEP} {\bfseries 11} (2011) 091} [\href{https://arxiv.org/abs/1107.1733}{{\ttfamily 1107.1733}}].

\bibitem{Coimbra:2012af}
A.~Coimbra, C.~Strickland-Constable and D.~Waldram, \emph{{Supergravity as Generalised Geometry II: $E_{d(d)} \times \mathbb{R}^+$ and M theory}}, \href{https://doi.org/10.1007/JHEP03(2014)019}{\emph{JHEP} {\bfseries 03} (2014) 019} [\href{https://arxiv.org/abs/1212.1586}{{\ttfamily 1212.1586}}].

\bibitem{Berkeley:2014nza}
J.~Berkeley, D.S.~Berman and F.J.~Rudolph, \emph{{Strings and Branes are Waves}}, \href{https://doi.org/10.1007/JHEP06(2014)006}{\emph{JHEP} {\bfseries 06} (2014) 006} [\href{https://arxiv.org/abs/1403.7198}{{\ttfamily 1403.7198}}].

\bibitem{Berman:2014jsa}
D.S.~Berman and F.J.~Rudolph, \emph{{Branes are Waves and Monopoles}}, \href{https://doi.org/10.1007/JHEP05(2015)015}{\emph{JHEP} {\bfseries 05} (2015) 015} [\href{https://arxiv.org/abs/1409.6314}{{\ttfamily 1409.6314}}].

\bibitem{Berman:2014hna}
D.S.~Berman and F.J.~Rudolph, \emph{{Strings, Branes and the Self-dual Solutions of Exceptional Field Theory}}, \href{https://doi.org/10.1007/JHEP05(2015)130}{\emph{JHEP} {\bfseries 05} (2015) 130} [\href{https://arxiv.org/abs/1412.2768}{{\ttfamily 1412.2768}}].

\bibitem{Bakhmatov:2016kfn}
I.~Bakhmatov, A.~Kleinschmidt and E.T.~Musaev, \emph{{Non-geometric branes are DFT monopoles}}, \href{https://doi.org/10.1007/JHEP10(2016)076}{\emph{JHEP} {\bfseries 10} (2016) 076} [\href{https://arxiv.org/abs/1607.05450}{{\ttfamily 1607.05450}}].

\bibitem{Bakhmatov:2017les}
I.~Bakhmatov, D.~Berman, A.~Kleinschmidt, E.~Musaev and R.~Otsuki, \emph{{Exotic branes in Exceptional Field Theory: the SL(5) duality group}}, \href{https://doi.org/10.1007/JHEP08(2018)021}{\emph{JHEP} {\bfseries 08} (2018) 021} [\href{https://arxiv.org/abs/1710.09740}{{\ttfamily 1710.09740}}].

\bibitem{Berman:2018okd}
D.S.~Berman, E.T.~Musaev and R.~Otsuki, \emph{{Exotic Branes in Exceptional Field Theory: $E_{7(7)}$ and Beyond}}, \href{https://doi.org/10.1007/JHEP12(2018)053}{\emph{JHEP} {\bfseries 12} (2018) 053} [\href{https://arxiv.org/abs/1806.00430}{{\ttfamily 1806.00430}}].

\bibitem{Berman:2019zcq}
D.S.~Berman, E.T.~Musaev and R.~Otsuki, \emph{{Exotic Branes in M-Theory}}, \href{https://doi.org/10.22323/1.347.0138}{\emph{PoS} {\bfseries CORFU2018} (2019) 138} [\href{https://arxiv.org/abs/1903.10247}{{\ttfamily 1903.10247}}].

\bibitem{Bergshoeff:1997gy}
E.~Bergshoeff, B.~Janssen and T.~Ortin, \emph{{Kaluza-Klein monopoles and gauged sigma models}}, \href{https://doi.org/10.1016/S0370-2693(97)00946-5}{\emph{Phys. Lett. B} {\bfseries 410} (1997) 131} [\href{https://arxiv.org/abs/hep-th/9706117}{{\ttfamily hep-th/9706117}}].

\bibitem{Kimura:2014upa}
T.~Kimura, S.~Sasaki and M.~Yata, \emph{{World-volume Effective Actions of Exotic Five-branes}}, \href{https://doi.org/10.1007/JHEP07(2014)127}{\emph{JHEP} {\bfseries 07} (2014) 127} [\href{https://arxiv.org/abs/1404.5442}{{\ttfamily 1404.5442}}].

\bibitem{Kimura:2016anf}
T.~Kimura, S.~Sasaki and M.~Yata, \emph{{World-volume Effective Action of Exotic Five-brane in M-theory}}, \href{https://doi.org/10.1007/JHEP02(2016)168}{\emph{JHEP} {\bfseries 02} (2016) 168} [\href{https://arxiv.org/abs/1601.05589}{{\ttfamily 1601.05589}}].

\bibitem{Tseytlin:1990nb}
A.A.~Tseytlin, \emph{{Duality Symmetric Formulation of String World Sheet Dynamics}}, \href{https://doi.org/10.1016/0370-2693(90)91454-J}{\emph{Phys. Lett.} {\bfseries B242} (1990) 163}.

\bibitem{Siegel:1993bj}
W.~Siegel, \emph{{Manifest duality in low-energy superstrings}},  in \emph{{International Conference on Strings 93 Berkeley, California, May 24-29, 1993}}, pp.~353--363, 1993 [\href{https://arxiv.org/abs/hep-th/9308133}{{\ttfamily hep-th/9308133}}].

\bibitem{Alekseev:2004np}
A.~Alekseev and T.~Strobl, \emph{{Current algebras and differential geometry}}, \href{https://doi.org/10.1088/1126-6708/2005/03/035}{\emph{JHEP} {\bfseries 03} (2005) 035} [\href{https://arxiv.org/abs/hep-th/0410183}{{\ttfamily hep-th/0410183}}].

\bibitem{Hatsuda:2012uk}
M.~Hatsuda and T.~Kimura, \emph{{Canonical approach to Courant brackets for D-branes}}, \href{https://doi.org/10.1007/JHEP06(2012)034}{\emph{JHEP} {\bfseries 06} (2012) 034} [\href{https://arxiv.org/abs/1203.5499}{{\ttfamily 1203.5499}}].

\bibitem{Hatsuda:2012vm}
M.~Hatsuda and K.~Kamimura, \emph{{SL(5) duality from canonical M2-brane}}, \href{https://doi.org/10.1007/JHEP11(2012)001}{\emph{JHEP} {\bfseries 11} (2012) 001} [\href{https://arxiv.org/abs/1208.1232}{{\ttfamily 1208.1232}}].

\bibitem{Hatsuda:2013dya}
M.~Hatsuda and K.~Kamimura, \emph{{M5 algebra and SO(5,5) duality}}, \href{https://doi.org/10.1007/JHEP06(2013)095}{\emph{JHEP} {\bfseries 06} (2013) 095} [\href{https://arxiv.org/abs/1305.2258}{{\ttfamily 1305.2258}}].

\bibitem{Osten:2021fil}
D.~Osten, \emph{{Currents, charges and algebras in exceptional generalised geometry}}, \href{https://doi.org/10.1007/JHEP06(2021)070}{\emph{JHEP} {\bfseries 06} (2021) 070} [\href{https://arxiv.org/abs/2103.03267}{{\ttfamily 2103.03267}}].

\bibitem{Osten:2024mjt}
D.~Osten, \emph{{On the universal exceptional structure of world-volume theories in string and M-theory}}, \href{https://doi.org/10.1016/j.physletb.2024.138814}{\emph{Phys. Lett. B} {\bfseries 855} (2024) 138814} [\href{https://arxiv.org/abs/2402.10269}{{\ttfamily 2402.10269}}].

\bibitem{Osten:2023iwc}
D.~Osten, \emph{{On exceptional QP-manifolds}}, \href{https://doi.org/10.1007/JHEP01(2024)028}{\emph{JHEP} {\bfseries 01} (2024) 028} [\href{https://arxiv.org/abs/2306.11093}{{\ttfamily 2306.11093}}].

\bibitem{Baguet:2016jph}
A.~Baguet and H.~Samtleben, \emph{{E$_{8(8)}$ Exceptional Field Theory: Geometry, Fermions and Supersymmetry}}, \href{https://doi.org/10.1007/JHEP09(2016)168}{\emph{JHEP} {\bfseries 09} (2016) 168} [\href{https://arxiv.org/abs/1607.03119}{{\ttfamily 1607.03119}}].

\bibitem{Bossard:2018utw}
G.~Bossard, F.~Ciceri, G.~Inverso, A.~Kleinschmidt and H.~Samtleben, \emph{{E$_{9}$ exceptional field theory. Part I. The potential}}, \href{https://doi.org/10.1007/JHEP03(2019)089}{\emph{JHEP} {\bfseries 03} (2019) 089} [\href{https://arxiv.org/abs/1811.04088}{{\ttfamily 1811.04088}}].

\bibitem{Bossard:2021jix}
G.~Bossard, F.~Ciceri, G.~Inverso, A.~Kleinschmidt and H.~Samtleben, \emph{{E$_{9}$ exceptional field theory. Part II. The complete dynamics}}, \href{https://doi.org/10.1007/JHEP05(2021)107}{\emph{JHEP} {\bfseries 05} (2021) 107} [\href{https://arxiv.org/abs/2103.12118}{{\ttfamily 2103.12118}}].

\bibitem{Bossard:2019ksx}
G.~Bossard, A.~Kleinschmidt and E.~Sezgin, \emph{{On supersymmetric E$_{11}$ exceptional field theory}}, \href{https://doi.org/10.1007/JHEP10(2019)165}{\emph{JHEP} {\bfseries 10} (2019) 165} [\href{https://arxiv.org/abs/1907.02080}{{\ttfamily 1907.02080}}].

\bibitem{Bossard:2021ebg}
G.~Bossard, A.~Kleinschmidt and E.~Sezgin, \emph{{A master exceptional field theory}}, \href{https://doi.org/10.1007/JHEP06(2021)185}{\emph{JHEP} {\bfseries 06} (2021) 185} [\href{https://arxiv.org/abs/2103.13411}{{\ttfamily 2103.13411}}].

\bibitem{Boulanger:2008nd}
N.~Boulanger and O.~Hohm, \emph{{Non-linear parent action and dual gravity}}, \href{https://doi.org/10.1103/PhysRevD.78.064027}{\emph{Phys. Rev. D} {\bfseries 78} (2008) 064027} [\href{https://arxiv.org/abs/0806.2775}{{\ttfamily 0806.2775}}].

\bibitem{Hohm:2018qhd}
O.~Hohm and H.~Samtleben, \emph{{The dual graviton in duality covariant theories}}, \href{https://doi.org/10.1002/prop.201900021}{\emph{Fortsch. Phys.} {\bfseries 67} (2019) 1900021} [\href{https://arxiv.org/abs/1807.07150}{{\ttfamily 1807.07150}}].

\bibitem{Berman:2012vc}
D.S.~Berman, M.~Cederwall, A.~Kleinschmidt and D.C.~Thompson, \emph{{The gauge structure of generalised diffeomorphisms}}, \href{https://doi.org/10.1007/JHEP01(2013)064}{\emph{JHEP} {\bfseries 01} (2013) 064} [\href{https://arxiv.org/abs/1208.5884}{{\ttfamily 1208.5884}}].

\bibitem{Cederwall:2015ica}
M.~Cederwall and J.A.~Rosabal, \emph{{E$_{8}$ geometry}}, \href{https://doi.org/10.1007/JHEP07(2015)007}{\emph{JHEP} {\bfseries 07} (2015) 007} [\href{https://arxiv.org/abs/1504.04843}{{\ttfamily 1504.04843}}].

\bibitem{Cederwall:2017fjm}
M.~Cederwall and J.~Palmkvist, \emph{{Extended geometries}}, \href{https://doi.org/10.1007/JHEP02(2018)071}{\emph{JHEP} {\bfseries 02} (2018) 071} [\href{https://arxiv.org/abs/1711.07694}{{\ttfamily 1711.07694}}].

\bibitem{Bossard:2017aae}
G.~Bossard, M.~Cederwall, A.~Kleinschmidt, J.~Palmkvist and H.~Samtleben, \emph{{Generalized diffeomorphisms for $E_9$}}, \href{https://doi.org/10.1103/PhysRevD.96.106022}{\emph{Phys. Rev. D} {\bfseries 96} (2017) 106022} [\href{https://arxiv.org/abs/1708.08936}{{\ttfamily 1708.08936}}].

\bibitem{Berman:2019efr}
D.S.~Berman and R.~Otsuki, \emph{{Reductions of Exceptional Field Theories}}, \href{https://doi.org/10.1007/JHEP03(2020)066}{\emph{JHEP} {\bfseries 03} (2020) 066} [\href{https://arxiv.org/abs/1911.06150}{{\ttfamily 1911.06150}}].

\bibitem{Abzalov:2015ega}
A.~Abzalov, I.~Bakhmatov and E.T.~Musaev, \emph{{Exceptional field theory: $SO(5,5)$}}, \href{https://doi.org/10.1007/JHEP06(2015)088}{\emph{JHEP} {\bfseries 06} (2015) 088} [\href{https://arxiv.org/abs/1504.01523}{{\ttfamily 1504.01523}}].

\bibitem{Musaev:2015ces}
E.T.~Musaev, \emph{{Exceptional field theory: $SL(5)$}}, \href{https://doi.org/10.1007/JHEP02(2016)012}{\emph{JHEP} {\bfseries 02} (2016) 012} [\href{https://arxiv.org/abs/1512.02163}{{\ttfamily 1512.02163}}].

\bibitem{Hohm:2015xna}
O.~Hohm and Y.-N.~Wang, \emph{{Tensor hierarchy and generalized Cartan calculus in SL(3) \texttimes{} SL(2) exceptional field theory}}, \href{https://doi.org/10.1007/JHEP04(2015)050}{\emph{JHEP} {\bfseries 04} (2015) 050} [\href{https://arxiv.org/abs/1501.01600}{{\ttfamily 1501.01600}}].

\bibitem{Wang:2015hca}
Y.-N.~Wang, \emph{{Generalized Cartan Calculus in general dimension}}, \href{https://doi.org/10.1007/JHEP07(2015)114}{\emph{JHEP} {\bfseries 07} (2015) 114} [\href{https://arxiv.org/abs/1504.04780}{{\ttfamily 1504.04780}}].

\bibitem{Palmkvist:2013vya}
J.~Palmkvist, \emph{{The tensor hierarchy algebra}}, \href{https://doi.org/10.1063/1.4858335}{\emph{J. Math. Phys.} {\bfseries 55} (2014) 011701} [\href{https://arxiv.org/abs/1305.0018}{{\ttfamily 1305.0018}}].

\bibitem{Cederwall:2018aab}
M.~Cederwall and J.~Palmkvist, \emph{{$L_{\infty }$ Algebras for Extended Geometry from Borcherds Superalgebras}}, \href{https://doi.org/10.1007/s00220-019-03451-2}{\emph{Commun. Math. Phys.} {\bfseries 369} (2019) 721} [\href{https://arxiv.org/abs/1804.04377}{{\ttfamily 1804.04377}}].

\bibitem{Cederwall:2019qnw}
M.~Cederwall and J.~Palmkvist, \emph{{Tensor hierarchy algebras and extended geometry. Part I. Construction of the algebra}}, \href{https://doi.org/10.1007/JHEP02(2020)144}{\emph{JHEP} {\bfseries 02} (2020) 144} [\href{https://arxiv.org/abs/1908.08695}{{\ttfamily 1908.08695}}].

\bibitem{Cederwall:2019bai}
M.~Cederwall and J.~Palmkvist, \emph{{Tensor hierarchy algebras and extended geometry. Part II. Gauge structure and dynamics}}, \href{https://doi.org/10.1007/JHEP02(2020)145}{\emph{JHEP} {\bfseries 02} (2020) 145} [\href{https://arxiv.org/abs/1908.08696}{{\ttfamily 1908.08696}}].

\bibitem{Bonezzi:2019ygf}
R.~Bonezzi and O.~Hohm, \emph{{Leibniz Gauge Theories and Infinity Structures}}, \href{https://doi.org/10.1007/s00220-020-03785-2}{\emph{Commun. Math. Phys.} {\bfseries 377} (2020) 2027} [\href{https://arxiv.org/abs/1904.11036}{{\ttfamily 1904.11036}}].

\bibitem{Lavau:2019oja}
S.~Lavau and J.~Palmkvist, \emph{{Infinity-enhancing of Leibniz algebras}}, \href{https://doi.org/10.1007/s11005-020-01324-7}{\emph{Lett. Math. Phys.} {\bfseries 110} (2020) 3121} [\href{https://arxiv.org/abs/1907.05752}{{\ttfamily 1907.05752}}].

\bibitem{Bonezzi:2019bek}
R.~Bonezzi and O.~Hohm, \emph{{Duality Hierarchies and Differential Graded Lie Algebras}},  \href{https://arxiv.org/abs/1910.10399}{{\ttfamily 1910.10399}}.

\bibitem{Arvanitakis:2018cyo}
A.S.~Arvanitakis, \emph{{Brane Wess-Zumino terms from AKSZ and exceptional generalised geometry as an $L_\infty$-algebroid}}, \href{https://doi.org/10.4310/ATMP.2019.v23.n5.a1}{\emph{Adv. Theor. Math. Phys.} {\bfseries 23} (2019) 1159} [\href{https://arxiv.org/abs/1804.07303}{{\ttfamily 1804.07303}}].

\bibitem{Park:2013mpa}
J.-H.~Park, \emph{{Comments on double field theory and diffeomorphisms}}, \href{https://doi.org/10.1007/JHEP06(2013)098}{\emph{JHEP} {\bfseries 06} (2013) 098} [\href{https://arxiv.org/abs/1304.5946}{{\ttfamily 1304.5946}}].

\bibitem{Lee:2013hma}
K.~Lee and J.-H.~Park, \emph{{Covariant action for a string in ''doubled yet gauged'' spacetime}}, \href{https://doi.org/10.1016/j.nuclphysb.2014.01.003}{\emph{Nucl. Phys. B} {\bfseries 880} (2014) 134} [\href{https://arxiv.org/abs/1307.8377}{{\ttfamily 1307.8377}}].

\bibitem{PiresPacheco:2008qik}
P.~Pires~Pacheco and D.~Waldram, \emph{{M-theory, exceptional generalised geometry and superpotentials}}, \href{https://doi.org/10.1088/1126-6708/2008/09/123}{\emph{JHEP} {\bfseries 09} (2008) 123} [\href{https://arxiv.org/abs/0804.1362}{{\ttfamily 0804.1362}}].

\bibitem{Bugden:2021wxg}
M.~Bugden, O.~Hulik, F.~Valach and D.~Waldram, \emph{{G-Algebroids: A Unified Framework for Exceptional and Generalised Geometry, and Poisson{\textendash}Lie Duality}}, \href{https://doi.org/10.1002/prop.202100028}{\emph{Fortsch. Phys.} {\bfseries 69} (2021) 2100028} [\href{https://arxiv.org/abs/2103.01139}{{\ttfamily 2103.01139}}].

\bibitem{Hulik:2023aks}
O.~Hulik, E.~Malek, F.~Valach and D.~Waldram, \emph{{Y-algebroids and $E_{7(7)} \times \mathbb{R}^+$-generalised geometry}},  \href{https://arxiv.org/abs/2308.01130}{{\ttfamily 2308.01130}}.

\bibitem{Grana:2008yw}
M.~Gra{\~{n}}a, R.~Minasian, M.~Petrini and D.~Waldram, \emph{{T-duality, Generalized Geometry and Non-Geometric Backgrounds}}, \href{https://doi.org/10.1088/1126-6708/2009/04/075}{\emph{JHEP} {\bfseries 04} (2009) 075} [\href{https://arxiv.org/abs/0807.4527}{{\ttfamily 0807.4527}}].

\bibitem{Coimbra:2016ydd}
A.~Coimbra and C.~Strickland-Constable, \emph{{Supersymmetric Backgrounds, the Killing Superalgebra, and Generalised Special Holonomy}}, \href{https://doi.org/10.1007/JHEP11(2016)063}{\emph{JHEP} {\bfseries 11} (2016) 063} [\href{https://arxiv.org/abs/1606.09304}{{\ttfamily 1606.09304}}].

\bibitem{Kimura:2016xzd}
T.~Kimura, \emph{{Supersymmetry projection rules on exotic branes}}, \href{https://doi.org/10.1093/ptep/ptw052}{\emph{PTEP} {\bfseries 2016} (2016) 053B05} [\href{https://arxiv.org/abs/1601.02175}{{\ttfamily 1601.02175}}].

\bibitem{Hull:1984vh}
C.M.~Hull, \emph{{Exact PP-Wave Solutions of 11-Dimensional Supergravity}}, \href{https://doi.org/10.1016/0370-2693(84)90030-3}{\emph{Phys. Lett. B} {\bfseries 139} (1984) 39}.

\bibitem{Chatzistavrakidis:2013jqa}
A.~Chatzistavrakidis, F.F.~Gautason, G.~Moutsopoulos and M.~Zagermann, \emph{{Effective actions of nongeometric five-branes}}, \href{https://doi.org/10.1103/PhysRevD.89.066004}{\emph{Phys. Rev.} {\bfseries D89} (2014) 066004} [\href{https://arxiv.org/abs/1309.2653}{{\ttfamily 1309.2653}}].

\bibitem{Chatzistavrakidis:2014sua}
A.~Chatzistavrakidis and F.F.~Gautason, \emph{{U-dual branes and mixed symmetry tensor fields}}, \href{https://doi.org/10.1002/prop.201400023}{\emph{Fortsch. Phys.} {\bfseries 62} (2014) 743} [\href{https://arxiv.org/abs/1404.7635}{{\ttfamily 1404.7635}}].

\bibitem{Bergshoeff:1998ef}
E.~Bergshoeff, E.~Eyras and Y.~Lozano, \emph{{The Massive Kaluza-Klein monopole}}, \href{https://doi.org/10.1016/S0370-2693(98)00501-2}{\emph{Phys. Lett. B} {\bfseries 430} (1998) 77} [\href{https://arxiv.org/abs/hep-th/9802199}{{\ttfamily hep-th/9802199}}].

\bibitem{Duff:1989tf}
M.J.~Duff, \emph{{Duality Rotations in String Theory}}, \href{https://doi.org/10.1016/0550-3213(90)90520-N}{\emph{Nucl. Phys.} {\bfseries B335} (1990) 610}.

\bibitem{Tseytlin:1990va}
A.A.~Tseytlin, \emph{{Duality symmetric closed string theory and interacting chiral scalars}}, \href{https://doi.org/10.1016/0550-3213(91)90266-Z}{\emph{Nucl. Phys.} {\bfseries B350} (1991) 395}.

\bibitem{Siegel:1993th}
W.~Siegel, \emph{{Superspace duality in low-energy superstrings}}, \href{https://doi.org/10.1103/PhysRevD.48.2826}{\emph{Phys. Rev.} {\bfseries D48} (1993) 2826} [\href{https://arxiv.org/abs/hep-th/9305073}{{\ttfamily hep-th/9305073}}].

\bibitem{Klimcik:1995ux}
C.~Klim{\v{c}}{\'{i}}k and P.~{\v{S}}evera, \emph{{Dual nonAbelian duality and the Drinfeld double}}, \href{https://doi.org/10.1016/0370-2693(95)00451-P}{\emph{Phys. Lett.} {\bfseries B351} (1995) 455} [\href{https://arxiv.org/abs/hep-th/9502122}{{\ttfamily hep-th/9502122}}].

\bibitem{Sfetsos:1999cc}
K.~Sfetsos, \emph{{Duality-invariant class of two-dimensional field theories}}, \href{https://doi.org/10.1016/S0550-3213(99)00485-X}{\emph{Nuclear Physics B} {\bfseries 561} (1999) 316} [\href{https://arxiv.org/abs/9904188}{{\ttfamily 9904188}}].

\bibitem{Demulder:2018lmj}
S.~Demulder, F.~Hassler and D.C.~Thompson, \emph{{Doubled aspects of generalised dualities and integrable deformations}}, \href{https://doi.org/10.1007/JHEP02(2019)189}{\emph{JHEP} {\bfseries 02} (2019) 189} [\href{https://arxiv.org/abs/1810.11446}{{\ttfamily 1810.11446}}].

\bibitem{Hassler:2020xyj}
F.~Hassler, \emph{{RG flow of integrable $\mathcal{E}$-models}}, \href{https://doi.org/10.1016/j.physletb.2021.136367}{\emph{Phys. Lett. B} {\bfseries 818} (2021) 136367} [\href{https://arxiv.org/abs/2012.10451}{{\ttfamily 2012.10451}}].

\bibitem{Borsato:2021gma}
R.~Borsato and S.~Driezen, \emph{{Supergravity solution-generating techniques and canonical transformations of $\sigma$-models from $O(D,D)$}}, \href{https://doi.org/10.1007/JHEP05(2021)180}{\emph{JHEP} {\bfseries 05} (2021) 180} [\href{https://arxiv.org/abs/2102.04498}{{\ttfamily 2102.04498}}].

\bibitem{Borsato:2021vfy}
R.~Borsato, S.~Driezen and F.~Hassler, \emph{{An algebraic classification of solution generating techniques}}, \href{https://doi.org/10.1016/j.physletb.2021.136771}{\emph{Phys. Lett. B} {\bfseries 823} (2021) 136771} [\href{https://arxiv.org/abs/2109.06185}{{\ttfamily 2109.06185}}].

\bibitem{Blair:2014kla}
C.D.A.~Blair, \emph{{Non-commutativity and non-associativity of the doubled string in non-geometric backgrounds}}, \href{https://doi.org/10.1007/JHEP06(2015)091}{\emph{JHEP} {\bfseries 06} (2015) 091} [\href{https://arxiv.org/abs/1405.2283}{{\ttfamily 1405.2283}}].

\bibitem{Osten:2019ayq}
D.~Osten, \emph{{Current algebras, generalised fluxes and non-geometry}}, \href{https://doi.org/10.1088/1751-8121/ab8f3d}{\emph{J. Phys. A} {\bfseries 53} (2020) 265402} [\href{https://arxiv.org/abs/1910.00029}{{\ttfamily 1910.00029}}].

\bibitem{Hassler:2024hgq}
F.~Hassler, O.~Hulik and D.~Osten, \emph{{Current algebra and generalized Cartan geometry}}, \href{https://doi.org/10.1103/PhysRevD.110.126022}{\emph{Phys. Rev. D} {\bfseries 110} (2024) 126022} [\href{https://arxiv.org/abs/2409.00176}{{\ttfamily 2409.00176}}].

\bibitem{Hassler:2025rag}
F.~{Hassler}, D.~{Osten} and A.~{Swash}, \emph{{Gauged extended field theory and generalized Cartan geometry}}, \href{https://doi.org/10.1103/pzq2-1rhv}{\emph{Phys. Rev. D} {\bfseries 113} (2026) 066017} [\href{https://arxiv.org/abs/2509.04595}{{\ttfamily 2509.04595}}].

\bibitem{Hatsuda:2022zpi}
M.~Hatsuda, H.~Mori, S.~Sasaki and M.~Yata, \emph{{Gauged double field theory, current algebras and heterotic sigma models}}, \href{https://doi.org/10.1007/JHEP05(2023)220}{\emph{JHEP} {\bfseries 05} (2023) 220} [\href{https://arxiv.org/abs/2212.06476}{{\ttfamily 2212.06476}}].

\bibitem{Osten:2023cza}
D.~Osten, \emph{{A heterotic integrable deformation of the principal chiral model}},  \href{https://arxiv.org/abs/2312.10149}{{\ttfamily 2312.10149}}.

\bibitem{Duff:1990hn}
M.J.~Duff and J.X.~Lu, \emph{{Duality Rotations in Membrane Theory}}, \href{https://doi.org/10.1016/0550-3213(90)90565-U}{\emph{Nucl. Phys.} {\bfseries B347} (1990) 394}.

\bibitem{Duff:2015jka}
M.J.~Duff, J.X.~Lu, R.~Percacci, C.N.~Pope, H.~Samtleben and E.~Sezgin, \emph{{Membrane Duality Revisited}}, \href{https://doi.org/10.1016/j.nuclphysb.2015.10.003}{\emph{Nucl. Phys.} {\bfseries B901} (2015) 1} [\href{https://arxiv.org/abs/1509.02915}{{\ttfamily 1509.02915}}].

\bibitem{Strickland-Constable:2021afa}
C.~Strickland-Constable, \emph{{Classical worldvolumes as generalised geodesics}},  \href{https://arxiv.org/abs/2102.00555}{{\ttfamily 2102.00555}}.

\bibitem{Arvanitakis:2021wkt}
A.S.~Arvanitakis, \emph{{Brane current algebras and generalised geometry from QP manifolds. Or, \textquotedblleft{}when they go high, we go low\textquotedblright{}}}, \href{https://doi.org/10.1007/JHEP11(2021)114}{\emph{JHEP} {\bfseries 11} (2021) 114} [\href{https://arxiv.org/abs/2103.08608}{{\ttfamily 2103.08608}}].

\bibitem{Arvanitakis:2022fvv}
A.S.~Arvanitakis, E.~Malek and D.~Tennyson, \emph{{Romans Massive QP Manifolds}}, \href{https://doi.org/10.3390/universe8030147}{\emph{Universe} {\bfseries 8} (2022) 147} [\href{https://arxiv.org/abs/2201.07807}{{\ttfamily 2201.07807}}].

\bibitem{Sakatani:2016sko}
Y.~Sakatani and S.~Uehara, \emph{{Branes in Extended Spacetime: Brane Worldvolume Theory Based on Duality Symmetry}}, \href{https://doi.org/10.1103/PhysRevLett.117.191601}{\emph{Phys. Rev. Lett.} {\bfseries 117} (2016) 191601} [\href{https://arxiv.org/abs/1607.04265}{{\ttfamily 1607.04265}}].

\bibitem{Sakatani:2017vbd}
Y.~Sakatani and S.~Uehara, \emph{{Exceptional M-brane sigma models and $\eta$-symbols}}, \href{https://doi.org/10.1093/ptep/pty021}{\emph{PTEP} {\bfseries 2018} (2018) 033B05} [\href{https://arxiv.org/abs/1712.10316}{{\ttfamily 1712.10316}}].

\bibitem{Sakatani:2020umt}
Y.~Sakatani and S.~Uehara, \emph{{Born sigma model for branes in exceptional geometry}}, \href{https://doi.org/10.1093/ptep/ptaa081}{\emph{PTEP} {\bfseries 2020} (2020) 073B05} [\href{https://arxiv.org/abs/2004.09486}{{\ttfamily 2004.09486}}].

\bibitem{Hatsuda:2023dwx}
M.~Hatsuda, O.~Hul\'\i{}k, W.D.~Linch, W.D.~Siegel, D.~Wang and Y.-P.~Wang, \emph{{$ \mathcal{A} $-theory \textemdash{} A brane world-volume theory with manifest U-duality}}, \href{https://doi.org/10.1007/JHEP10(2023)087}{\emph{JHEP} {\bfseries 10} (2023) 087} [\href{https://arxiv.org/abs/2307.04934}{{\ttfamily 2307.04934}}].

\bibitem{Arvanitakis:2021lwo}
A.S.~Arvanitakis, C.D.A.~Blair and D.C.~Thompson, \emph{{A QP perspective on topology change in Poisson\textendash{}Lie T-duality}}, \href{https://doi.org/10.1088/1751-8121/acd503}{\emph{J. Phys. A} {\bfseries 56} (2023) 255205} [\href{https://arxiv.org/abs/2110.08179}{{\ttfamily 2110.08179}}].

\bibitem{Arvanitakis:2023dud}
A.S.~Arvanitakis and D.~Tennyson, \emph{{Brane wrapping, Alexandrov-Kontsevich-Schwarz-Zaboronsky sigma models, and QP manifolds}}, \href{https://doi.org/10.1103/PhysRevD.108.086024}{\emph{Phys. Rev. D} {\bfseries 108} (2023) 086024} [\href{https://arxiv.org/abs/2301.02670}{{\ttfamily 2301.02670}}].

\bibitem{Fernandez-Melgarejo:2019mgd}
J.J.~Fern{\'a}ndez-Melgarejo, Y.~Sakatani and S.~Uehara, \emph{{Exotic branes and mixed-symmetry potentials I: Predictions from $E_{11}$ symmetry}}, \href{https://doi.org/10.1093/ptep/ptaa021}{\emph{PTEP} {\bfseries 2020} (2020) 053B02} [\href{https://arxiv.org/abs/1907.07177}{{\ttfamily 1907.07177}}].

\bibitem{Brown:2004jb}
J.~Brown, O.J.~Ganor and C.~Helfgott, \emph{{M theory and E(10): Billiards, branes, and imaginary roots}}, \href{https://doi.org/10.1088/1126-6708/2004/08/063}{\emph{JHEP} {\bfseries 08} (2004) 063} [\href{https://arxiv.org/abs/hep-th/0401053}{{\ttfamily hep-th/0401053}}].

\bibitem{West:2004kb}
P.C.~West, \emph{{E(11) origin of brane charges and U-duality multiplets}}, \href{https://doi.org/10.1088/1126-6708/2004/08/052}{\emph{JHEP} {\bfseries 08} (2004) 052} [\href{https://arxiv.org/abs/hep-th/0406150}{{\ttfamily hep-th/0406150}}].

\bibitem{Englert:2007qb}
F.~Englert, L.~Houart, A.~Kleinschmidt, H.~Nicolai and N.~Tabti, \emph{{An E(9) multiplet of BPS states}}, \href{https://doi.org/10.1088/1126-6708/2007/05/065}{\emph{JHEP} {\bfseries 05} (2007) 065} [\href{https://arxiv.org/abs/hep-th/0703285}{{\ttfamily hep-th/0703285}}].

\bibitem{Cook:2008bi}
P.P.~Cook and P.C.~West, \emph{{Charge multiplets and masses for E(11)}}, \href{https://doi.org/10.1088/1126-6708/2008/11/091}{\emph{JHEP} {\bfseries 11} (2008) 091} [\href{https://arxiv.org/abs/0805.4451}{{\ttfamily 0805.4451}}].

\bibitem{Cook:2009ri}
P.P.~Cook, \emph{{Exotic E(11) branes as composite gravitational solutions}}, \href{https://doi.org/10.1088/0264-9381/26/23/235023}{\emph{Class. Quant. Grav.} {\bfseries 26} (2009) 235023} [\href{https://arxiv.org/abs/0908.0485}{{\ttfamily 0908.0485}}].

\bibitem{Houart:2009ya}
L.~Houart, A.~Kleinschmidt and J.~Lindman~Hornlund, \emph{{Some Algebraic Aspects of Half-BPS Bound States in M-Theory}}, \href{https://doi.org/10.1007/JHEP03(2010)022}{\emph{JHEP} {\bfseries 03} (2010) 022} [\href{https://arxiv.org/abs/0911.5141}{{\ttfamily 0911.5141}}].

\bibitem{Houart:2011sk}
L.~Houart, A.~Kleinschmidt and J.~Lindman~Hornlund, \emph{{An M-theory solution from null roots in $E_{11}$}}, \href{https://doi.org/10.1007/JHEP01(2011)154}{\emph{JHEP} {\bfseries 01} (2011) 154} [\href{https://arxiv.org/abs/1101.2816}{{\ttfamily 1101.2816}}].

\bibitem{West:2018lfn}
P.~West, \emph{{E$_{11}$, Brane Dynamics and Duality Symmetries}}, \href{https://doi.org/10.1142/S0217751X1850080X}{\emph{Int. J. Mod. Phys. A} {\bfseries 33} (2018) 1850080} [\href{https://arxiv.org/abs/1801.00669}{{\ttfamily 1801.00669}}].

\bibitem{Bossard:2017wxl}
G.~Bossard, A.~Kleinschmidt, J.~Palmkvist, C.N.~Pope and E.~Sezgin, \emph{{Beyond E$_{11}$}}, \href{https://doi.org/10.1007/JHEP05(2017)020}{\emph{JHEP} {\bfseries 05} (2017) 020} [\href{https://arxiv.org/abs/1703.01305}{{\ttfamily 1703.01305}}].

\bibitem{Kirillov:1962KKS}
A.A.~Kirillov, \emph{{Unitary representations of nilpotent Lie groups}}, \href{https://doi.org/10.1070/RM1962v017n04ABEH004118}{\emph{Russ. Math. Surveys} {\bfseries 17} (1962) 53}.

\bibitem{Kostant:1970KKS}
B.~Kostant, \emph{{Quantization and unitary representations}},  in \emph{{Lectures in Modern Analysis and Applications III}}, C.T.~Taam, ed., vol.~170 of \emph{Lecture Notes in Mathematics}, (Berlin and Heidelberg), pp.~87--208, Springer, 1970, \href{https://doi.org/10.1007/BFb0079068}{DOI}.

\bibitem{Souriau:1970KKS}
J.-M.~Souriau, \emph{{Structure des syst{\`e}mes dynamiques: Ma{\^\i}trises de math{\'e}matiques}}, Collection Dunod Universit{\'e}, Dunod, Paris (1970).

\bibitem{Wong:1970fu}
S.K.~Wong, \emph{{Field and particle equations for the classical Yang-Mills field and particles with isotopic spin}}, \href{https://doi.org/10.1007/BF02892134}{\emph{Nuovo Cim. A} {\bfseries 65} (1970) 689}.

\bibitem{Alekseev:2015hda}
A.~Alekseev, O.~Chekeres and P.~Mnev, \emph{{Wilson surface observables from equivariant cohomology}}, \href{https://doi.org/10.1007/JHEP11(2015)093}{\emph{JHEP} {\bfseries 11} (2015) 093} [\href{https://arxiv.org/abs/1507.06343}{{\ttfamily 1507.06343}}].

\bibitem{Alekseev:2018pbv}
A.~Alekseev and S.L.~Shatashvili, \emph{{Coadjoint Orbits, Cocycles and Gravitational Wess{\textendash}Zumino}},  \href{https://arxiv.org/abs/1801.07963}{{\ttfamily 1801.07963}}.

\bibitem{Curtright:1980yk}
T.~Curtright, \emph{{Generalised gauge fields}}, \href{https://doi.org/10.1016/0370-2693(85)91235-3}{\emph{Phys. Lett. B} {\bfseries 165} (1985) 304}.

\bibitem{Hull:2000zn}
C.M.~Hull, \emph{{Strongly coupled gravity and duality}}, \href{https://doi.org/10.1016/S0550-3213(00)00323-0}{\emph{Nucl. Phys. B} {\bfseries 583} (2000) 237} [\href{https://arxiv.org/abs/hep-th/0004195}{{\ttfamily hep-th/0004195}}].

\bibitem{Hull:2001iu}
C.M.~Hull, \emph{{Duality in gravity and higher spin gauge fields}}, \href{https://doi.org/10.1088/1126-6708/2001/09/027}{\emph{JHEP} {\bfseries 09} (2001) 027} [\href{https://arxiv.org/abs/hep-th/0107149}{{\ttfamily hep-th/0107149}}].

\bibitem{Boulanger:2003vs}
N.~Boulanger, S.~Cnockaert and M.~Henneaux, \emph{{A note on spin s duality}}, \href{https://doi.org/10.1088/1126-6708/2003/06/060}{\emph{JHEP} {\bfseries 06} (2003) 060} [\href{https://arxiv.org/abs/hep-th/0306023}{{\ttfamily hep-th/0306023}}].

\bibitem{Blau:2001ne}
M.~Blau, J.M.~Figueroa-O'Farrill, C.~Hull and G.~Papadopoulos, \emph{{A New maximally supersymmetric background of IIB superstring theory}}, \href{https://doi.org/10.1088/1126-6708/2002/01/047}{\emph{JHEP} {\bfseries 01} (2002) 047} [\href{https://arxiv.org/abs/hep-th/0110242}{{\ttfamily hep-th/0110242}}].

\bibitem{Breitenlohner:1987dg}
P.~Breitenlohner, D.~Maison and G.W.~Gibbons, \emph{{Four-Dimensional Black Holes from Kaluza-Klein Theories}}, \href{https://doi.org/10.1007/BF01217967}{\emph{Commun. Math. Phys.} {\bfseries 120} (1988) 295}.

\bibitem{Galtsov:2008zz}
D.V.~Galtsov, \emph{{Generating solutions via sigma-models}}, \href{https://doi.org/10.1143/PTPS.172.121}{\emph{Prog. Theor. Phys. Suppl.} {\bfseries 172} (2008) 121} [\href{https://arxiv.org/abs/0901.0098}{{\ttfamily 0901.0098}}].

\bibitem{Hohm:2013jma}
O.~Hohm and H.~Samtleben, \emph{{U-duality covariant gravity}}, \href{https://doi.org/10.1007/JHEP09(2013)080}{\emph{JHEP} {\bfseries 09} (2013) 080} [\href{https://arxiv.org/abs/1307.0509}{{\ttfamily 1307.0509}}].

\bibitem{Ernst:1967wx}
F.J.~Ernst, \emph{{New formulation of the axially symmetric gravitational field problem}}, \href{https://doi.org/10.1103/PhysRev.167.1175}{\emph{Phys. Rev.} {\bfseries 167} (1968) 1175}.

\bibitem{deBoer:2010ud}
J.~de~Boer and M.~Shigemori, \emph{{Exotic branes and non-geometric backgrounds}}, \href{https://doi.org/10.1103/PhysRevLett.104.251603}{\emph{Phys. Rev. Lett.} {\bfseries 104} (2010) 251603} [\href{https://arxiv.org/abs/1004.2521}{{\ttfamily 1004.2521}}].

\bibitem{Kimura:2023nvt}
T.~Kimura, S.~Sasaki and K.~Shiozawa, \emph{{On geometries and monodromies for branes of codimension two}}, \href{https://doi.org/10.1016/j.physletb.2023.138425}{\emph{Phys. Lett. B} {\bfseries 849} (2024) 138425} [\href{https://arxiv.org/abs/2312.03358}{{\ttfamily 2312.03358}}].

\bibitem{Damour:2002cu}
T.~Damour, M.~Henneaux and H.~Nicolai, \emph{{E(10) and a `small tension expansion' of M theory}}, \href{https://doi.org/10.1103/PhysRevLett.89.221601}{\emph{Phys. Rev. Lett.} {\bfseries 89} (2002) 221601} [\href{https://arxiv.org/abs/hep-th/0207267}{{\ttfamily hep-th/0207267}}].

\bibitem{Damour:2004zy}
T.~Damour and H.~Nicolai, \emph{{Eleven dimensional supergravity and the E(10)/E(E10) sigma-model at low A(9) levels}},  in \emph{{25th International Colloquium on Group Theoretical Methods in Physics}}, 10, 2004 [\href{https://arxiv.org/abs/hep-th/0410245}{{\ttfamily hep-th/0410245}}].

\bibitem{Henneaux:2007ej}
M.~Henneaux, D.~Persson and P.~Spindel, \emph{{Spacelike Singularities and Hidden Symmetries of Gravity}}, \href{https://doi.org/10.12942/lrr-2008-1}{\emph{Living Rev. Rel.} {\bfseries 11} (2008) 1} [\href{https://arxiv.org/abs/0710.1818}{{\ttfamily 0710.1818}}].

\bibitem{Damour:2007dt}
T.~Damour, A.~Kleinschmidt and H.~Nicolai, \emph{{Constraints and the E10 coset model}}, \href{https://doi.org/10.1088/0264-9381/24/23/025}{\emph{Class. Quant. Grav.} {\bfseries 24} (2007) 6097} [\href{https://arxiv.org/abs/0709.2691}{{\ttfamily 0709.2691}}].

\bibitem{West:2001as}
P.C.~West, \emph{{E(11) and M theory}}, \href{https://doi.org/10.1088/0264-9381/18/21/305}{\emph{Class. Quant. Grav.} {\bfseries 18} (2001) 4443} [\href{https://arxiv.org/abs/hep-th/0104081}{{\ttfamily hep-th/0104081}}].

\bibitem{Tumanov:2015yjd}
A.G.~Tumanov and P.~West, \emph{{E$_{11}$ must be a symmetry of strings and branes}}, \href{https://doi.org/10.1016/j.physletb.2016.06.011}{\emph{Phys. Lett. B} {\bfseries 759} (2016) 663} [\href{https://arxiv.org/abs/1512.01644}{{\ttfamily 1512.01644}}].

\bibitem{Tumanov:2016abm}
A.G.~Tumanov and P.~West, \emph{{E11 in 11D}}, \href{https://doi.org/10.1016/j.physletb.2016.04.058}{\emph{Phys. Lett. B} {\bfseries 758} (2016) 278} [\href{https://arxiv.org/abs/1601.03974}{{\ttfamily 1601.03974}}].

\bibitem{Sugawara:2017fds}
H.~Sugawara, \emph{{Current Algebra Formulation of M-theory based on E11 Kac-Moody Algebra}}, \href{https://doi.org/10.1142/S0217751X17500245}{\emph{Int. J. Mod. Phys. A} {\bfseries 32} (2017) 1750024} [\href{https://arxiv.org/abs/1701.06894}{{\ttfamily 1701.06894}}].

\bibitem{Shiba:2017oaa}
S.~Shiba and H.~Sugawara, \emph{{M2- and M5-branes in E11 Current Algebra Formulation of M-theory}}, \href{https://doi.org/10.1142/S0217751X18500513}{\emph{Int. J. Mod. Phys. A} {\bfseries 33} (2018) 1850051} [\href{https://arxiv.org/abs/1709.07169}{{\ttfamily 1709.07169}}].

\bibitem{Glennon:2026tuk}
K.~Glennon, \emph{{On the Sugawara Current Algebra Proposal for M-Theory}},  \href{https://arxiv.org/abs/2603.11923}{{\ttfamily 2603.11923}}.

\bibitem{Hull:2004in}
C.M.~Hull, \emph{{A Geometry for non-geometric string backgrounds}}, \href{https://doi.org/10.1088/1126-6708/2005/10/065}{\emph{JHEP} {\bfseries 10} (2005) 065} [\href{https://arxiv.org/abs/hep-th/0406102}{{\ttfamily hep-th/0406102}}].

\bibitem{Hull:2006va}
C.M.~Hull, \emph{{Doubled Geometry and T-Folds}}, \href{https://doi.org/10.1088/1126-6708/2007/07/080}{\emph{JHEP} {\bfseries 07} (2007) 080} [\href{https://arxiv.org/abs/hep-th/0605149}{{\ttfamily hep-th/0605149}}].

\bibitem{Bergshoeff:1998bs}
E.~Bergshoeff and J.P.~van~der Schaar, \emph{{On M nine-branes}}, \href{https://doi.org/10.1088/0264-9381/16/1/002}{\emph{Class. Quant. Grav.} {\bfseries 16} (1999) 23} [\href{https://arxiv.org/abs/hep-th/9806069}{{\ttfamily hep-th/9806069}}].

\bibitem{Sato:1999bu}
T.~Sato, \emph{{A Ten form gauge potential and an M9 brane Wess-Zumino action in massive 11-D theory}}, \href{https://doi.org/10.1016/S0370-2693(00)00229-X}{\emph{Phys. Lett. B} {\bfseries 477} (2000) 457} [\href{https://arxiv.org/abs/hep-th/9912030}{{\ttfamily hep-th/9912030}}].

\bibitem{Sato:2000mw}
T.~Sato, \emph{{On dimensional reductions of the M-nine-brane}},  \href{https://arxiv.org/abs/hep-th/0003240}{{\ttfamily hep-th/0003240}}.

\bibitem{Bergshoeff:2006qw}
E.A.~Bergshoeff, M.~de~Roo, S.F.~Kerstan, T.~Ortin and F.~Riccioni, \emph{{IIA ten-forms and the gauge algebras of maximal supergravity theories}}, \href{https://doi.org/10.1088/1126-6708/2006/07/018}{\emph{JHEP} {\bfseries 07} (2006) 018} [\href{https://arxiv.org/abs/hep-th/0602280}{{\ttfamily hep-th/0602280}}].

\bibitem{Tumanov:2016dxc}
A.G.~Tumanov and P.~West, \emph{{E11, Romans theory and higher level duality relations}}, \href{https://doi.org/10.1142/S0217751X17500233}{\emph{Int. J. Mod. Phys. A} {\bfseries 32} (2017) 1750023} [\href{https://arxiv.org/abs/1611.03369}{{\ttfamily 1611.03369}}].

\bibitem{Cesaro:2024ipq}
M.~Ces{\`a}ro, A.~Kleinschmidt and D.~Osten, \emph{{Integrable auxiliary field deformations of coset models}}, \href{https://doi.org/10.1007/JHEP11(2024)028}{\emph{JHEP} {\bfseries 11} (2024) 028} [\href{https://arxiv.org/abs/2409.04523}{{\ttfamily 2409.04523}}].

\bibitem{Cesaro:2025msv}
M.~Ces{\`a}ro and D.~Osten, \emph{{Integrable deformations of dimensionally reduced gravity}}, \href{https://doi.org/10.1007/JHEP06(2025)064}{\emph{JHEP} {\bfseries 06} (2025) 064} [\href{https://arxiv.org/abs/2502.01750}{{\ttfamily 2502.01750}}].

\end{thebibliography}
\end{document}